\documentclass[11pt]{article}
\usepackage[english]{babel}

\usepackage{cite}
\usepackage{amsfonts,amsmath,amssymb}
\usepackage{enumerate}
\usepackage{hyperref}
\usepackage{bbm}
\usepackage{nicefrac}
\usepackage[all]{xy}
\usepackage{graphicx}
\usepackage{bm}
\usepackage{makecell}
\usepackage[table]{xcolor}
\usepackage{upgreek}
\usepackage{cancel}		
\usepackage{multicol}	
\usepackage{float}		
\usepackage{framed}
\usepackage{longtable}  
\usepackage{booktabs}
\usepackage{stmaryrd}
\usepackage{arydshln}

\usepackage{ytableau}
\ytableausetup{boxsize=0.8em,centertableaux}

\usepackage{titlesec}
\titleformat{\section}[block]{\Large\boldmath\bfseries}{\thesection}{1em}{}
\titleformat{\subsection}[block]{\large\boldmath\bfseries}{\thesubsection}{0.5em}{}

\usepackage{titletoc}
\titlecontents{section}
    [0pt]             										
    {\vspace{0.5cm}}                  					
    {\boldmath\bfseries \thecontentslabel\quad  }  
    {\bfseries\boldmath}         						
    {\hfill\contentspage}  

\usepackage[bottom]{footmisc}

\usepackage[top=2.5cm,bottom=2.5cm,inner=2.3cm,outer=2.3cm]{geometry}

\newcommand{\cN}{{\cal N}}

\newcommand{\tr}{\mbox{tr}}

\def\sst#1{{\scriptscriptstyle #1}}

\def\0{{\sst{(0)}}}
\def\1{{\sst{(1)}}}
\def\2{{\sst{(2)}}}
\def\3{{\sst{(3)}}}
\def\4{{\sst{(4)}}}
\def\5{{\sst{(5)}}}
\def\6{{\sst{(6)}}}
\def\7{{\sst{(7)}}}
\def\8{{\sst{(8)}}}

\newcommand{\bK}{{\bar{K}}}
\newcommand{\bL}{{\bar{L}}}
\newcommand{\bM}{{\bar{M}}}
\newcommand{\bN}{{\bar{N}}}
\newcommand{\bP}{{\bar{P}}}
\newcommand{\bQ}{{\bar{Q}}}
\newcommand{\bR}{{\bar{R}}}
\newcommand{\bS}{{\bar{S}}}

\newcommand{\bzero}{{\bar{0}}}
\newcommand{\balpha}{{\bar{\alpha}}}
\newcommand{\bbeta}{{\bar{\beta}}}

\def\dA{{\dot{A}}}
\def\dB{{\dot{B}}}

\newcommand{\vol}{\textrm{vol}}

\usepackage{xspace}
\usepackage{tikz}
\usetikzlibrary{calc}
\usepackage{dsfont}
\definecolor{JoliBleu}{rgb}{0,0.55,0.55}
\definecolor{JoliVert}{rgb}{0.15,0.6,0}
\definecolor{JoliRouge}{rgb}{0.86,0.08,0}
\definecolor{JoliJaune}{rgb}{1,0.75,0}
\definecolor{JoliGris}{rgb}{0.52,0.52,0.51}
\definecolor{myred}{RGB}{181, 73, 30}
\definecolor{myblack}{RGB}{43, 65, 82}
\definecolor{myblue}{RGB}{26, 77, 116}

\usepackage[font=small,labelfont=bf,labelsep=space]{caption}
\addto\captionsenglish{}
\addto\captionsenglish{}
\definecolor{darkred}{rgb}{0.65,0.15,0}
\hypersetup{pdfborder={0 0 0},colorlinks=true,urlcolor=darkred,citecolor=darkred,linkcolor=darkred,linktocpage=true}

\newcommand{\SO}{\ensuremath{\mathrm{SO}}\xspace}

\newcommand{\GL}{\ensuremath{\mathrm{GL}}\xspace}

\newcommand{\SU}{\ensuremath{\mathrm{SU}}\xspace}

\renewcommand{\S}{\ensuremath{\mathcal{S}}\xspace}

\renewcommand{\d}{\ensuremath{\mathrm{d}}\xspace}

\allowdisplaybreaks  

\usepackage{multirow}
\usepackage{rotating}
\usepackage{changepage}

\numberwithin{equation}{section}

\newcommand{\underdashed}[1]{
    \tikz[baseline=(todotted.base)]{
        \node[inner sep=1pt,outer sep=0pt] (todotted) {#1};
        \draw[dashed] (todotted.south west) -- (todotted.south east);
    }
}

\newcommand{\revA}{\protect\rotatebox[origin=c]{180}{A}\xspace}

\begin{document}

\begin{titlepage}

\begin{flushright}

IFT-UAM/CSIC-21-112 \\
LCTP-21-28\\
\today
\end{flushright}

\vspace{25pt}

   \begin{center}
   \baselineskip=16pt

   \begin{Large}

\mbox{ \bfseries \boldmath  Triality and the consistent reductions on AdS$_3\times S^3$}
   \end{Large}

\vspace{25pt}

{\large  Camille Eloy$^{1}$,\ Gabriel Larios$^{2,3}$ \,and\, Henning Samtleben$^{4,5}$}
		
\vspace{25pt}

	\begin{small}

	{\it $^{1}$ Theoretische Natuurkunde, Vrije Universiteit Brussel, and\\
	 the International Solvay Institutes, Pleinlaan 2, B-1050 Brussels, Belgium}  \\

	\vspace{10pt}
	
	{\it $^{2}$ Departamento de F\'\i sica Te\'orica and Instituto de F\'\i sica Te\'orica UAM/CSIC, \\
   Universidad Aut\'onoma de Madrid, Cantoblanco, 28049 Madrid, Spain}     \\

	\vspace{10pt}
	
	{\it $^{3}$ Leinweber Center for Theoretical Physics, University of Michigan, Ann Arbor, MI 48109, USA}     \\	

	\vspace{10pt}

	{\it $^{4}$ Univ Lyon, Ens de Lyon, CNRS, Laboratoire de Physique, F-69342 Lyon, France}  \\

	\vspace{10pt}

	{\it $^{5}$  Institut Universitaire de France (IUF)}  \\

	\end{small}

\vskip 50pt

\end{center}

\begin{center}
\textbf{Abstract}
\end{center}

\begin{quote}

We study compactifications on AdS$_3\times S^3$ and deformations thereof. 
We exploit the triality symmetry of the underlying duality group SO(4,4) of three-dimensional supergravity
in order to construct and relate new consistent truncations. For non-chiral $D=6$, ${\cal N}_{\rm 6d}=(1,1)$ supergravity,
we find two different consistent truncations to three-dimensional supergravity, retaining different subsets of
Kaluza-Klein modes, thereby offering access to different subsectors of the full nonlinear dynamics.
As an application, we construct a two-parameter family of AdS$_3\times M^3$ backgrounds on squashed spheres
preserving ${\rm U}(1)^2$ isometries. For generic value of the parameters, these solutions break all supersymmetries,
yet they remain perturbatively stable within a non-vanishing region in parameter space. They also contain 
a one-parameter family of ${\cal N}=(0,4)$ supersymmetric AdS$_3\times M^3$ backgrounds on 
squashed spheres with U(2) isometries. Using techniques from exceptional field theory, we determine the
full Kaluza-Klein spectrum around these backgrounds.

\end{quote}

\vfill

\end{titlepage}

\tableofcontents



\section{Introduction}

\addtocontents{toc}{\setcounter{tocdepth}{2}}


Consistent truncations are reductions of higher-dimensional supergravity to a finite number of Kaluza-Klein fluctuations whose full non-linear dynamics is consistently described by a lower-dimensional theory. In particular,  every solution to the lower-dimensional field equations allows an uplift to a solution of the higher-dimensional theory by virtue of the exact nonlinear embedding. 
The existence and the explicit construction of consistent truncations remains a powerful tool for the construction of higher-dimensional supergravity solutions. In the context of holographic dualities, consistent truncations ensure the validity of lower-dimensional supergravity computations, such as holographic correlation functions, marginal deformations, and renormalization group (RG) flows.

For extended  supergravity theories, consistent truncations used to be rare~\cite{deWit:1986iy,Nastase:1999kf,Cvetic:2000dm}, but the advent of new techniques from generalized and exceptional geometry~\cite{Aldazabal:2011nj,Geissbuhler:2011mx,Berman:2012uy,Lee:2014mla,Hohm:2014qga,Malek:2017njj,Hohm:2017wtr,Hassler:2019wvn} has revealed their existence for a wealth of supergravity backgrounds. 
In this framework, a consistent Kaluza-Klein reduction Ansatz is identified within the duality covariant formulation of the
higher-dimensional theory. Specifically, it is encoded in a group-valued Scherk-Schwarz twist matrix living on the duality group of the 
lower-dimensional theory. This formulation moreover allows a straightforward access to computing the entire Kaluza-Klein spectrum around any background living within the consistent truncation~\cite{Malek:2019eaz,Malek:2020yue}.
\medskip

\begin{figure}[bt]
	 \centering
	 \includegraphics[scale=0.4]{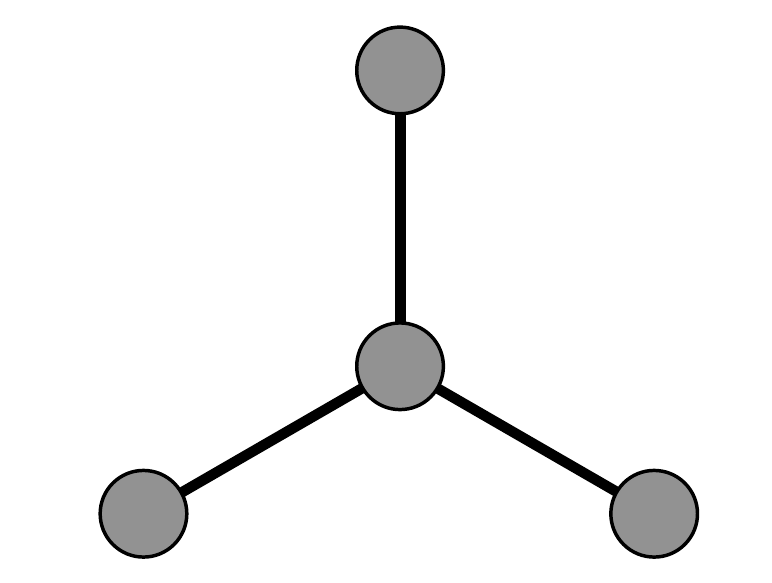}
	 \caption{${\rm SO}(8)$ triality.
	 }
	 \label{fig:so8}
\end{figure}

In this paper, we consider consistent truncations of chiral and non-chiral six-dimensional supergravity on AdS$_3\times M^3$ backgrounds, where here and in the following, $M^3$ denotes deformations of the three-sphere $S^3$. Their construction is based on the reformulation of $D=6$ supergravities as an exceptional field theory (ExFT) based on the group ${\rm SO}(4+m,4+n)$ \cite{Hohm:2017wtr}, with the integers $m, n$ depending on the matter content of the $D=6$ theory. 
In particular, the ${\rm SO}(4,4)$ theory describes the minimal  $D=6$, ${\cal N}_{\rm 6d}=(1,0)$ supergravity coupled to a single tensor multiplet.
Consistent truncations of these theories on a sphere $S^3$ are encoded in a Scherk-Schwarz twist matrix
\begin{equation}
{\cal U} ~\in~
{\rm GL}(4) \sim {\rm SO}(3,3)\times  {\rm SO}(1,1) \subset {\rm SO}(4,4) \subset  {\rm SO}(4+m,4+n)
\;,
\label{Utwisttrial}
\end{equation}
constructed in \cite{Hohm:2017wtr}. Here, we will exploit the triality structure of the group ${\rm SO}(4,4)$ and use its outer automorphisms (identified with the symmetries of the Dynkin diagram of figure~\ref{fig:so8}) to map the given twist matrix (\ref{Utwisttrial}) into new consistent twist matrices. While the resulting consistent truncations are equivalent within the group ${\rm SO}(4,4)$, i.e.\
within minimal $D=6$, ${\cal N}_{\rm 6d}=(1,0)$ supergravity, they give rise to inequivalent embeddings into ${\rm SO}(4+m,4+n)$, corresponding to inequivalent higher-dimensional origin. In particular, upon adding six-dimensional tensor and vector multiplets, respectively, minimal ${\cal N}_{\rm 6d}=(1,0)$ supergravity enhances to chiral ${\cal N}_{\rm 6d}=(2,0)$ 
and non-chiral ${\cal N}_{\rm 6d}=(1,1)$ supergravity, respectively. 
A triality flip of ${\rm SO}(4,4)$ within ${\rm SO}(4+m,4+n)$ relates the twist matrices describing the consistent truncation of the ${\cal N}_{\rm 6d}=(1,1)$ and the ${\cal N}_{\rm 6d}=(2,0)$ theory on $S^3$ to inequivalent three-dimensional supergravities~\cite{Hohm:2017wtr,Samtleben:2019zrh}, which we will refer to as A and B, respectively.\footnote{A similar duality has been analyzed for IIA/IIB compactifications to maximal $D=7$ supergravites \cite{Malek:2015hma}.} 

\begin{figure}[bt]
	 \centering
	 \includegraphics[scale=0.85]{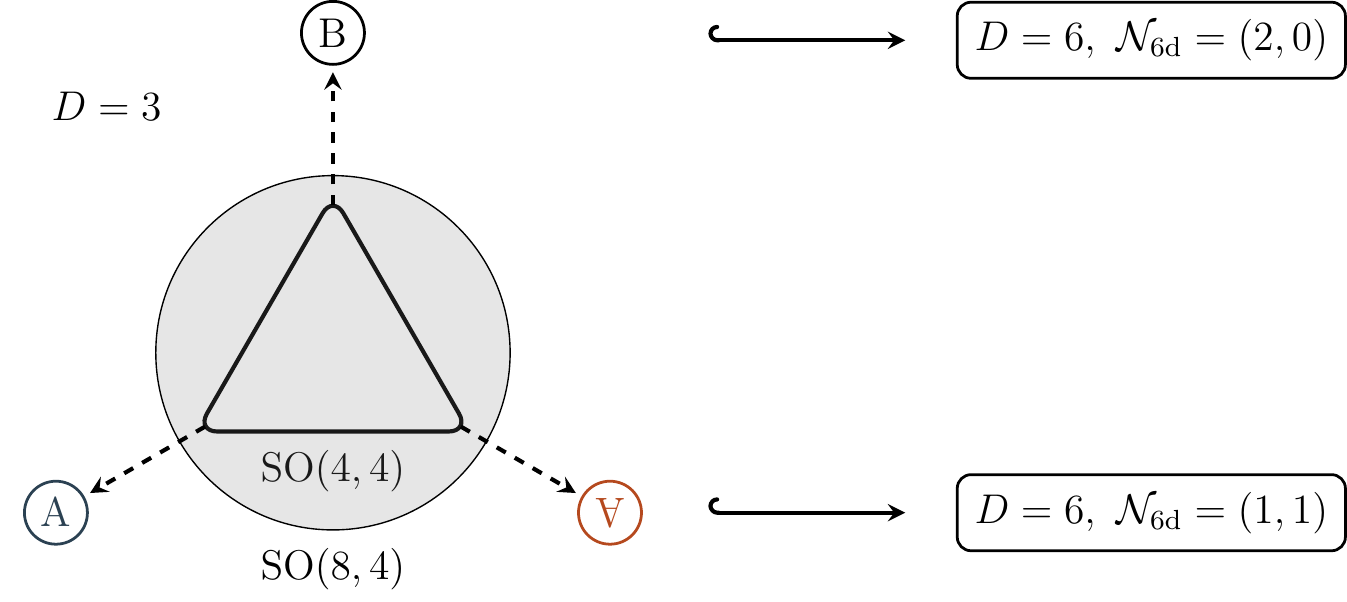}
	 \caption{Triality structure of the $D=3$ theories. The different embeddings of $\SO(4,4)$ are represented with dashed arrows, and the associated three-dimensional theories named A, B and \revA. These theories enjoy uplifts to $D=6$, with ${\cal N}_{\rm 6d}=(2,0)$ supersymmetries for B, and ${\cal N}_{\rm 6d}=(1,1)$ for A and \revA.
	 }
	 \label{fig:triality}
\end{figure}

Triality however allows for a third inequivalent embedding of ${\rm SO}(4,4)$ which we will explore in this paper. 
Remarkably, it describes a consistent truncation of the ${\cal N}_{\rm 6d}=(1,1)$ theory to a {\em different} set of Kaluza-Klein fluctuations. Accordingly, it results in a third, yet inequivalent, $D=3$ gauged supergravity (which we label by \revA), as sketched in figure~\ref{fig:triality}, which we will analyze here.

\begin{figure}[b]
	 \centering
	 \includegraphics[scale=1]{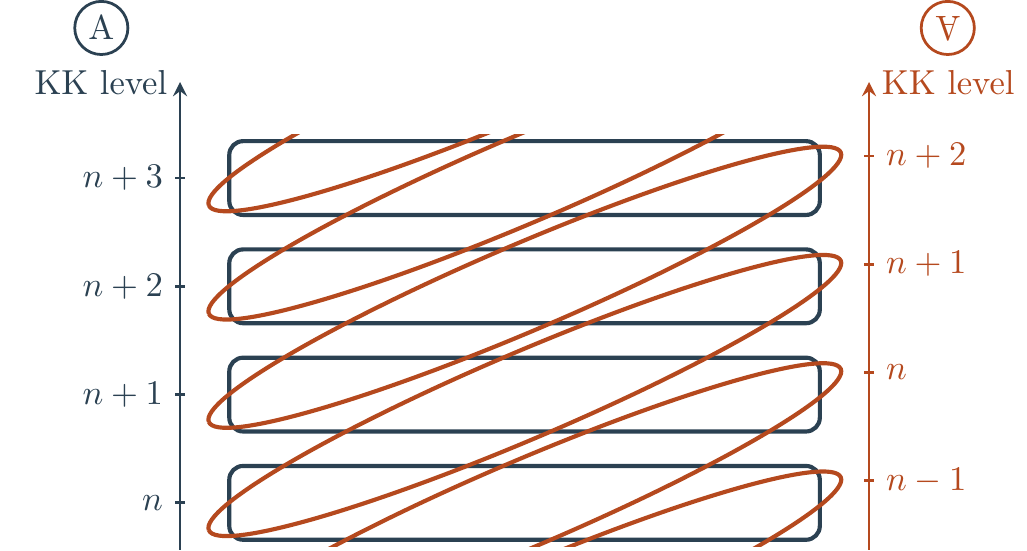}
	 \caption{Organization of the Kaluza-Klein spectrum according to different Kaluza-Klein levels.}
	 \label{fig:matching_schematic}
\end{figure}

The existence of different consistent truncations around a given background is a somewhat unusual phenomenon. It has previously been observed for the maximally supersymmetric backgrounds AdS$_4\times S^7$ \cite{Gauntlett:2009zw} and AdS$_5\times S^5$ \cite{Cassani:2010uw,Liu:2010sa,Gauntlett:2010vu,Skenderis:2010vz}, respectively. In both cases, the Sasaki-Einstein structure of the sphere gives rise to a second consistent truncation, not contained in the maximally supersymmetric one, to a gauged supergravity that preserves 
one-quarter (for $S^7$) and one-half (for $S^5$) of the supersymmetries, respectively. 
The two consistent truncations around AdS$_3 \times S^3$ which we study in this paper are related by triality and accordingly preserve the same amount of supersymmetry.
This implies that the full Kaluza-Klein spectrum around this background can be organized according to different gradings (or Kaluza-Klein levels) such that in each grading the full non-linear dynamics of the respective level $n=0$ fluctuations is described by a different lower-dimensional half-maximal theory. This is schematically sketched in figure~\ref{fig:matching_schematic}. Accordingly, the different consistent truncations give access to different subsectors of the full nonlinear dynamics.
 
\medskip

In our case, the full Kaluza-Klein spectrum of $D=6$, ${\cal N}_{\rm 6d}=(1,1)$ supergravity on AdS$_3\times S^3$ has been determined in \cite{Eloy:2020uix}. Its consistent truncation on $S^3$ to a three-dimensional ${\rm SO}(4)$ gauged supergravity corresponding to theory A of figure~\ref{fig:triality} has been studied extensively in \cite{Cvetic:2000dm,Deger:2014ofa,Samtleben:2019zrh}. This truncation retains 22 massive vector fluctuations and 10 of the scalar modes whose dynamics is described by a scalar potential in three dimensions. All stationary points of this potential have been identified in \cite{Samtleben:2019zrh}, each of which giving rise to an AdS$_3\times M^3$ solution of the six-dimensional theory. Apart from the supersymmetric AdS$_3\times S^3$ solution corresponding to the scalar origin, this analysis has revealed a number of discrete, non-supersymmetric solutions, all of which are unstable featuring scalar modes which violate the Breitenlohner-Freedman bound. In addition, the scalar potential exhibits a flat direction corresponding to a one-parameter family of non-supersymmetric AdS$_3\times M^3$ solutions~\cite{Deger:2019jtl}. In the holographic context such deformations correspond to marginal deformations of the dual conformal field theory. 

It is here, that the new consistent truncation \revA offers new insights. With the Kaluza-Klein spectrum reorganized as sketched in figure~\ref{fig:matching_schematic}, the consistent truncation \revA retains 10 massive vector fluctuations and 22 scalar modes. The additional scalar modes enhance the three-dimensional scalar potential which results in a richer vacuum structure. In particular, this potential exhibits two flat directions giving rise to a new two-parameter family of AdS$_3\times M_{\omega,\zeta}^3$ solutions, which we work out explicitly. For generic value of the parameters, these solutions break all supersymmetries. They contain as a subfamily the above discussed one-parameter family of theory A, but also a new one-parameter family of {\em supersymmetric} solutions AdS$_3\times M_\omega^3$, with a squashed sphere $M_\omega$ preserving ${\rm U}(2)$  isometries.

This supersymmetric family of six-dimensional AdS$_3$ vacua is controlled by a Sasaki-Einstein structure, in terms of which it has the simple form
\begin{equation}	\label{eq: susyconfiguration}
	\begin{aligned}
		&\d s^2=e^{\omega/2}\d s^2({\rm AdS}_3)+ e^{\omega/2}\d s^2(\mathbb{CP}^1)+e^{-3\omega/2}\,\bm{\eta}^2\,,	\qquad				
		&&e^{\phi}=e^{-\omega/2}	\,,															\\
		&G_{(3)}=2\,\vol({\rm AdS}_3)+2\,e^{-7\omega/4}\, \vol(M_{\omega}^3)\,,
		&&F_{(2)}= 2\sqrt{2}\, \sqrt{1-e^{-2\omega}}\;\bm{J}\,,	\qquad					
	\end{aligned}
\end{equation}
displaying a stretching of the Hopf fibre away from the scalar origin, and a constant dilaton $e^\phi$, with deformation parameter~$\omega>0$. Here, $\bm{\eta}$ and ${\bm J}=\frac12{\d \bm{\eta}}$ denote the contact 1-form and K\"ahler form on the base, respectively.
In particular, the solution crucially depends on a non-vanishing flux $F_{(2)}$ of the six-dimensional vector fields. As such, these solutions are not visible within the consistent truncation captured by the three-dimensional theory A, nor do they induce solutions of chiral ${\cal N}_{\rm 6d}=(2,0)$ supergravity in six dimensions.

\medskip

Using the work of \cite{Eloy:2020uix}, we compute the Kaluza-Klein spectrum around the two-parameter family of AdS$_3\times M_{\omega,\zeta}^3$ backgrounds within three-dimensional supergravity and at higher Kaluza-Klein levels, and determine its moduli dependence. Remarkably, the analysis exhibits a region in the two-dimensional parameter space $\{\omega, \zeta\}$ in which the spectrum remains perturbatively stable, although supersymmetry is entirely broken, see figure~\ref{fig:stabarea}. The resulting pattern of masses indicates that this stability persists to all Kaluza-Klein levels. We identify the modes that turn unstable as the parameters approach the boundary of the stability region, which all live at the lowest Kaluza-Klein level $n=0$, i.e.\ are already visible within three-dimensional supergravity.

\medskip

The rest of this paper is organized as follows. 
In section~2 we review the structure of three-dimensional gauged supergravities and give the details of the three theories A, B, and \revA, related by a triality rotation of the respective embedding tensors. We work out the scalar potential of theory~\revA as a function of its 22 scalar fields, and identify a two-parameter family of AdS$_3$ vacua connected to the scalar origin. For generic value of the parameters, these vacua break all supersymmetries and break the gauge symmetry down to ${\rm SO}(2)^2$. We determine the parameter-dependent spectrum of the three-dimensional theory around these vacua and identify the region in which all its scalar fluctuations remain perturbatively stable, i.e.\ exhibit masses above the Breitenlohner-Freedman bound. Inspection of the gravitino and vector masses shows that a one-parameter subfamily of AdS$_3$ vacua preserves chiral ${\cal N}=(0,4)$ supersymmetry with the unbroken gauge symmetry enhanced to ${\rm SO(2)}\times{\rm SO}(3)$.

In section~3, we work out the uplift of the two-parameter family of AdS$_3$ vacua to solutions of $D=6$, ${\cal N}_{\rm 6d}=(1,1)$ supergravity. To this end, we first review the reformulation of $D=6$, ${\cal N}_{\rm 6d}=(1,1)$ supergravity as an exceptional field theory based on the group ${\rm SO}(8,4)$. In this framework, the embedding of the three-dimensional supergravity is efficiently described as a generalized Scherk-Schwarz reduction. We present the relevant twist matrix and derive the explicit uplift formulas. In six dimensions, the solutions correspond to AdS$_3\times M_{\omega,\zeta}^3$ backgrounds with non-vanishing flux for vector and tensor fields.
Finally, in section~4, we use methods from exceptional field theory in order to extend the analysis of the Kaluza-Klein spectrum around these backgrounds to higher Kaluza-Klein levels. We exhibit the pattern indicating that within the stability region identified within three-dimensional supergravity, all higher Kaluza-Klein scalar modes remain also stable. For the supersymmetric family of vacua, we determine the ${\cal N}=(0,4)$ supermultiplet structure of the spectrum. We also compare the full Kaluza-Klein spectrum around the undeformed AdS$_3\times S^3$ backgrounds of theories A and \revA and show that they coincide upon reshuffling of the Kaluza-Klein towers as depicted in figure~\ref{fig:matching_schematic} and worked out in detail in figure~\ref{fig:matching} below. 
We close with concluding remarks and a few appendices that collect some of the technical details of our discussion.


\section{$\mathcal{N}=8$ supergravity in $D=3$}


Three-dimensional $\cN=8$ (half-maximal) gauged supergravity is described by the lagrangian \cite{Nicolai:2001ac,deWit:2003ja}
\begin{equation}	\label{eq: lagrangian} 
	e^{-1}\mathcal{L}=R+\frac1{8}g^{\mu\nu}D_\mu M^{\bM\bN}D_\nu M_{\bM\bN}+e^{-1}\mathcal{L}_{\text{CS}}-V\,,
\end{equation}
with the gauging specified by an embedding tensor of the form
\begin{equation}	\label{eq: embtensor}
	\Theta_{\bK\bL\vert\bM\bN}=\theta_{\bK\bL\bM\bN}+\frac12\Big(\eta_{\bM[\bK}\theta_{\bL]\bN}-\eta_{\bN[\bK}\theta_{\bL]\bM}\Big)+\theta\,\eta_{\bM[\bK}\eta_{\bL]\bN}\,,
\end{equation}
with antisymmetric $\theta_{\bK\bL\bM\bN}=\theta_{[\bK\bL\bM\bN]}$, symmetric and traceless $\theta_{\bL\bK}=\theta_{(\bL\bK)}$ and $\eta_{\bK\bL}$ the SO(8,4)-invariant metric.\footnote{This construction naturally generalizes to $\SO(8,n)$.} This metric is used to raise and lower indices $\bK,\bL,\dots$ in the vector representation of SO(8,4). The symmetric matrix $M_{\bK\bL}=\mathcal{V}_{\bK}{}^{\bM}\mathcal{V}_{\bL}{}^{\bN}\delta_{\bM\bN}$ parametrises the scalar coset
\begin{equation}	\label{eq: scalarcoset}
	\text{G}/\text{H}=\text{SO}(8,4)/\big(\text{SO}(8)\times\text{SO}(4)\big)\,,
\end{equation}
and the gauging requires vectors in the adjoint of SO(8,4) obeying Chern-Simons equations, and covariant derivatives given by
\begin{equation}
	D_\mu =\partial_\mu + A_\mu{}^{\bM\bN}\,\Theta_{\bM\bN\vert\bP\bQ}\, T^{\bP\bQ}\,,
\end{equation}
with the $\mathfrak{so}(8,4)$ generators
\begin{equation} \label{eq:so84gen}
\big(T^{\bar M\bar N}\big){}_{\bar P}{}^{\bar Q} = 2\,\delta_{\bar P}{}^{[\bar M}\,\eta^{\bar N]\bar Q}\,.
\end{equation}
The covariant derivative on $M_{\bM\bN}$ is then
\begin{equation}
	D_\mu M_{\bM\bN}=\partial_\mu M_{\bM\bN}+4\,A_\mu{}^{\bP\bQ}\,\Theta_{\bP\bQ\vert(\bM}{}^{\bK}\, M_{\bN)\bK}\,.
\end{equation}
The Chern-Simons term is given by
\begin{equation}
{\cal L}_{\rm CS}= -\varepsilon^{\,\mu\nu\rho}\,\Theta_{\bar M\bar N|\bar P\bar Q}\,A_{\mu}{}^{\bar M\bar N}\left(\partial_{\nu}\,A_{\rho}{}^{\bar P\bar Q}  + \frac{1}{3}\, \Theta_{\bar R\bar S|\bar U\bar V}\,f^{\bar P\bar Q,\bar R\bar S}{}_{\bar X\bar Y}\, A_{\nu}{}^{\bar U\bar V} A_{\rho}{}^{\bar X\bar Y} \right)\,,
\end{equation}
with $f^{\bar M\bar N,\bar P\bar Q}{}_{\bar K\bar L} = 4\,\delta_{[\bar K}{}^{[\bar M}\eta^{\bar N][\bar P}\delta_{\bar L]}{}^{\bar Q]}$ the structure constants of $\mathfrak{so}(8,4)$. Finally, the scalar potential in \eqref{eq: lagrangian} in terms of the parametrisation of $\Theta_{\bK\bL\vert\bM\bN}$ in \eqref{eq: embtensor} is given by~\cite{Schon:2006kz,Samtleben:2019zrh}\footnote{The additional constraint $\theta_{[\bK\bL\bM\bN}\theta_{\bP\bQ\bR\bS]}=0$ has been taken into account to guarantee that the $D=3$ theory can arise from a generalised Scherk-Schwarz reduction \cite{Hohm:2017wtr}.}
{\setlength\arraycolsep{1.2pt}
\begin{equation}	\label{eq: scalarpot}
	\begin{aligned}
	 V	&=	\frac1{12}\,\theta_{\bK\bL\bM\bN}\theta_{\bP\bQ\bR\bS}\Big(M^{\bK\bP}M^{\bL\bQ}M^{\bM\bR}M^{\bN\bS}-6\,M^{\bK\bP}M^{\bL\bQ}\eta^{\bM\bR}\eta^{\bN\bS}\\
	&\qquad\qquad\qquad\qquad\quad+8\,M^{\bK\bP}\eta^{\bL\bQ}\eta^{\bM\bR}\eta^{\bN\bS}-3\,\eta^{\bK\bP}\eta^{\bL\bQ}\eta^{\bM\bR}\eta^{\bN\bS}\Big)\\
	&\quad +\frac1{8}\,\theta_{\bK\bL}\theta_{\bP\bQ}\Big(2\,M^{\bK\bP}M^{\bL\bQ}-2\,\eta^{\bK\bP}\eta^{\bL\bQ}-M^{\bK\bL}M^{\bP\bQ}\Big)+4\,\theta\theta_{\bK\bL}M^{\bK\bL}-32\,\theta^2\,.
	\end{aligned}
\end{equation}
}

The fermions in this theory transform covariantly under the SO(8)$\times$SO(4) subgroup of SO(8,4) in~\eqref{eq: scalarcoset}, with the gravitini $\psi_\mu^A$ in the spinorial of SO(8), and the spin-$\nicefrac12$ fields $\chi^{\dA r}$ in the product of the cospinorial of SO(8) and the vector of SO(4). The remaining representation which will be relevant in our discussion is the vector of SO(8), which will be indexed by $I,J\in \llbracket1,8\rrbracket$, so that the $\SO(8,4)$ indices decompose into $\bar M=\{I,r\}$. The couplings of fermions among themselves and to the bosons are conveniently parametrised in terms of the \emph{dressed} embedding tensor
\begin{equation}	\label{eq: dressedembtens}
	T_{\bK\bL\vert\bM\bN}=(\mathcal{V}^{-1})_{\bK}{}^{\bP}(\mathcal{V}^{-1})_{\bL}{}^{\bQ}(\mathcal{V}^{-1})_{\bM}{}^{\bR}(\mathcal{V}^{-1})_{\bN}{}^{\bS}\,\Theta_{\bP\bQ\vert\bR\bS}\,,
\end{equation}
and the associated $T_{\bar K\bar L\bar M\bar N}, T_{\bar K\bar L}$ and $T$, following \eqref{eq: embtensor}.
Then, the vector and scalar masses are simply given by~\cite{deWit:2003ja,Deger:2019tem}\footnote{The factors in~\eqref{eq: bosonmasses}--\eqref{eq: fermionmasses} differ from the ones in \cite{deWit:2003ja,Deger:2019tem}, in line with the conventions we adopted for the lagrangian.}%
\begin{subequations}{\label{eq: bosonmasses}}
\begin{align}
	&\ \ \, M_{(1)\,Ir\vert Js}=-4\,T_{Ir\vert Js},	\\
	&\begin{aligned} 
    &M_{(0)\,{L}r\vert{M}s}^{2}=\delta_{LM}\bigg(-\frac{8}{3}\,T_{{IJK}r}T_{{IJK}s}+8\,T_{{IJ}rp}T_{{IJ}sp}-T_{II}T_{rs}+8\,T\,T_{rs}+T_{{I}r}T_{{I}s}+T_{pr}T_{ps}\bigg) \\
    &\quad{}+ \delta_{rs}\bigg(-\frac{8}{3}\,T_{IJKL}T_{IJKM}+8\,T_{{IJL}p}T_{{IJM}p} - T_{II}T_{LM}+\,T_{IL}T_{IM} +\,T_{{L}p}T_{{M}p}+8\,T\,T_{LM}\bigg) \\
    &\quad +16\, T_{{ILM}p}T_{{I}rsp}-16\,T_{{IL}sp}T_{{IM}rp} -2\,T_{{L}r}T_{{M}s}+2\,T_{{L}s}T_{{M}r}+2\,T_{LM}T_{rs}\,,
  \end{aligned}
\end{align}
\end{subequations}
and the fermion masses and gravitino--spin-$\nicefrac12$ couplings are given by the fermion shifts $A_{1,2,3}$
\begin{equation}	\label{eq: Ashifts}
	\begin{cases}
		A^{AB}_{1} = {-}\dfrac{1}{12}\,\Gamma^{IJKL}_{AB}\,T_{IJKL}{-}\,\dfrac1{4}\delta^{AB}\,T_{II}+2\,\delta^{AB}\,T\,,		\\[5pt]
		A^{A\dot{A}r}_{2} = -\dfrac{1}{3}\,\Gamma^{IJK}_{A\dA}\,T_{IJKr}-\dfrac12\,\Gamma^{I}_{A\dA}\,T_{Ir}\,,		\\[5pt]
		A^{\dA r\dB s}_{3} = \dfrac{1}{12}\,\delta^{rs}\Gamma^{IJKL}_{\dA\dB}\,T_{IJKL}+2\,\Gamma^{IJ}_{\dA\dB}\,T_{IJrs}-4\,\delta^{\dA\dB} \delta^{rs}T-2\,\delta^{\dot{A}\dot{B}}\,T_{rs}+\dfrac{1}{4}\,\delta^{\dot{A}\dot{B}} \delta^{rs}\,T_{II}\,,
	\end{cases}
\end{equation}
as
\begin{equation}	\label{eq: fermionmasses}
	M^{AB}_{(\nicefrac{3}{2})}= -A^{AB}_{1}\,,		\qquad\qquad
	M^{\dot{A}r\dot{B}s}_{(\nicefrac{1}{2})} = -A^{\dot{A}r\dot{B}s}_{3}\,.
\end{equation}
On a solution of the theory with negative constant potential $V_{0}$, it is convenient to normalise these masses with respect to the AdS length, defined as
\begin{equation}
	\ell^2_{\text{AdS}}=-\frac2{V_0}\,.
\end{equation}

\subsection{Triality and the different Pauli gaugings}

Minimal $D=6$, ${\cal N}_{\rm 6d}=(1,0)$ supergravity coupled to a tensor multiplet admits a reduction on $S^{3}$~\cite{Cvetic:2000dm}. The resulting three-dimensional theory is quarter-maximal (${\cal N}=4$) and enjoys a scalar coset space $\SO(4,4)/\left(\SO(4)\times\SO(4)\right)$. Its scalar potential does not have a ground state. The lagrangian is described in terms of an embedding tensor $\Theta_{\underline{K}\underline{L}\vert\underline{M}\underline{N}}$, which decomposes as in \eqref{eq: embtensor} and where $\underline{M}, \underline{N}$ denote the vectorial index of $\SO(4,4)$. In this case,
\begin{equation}\label{eq:repSO44}
	\theta_{\underline{M}\underline{N}} \subset \begin{ytableau} ~ & ~  \end{ytableau} = \bm{35_{\rm v}}, \quad \theta_{\underline{K}\underline{L}\underline{M}\underline{N}} \subset  \begin{ytableau} ~ \\ ~ \\ ~ \\ ~ \end{ytableau} = \bm{35_{\rm s}\oplus35_{\rm c}}\,,
\end{equation}
and we also take $\theta=0$. There exist three equivalent embedding tensors describing this $S^{3}$ reduction~\cite{Hohm:2017wtr,Baguet:2015iou}. They all trigger different representations within~\eqref{eq:repSO44}, as follows\footnote{The embedding tensors a and b were first given in \cite{Hohm:2017wtr}, and \rotatebox[origin=c]{180}{a} can be constructed from \cite{Baguet:2015iou}. See the explicit expressions~\eqref{eq:SO84embeddings} below.}
\begin{equation}
	\begin{array}{rccc}
	& \bm{35_{\rm v}} & \bm{35_{\rm s}} & \bm{35_{\rm c}} \\
	{\rm a} & \times & & \\
	{\rm b} & & & \times \\
	\rotatebox[origin=c]{180}{\rm a} & & \times &
	\end{array}
\end{equation}
These different possibilities are related by $\SO(4,4)$ triality flips and generate physically equivalent theories.

As shown in \cite{Deger:2014ofa}, the non-dynamical two-form gauge potential in $D=3$ can be integrated out from the three-dimensional theory, giving rise to a new parameter $\lambda$ in the embedding tensor which can be tuned so that the scalar potential admits a supersymmetric ${\rm AdS}_{3}$ solution at the origin. For the three embedding tensors just mentioned, this amounts to turning on new representations~\cite{Hohm:2017wtr,Eloy:2020uix}:
\begin{equation} \label{eq:schematicembeddingSO44}
	\begin{array}{rccc}
	& \bm{35_{\rm v}} & \bm{35_{\rm s}} & \bm{35_{\rm c}} \\
	{\rm a} & \times & \lambda & \\
	{\rm b} & & \lambda & \times \\
	\rotatebox[origin=c]{180}{\rm a} & \lambda & \times &
	\end{array}
\end{equation}
All possibilities are again related by $\SO(4,4)$ triality transformations and describe the same physics.

Upon embedding $\SO(4,4)$ into $\SO(8,4)$, \textit{i.e.} upon embedding the quarter-maximal theories into half-maximal three-dimensional supergravity, the same gaugings can be employed to describe consistent truncation of six-dimensional ${\cal N}_{\rm 6d}=(1,1)$ and ${\cal N}_{\rm 6d}=(2,0)$ supergravities, respectively. This embedding breaks the triality symmetry, as $\bm{8_{\rm v}}$ is singled out and embedded into the vector representation of $\SO(8,4)$. Then, the three embedding tensors schematically described in \eqref{eq:schematicembeddingSO44} give rise to three different half-maximal supergravities in three dimensions, which we will denote A, B and \rotatebox[origin=c]{180}{A}.\footnote{There exist three other possible theories that one can generate upon applying the triality flip $\bm{35_{\rm s}}\leftrightarrow\bm{35_{\rm c}}$ in \eqref{eq:schematicembeddingSO44}, but they are physically equivalent to A, B and \rotatebox[origin=c]{180}{A}.}
Their explicit embedding into six dimensions shows that theories A and $\revA$ descend from the non-chiral ${\cal N}_{\rm 6d}=(1,1)$ supergravity, whereas theory B is embedded into chiral ${\cal N}_{\rm 6d}=(2,0)$ supergravity, see~figure~\ref{fig:triality}.

To describe these gaugings, it is useful to break indices following
\begin{equation}	\label{eq: gl3gradingbar}
	\begin{aligned}
	\text{SO}(8,4)	&\longrightarrow	\enspace \text{GL}(3,\mathbb{R})\times\text{SO}(1,1)\times\text{SO}(4)_{\rm global}\,,	\\
	X^{\bM}		&\longrightarrow	\quad \{X^{\bar m},\; X_{\bar m},\; X^\bzero,\; X_\bzero,\; X^\balpha\}\,,
	\end{aligned}
\end{equation}
where $\bar m\in\llbracket1,3\rrbracket$ and $\balpha\in\llbracket9,12\rrbracket$ label the SL(3, $\mathbb{R}$) and SO(4)$_{\rm global}$ vector representations, respectively. In this basis, the SO(8,4)-invariant tensor reads
\begin{equation}	\label{eq: Paulieta}
		\eta_{\bM\bN}=
		\begin{pmatrix}
			0 & \delta_{\bar m}{}^{\bar n} & 0 & 0 &0\\
			\delta^{\bar m}{}_{\bar n} & 0 & 0 & 0 &0\\
			0 & 0 & 0 & 1 & 0\\
			0 & 0 & 1 & 0 & 0\\
			0 & 0 & 0 & 0 & -\delta_{\balpha\bbeta}
		\end{pmatrix}\,.
\end{equation}
The different embedding tensors have then the following expressions
\begin{subequations}{\label{eq:SO84embeddings}}
	\begin{align}
	({\rm A})&:\qquad \theta_{\bar m \bar n \bar p\,\bzero}=-2\,\lambda\,\varepsilon_{\bar m \bar n \bar p}\,,		\qquad	\theta_{\bar m \bar n}=4\,\delta_{\bar m \bar n}\,, \qquad	\theta_{\bzero\bzero}=4\,, \\
	({\rm B})&:\qquad\theta_{\bar m \bar n \bar p\,\bzero}=-2\,\lambda\,\varepsilon_{\bar m \bar n \bar p}\,,		\qquad	\theta_{\bar m \bar n \bar p}{}^{\bzero}=\varepsilon_{\bar m \bar n \bar p}\,, \qquad	\theta_{\bar m \bar n}{}^{\bar p}{}_{\bzero}=\varepsilon_{\bar m \bar n \bar p}\,, \\
	(\rotatebox[origin=c]{180}{\rm A})&:\qquad\theta_{\bM\bN\bP\,\bzero}=-\frac{1}{\sqrt{2}}\,X_{\bM\bN\bP}\,,		\qquad	\theta_{\bzero\bzero}=4\sqrt{2}\,\lambda\,,
	\label{eq:SO84embeddingsC}
	\end{align}
\end{subequations}
with 
\begin{equation}
	X_{\bar m\bar n \bar p} = \varepsilon_{\bar m \bar n \bar p}\,,		\qquad
	X_{\bar m}{}^{\bar n \bar p} = \varepsilon_{\bar m \bar n \bar p}\,,	\qquad
	X^{\bar m}{}_{\bar n}{}^{\bar p} = \varepsilon_{\bar m \bar n \bar p}\,,	\qquad
	X^{\bar m\bar n}{}_{\bar p} = \varepsilon_{\bar m \bar n \bar p}\,,
\end{equation}
and $\theta=0$ in all cases. Theories A and B have been described in \cite{Hohm:2017wtr} and further analysed in \cite{Samtleben:2019zrh,Eloy:2020uix}. The study of theory~\rotatebox[origin=c]{180}{A} is the main subject of this paper. 
Their scalar potentials have an extremal point at the origin for $\lambda=-1$, which is the value we will adopt in the following.

For theory \rotatebox[origin=c]{180}{A}, the embedding tensor (\ref{eq:SO84embeddingsC}) induces a gauge group
\begin{equation}	\label{eq: ggauge}
	\text{G}_{\text{gauge}}=(T^1)^4\times[\text{SO}(4)\ltimes(T^3\times T^3)]\,,
\end{equation}
with the SO(4) factor referred to in the following as SO(4)$_{\rm gauge}\simeq {\rm SO(3)}_{\text{L}}\times$ SO(3)$_{\text{R}}$, as detailed in appendix~\ref{sec: grouptheory}.

\subsection{Parametrisation of the scalar matrix}

Decomposing SO(8,4) according to \eqref{eq: gl3gradingbar}, we find
\begin{equation} \label{eq: branching}
	\begin{aligned}	
		\text{SO}(8,4) 	&\rightarrow
				[\text{SO}(1,1)+\text{GL}(3,\mathbb{R})+\text{SO}(4)_{\rm global}]_{(0,0)}	\\[3pt]
				&\quad+(\bm3,\bm1)_{(1,1)}+(\bm{3'},\bm1)_{(1,-1)}+(\bm1,\bm4)_{(1,0)}+(\bm{3'},\bm1)_{(0,2)}+(\bm3,\bm4)_{(0,1)} 	\\[3pt]
				&\quad+(\bm3',\bm4)_{(0,-1)}+(\bm3,\bm1)_{(0,-2)}+(\bm1,\bm4)_{(-1,0)}+(\bm3,\bm1)_{(-1,1)}+(\bm3',\bm1)_{(-1,-1)}\,,
	\end{aligned}
\end{equation}
with the elements of these representations in terms of the SO(8,4) generators given in appendix~\ref{sec: bosonicgrouptheory}. The compact part of the gauge group \eqref{eq: ggauge} does not sit at level zero of the grading, but involves the $(\bm3',\bm1)_{(0,2)}$ and $(\bm3,\bm1)_{(0,-2)}$ representations that together with GL(3$,\mathbb{R})$  comprise SO(3,3). In terms of the decomposition \eqref{eq: branching}, we can parametrise the coset \eqref{eq: scalarcoset} in triangular gauge as
\begin{equation}	\label{eq: cosetrepresentative}
	\mathcal{V}=\mathcal{V}_{10}\,\mathcal{V}_{(3,1)}\,\mathcal{V}_{(3,4)}\,\mathcal{V}_{\text{Gl(3)}}\,\mathcal{V}_{\text{SO(1,1)}}\,,
\end{equation}
where $\mathcal{V}_{10}=	\exp\big(\chi_{\bar{m}}\, T\,{}^{\bar{m}\bar{0}}+\chi^{\bar{m}}\, T\,{}_{\bar{m}}{}^{\bar{0}}+\chi_\balpha\, T\,{}^{\balpha\bar{0}}\big)$ can be gauge fixed to the identity employing the translations in \eqref{eq: ggauge}. Correspondingly, \eqref{eq: cosetrepresentative} can be written as
{\setlength\arraycolsep{2pt}
\begin{align}
	\mathcal{V}_{\bM}{}^{\bN}\vert_{\chi\to0}	
		&=	\exp\big(\phi_{\bar{m}\bar{n}} T\,{}^{\bar{m}\bar{n}}\big)\exp\big(\sqrt2\,\xi_{\bar{m}\balpha} T\,{}^{\bar{m}\balpha}\big)\mathcal{V}_{\text{Gl}(3)}\mathcal{V}_{\text{SO(1,1)}}			\nonumber\\[7pt]
		&=
		\begin{pmatrix}
			\nu_{\bar{m}}{}^{\bar{n}}	& [(\xi^2+\phi)\tilde{\nu}]_{\bar{m}\bar{n}}	&  0  &  0  &	-\sqrt2\,\xi_{\bar{m}\bbeta}	\\
			0					& \tilde{\nu}^{\bar{m}}{}_{\bar{n}}				&  0  &  0  &	0	\\
			0					&	0		&	e^{\tilde\varphi}		&	0				&	0	\\
			0	&	0	&	0				&	e^{-\tilde\varphi}	&	0	\\
			0	&	-\sqrt2\,[\xi^T\,\tilde\nu]_{\bar{n}\balpha}			&	0	&	0	&	\delta_\balpha^\bbeta	\\
		\end{pmatrix}\,,
\end{align}
}%
in terms of a GL(3$,\mathbb{R}$)/SO(3) representative, $\nu$ and $\tilde{\nu}=(\nu^{-1})^T$, together with the matrix product $(\xi^2)_{\bar{m}\bar{n}}=\xi_{\bar{m}\balpha}\, \xi_{\bar{n}\balpha}$.
The symmetric positive definite scalar matrix $M_{\bK\bL}=(\mathcal{V}\,\mathcal{V}^T)_{\bK\bL}$ is then given by
\begin{equation}	\label{eq: scalarmatrix}
	M_{\bM\bN}
		=
		\begin{pmatrix}
			m+(\xi^2+\phi)m^{-1}(\xi^2-\phi)+2\xi^2	& 	(\xi^2+\phi)m^{-1}		&	0	&  0  &	-\sqrt2\,[1+(\xi^2+\phi)m^{-1}]\xi	\\
			m^{-1}(\xi^2-\phi)	& 	m^{-1}	&	0  	&  0  &	-\sqrt2\,m^{-1}\xi	\\
			0		&	0		&	e^{2\tilde\varphi}	&	0				&	0		\\
			0		&	0		&	0				&	e^{-2\tilde\varphi}	&	0		\\
			-\sqrt2\,\xi^T[1+m^{-1}(\xi^2-\phi)]	&	-\sqrt2\,\xi^Tm^{-1}	&	0	&	0	&	1+2\,\xi^Tm^{-1}\xi	\\
		\end{pmatrix}\,,
\end{equation}
where $m_{\bar{m}\bar{n}}=(\nu\nu^T)_{\bar{m}\bar{n}}$.  With this parametrisation of the scalar coset and the embedding tensor in~\eqref{eq:SO84embeddingsC} with $\lambda=-1$, the scalar potential \eqref{eq: scalarpot} becomes
\begin{equation}
	\begin{aligned}
 		V & = 4\,e^{-4\tilde\varphi}+2\,e^{-2\tilde\varphi}\Big[-\tr\left(m+m^{-1}\right)+\tr\left(\phi m^{-1}\phi\right) -2\,\tr\left(\phi m^{-1}\xi^{2}\right)-2\,\tr\left(\xi^{2}\right)\\
 		&\qquad\quad-\tr\left(\xi^{2}m^{-1}\xi^{2}\right)  +\frac{1}{2}\,\det\left(m^{-1}\right)\left(1-\tr\left(\phi^{2}\right)-\tr\left(\xi^{4}\right)+\tr\left(\xi^{2}\right)^{2}\right) \\
 		&\qquad\quad +\frac{1}{2}\,{\rm T}\left(m^{-1}(\xi^{2}-\phi),(\xi^{2}+\phi)m^{-1},m+(\xi^{2}+\phi)m^{-1}(\xi^{2}-\phi)+2\,\xi^{2}\right)\\
 		&\qquad\quad +\frac{1}{4}\,{\rm T}\left(m^{-1},m+(\xi^{2}+\phi)m^{-1}(\xi^{2}-\phi)+2\,\xi^{2},m+(\xi^{2}+\phi)m^{-1}(\xi^{2}-\phi)+2\,\xi^{2}\right)\Big]\,, 
 	\end{aligned}
\end{equation}
where ${\rm T}\left(A,B,C\right)=\varepsilon_{mnp}\,\varepsilon_{qrs}\,A^{mq}B^{nr}C^{ps}$\,.

An interesting consistent truncation is achieved by requiring invariance under SO(3)$_{\text{diag}}\subset$ SO(4)$_\text{global}$. This truncation retains all SO(4)$_\text{global}$ singlets in \eqref{eq: cosetrepresentative} together with $\xi_{1\, 12}$, $\xi_{2\, 12}$ and $\xi_{3\, 12}$, thus keeping 13 scalars. In this sector we identify a bi-parametric family of extrema connected to the origin on which the scalars take the following vevs
\begin{equation}	\label{eq: etazetafamily}
	\tilde\varphi=0\,,
	\qquad
	m=\text{diag}(1,\,1,\,e^{-2\omega})\,,
	\qquad
	\phi_{\bar m\bar n}=\xi_{1\, 12}=\xi_{2\, 12}=0\,,
	\qquad
	\xi_{3\, 12}=\zeta\,.
\end{equation}
The coset representative corresponding to this family takes on the form
{\setlength\arraycolsep{2pt}
\begin{align}
	\mathcal{V}
		&=	\exp\Big(\sqrt2\,\zeta\, T\,{}^{\bar3\bar{12}}\Big)\exp\Big(-\omega\, T\,{}^{\bar3}{}_{\bar3}\Big)
		=\exp\Big[\Big(\frac{\sqrt2\,\omega\,\zeta}{1-e^{-\omega}}\Big)\,T\,{}^{\bar3\bar{12}}-\omega\, T\,{}^{\bar3}{}_{\bar3}\Big]	\nonumber\\[7pt]
		&=
		\left(
		\begin{array}{ccc;{2pt/2pt}ccc;{2pt/2pt}cc}
			1	&	0	&	0		&	0	&	0	&	0				&	0			&	0			\\
			0	&	1	&	0		&	0	&	0	&	0				&	0			&	0			\\
			0	&	0	&	e^{-\omega}	&	0	&	0	&	e^\omega\,\zeta^2		&	0			&	-\sqrt2\,\zeta	\\\hdashline[2pt/2pt]
			0	&	0	&	0		&	1	&	0	&	0				&	0			&	0			\\
			0	&	0	&	0		&	0	&	1	&	0				&	0			&	0			\\
			0	&	0	&	0		&	0	&	0	&	e^\omega			&	0			&	0			\\\hdashline[2pt/2pt]
			0	&	0	&	0		&	0	&	0	&	0				&	\mathbbm{1}_5	&	0			\\
			0	&	0	&	0		&	0	&	0	&	-\sqrt2\,e^\omega\,\zeta	&	0			&	1			\\
		\end{array}
		\right)\,,
\end{align}
}%
and the preserved gauge symmetry is the Cartan subgroup SO(2)$_{\text{L}}\times\SO(2)_{\text{R}}\subset$ SO(4)$_{\rm gauge}$. 
This solution describes the holographic conformal manifold of the CFT$_2$ at the scalar origin, whose leading order Zamolodchikov metric in the large-$N$ limit can be taken to be
\begin{equation}
	\d s^2_{\text{Zam.}}=\d \omega^2+e^{2\omega}\d\zeta^2\;.
\end{equation}

Similarly, we can truncate out all scalars but the SO(1,1)$\,\times\,$GL(4$,\mathbb{R}$)/SO(4) factor, where the field content of theories A and \revA coincide. Then, the potential reduces to
\begin{equation} \label{eq: Paulipot}
	\begin{aligned}	
	V	&=4\,e^{-4\tilde\varphi}+e^{-2\tilde\varphi}\Big[\det\left(m^{-1}\right)+\det(m)\,\tr\left(m^{-2}\right)-2\,\tr\big(m+m^{-1}\big) \\
	& +2\,\det\left(m^{-1}\right) \vec{\phi}\cdot\vec{\phi}-2\,\tr\left(m^{-1}\right)\vec{\phi}\cdot\vec{\phi} +4\,\vec{\phi}\,m^{-1}\vec{\phi}+\det\left(m^{-1}\right) \left(\vec{\phi}\cdot\vec{\phi}\right)^{2}\Big]\,.
\end{aligned}
\end{equation}
with $\vec{\phi}_{\bar p}\equiv\frac12\,\varepsilon_{\bar p\bar m\bar n}\,\phi_{\bar{m}\bar{n}}$\,.
This potential and the associated kinetic term in \eqref{eq: lagrangian} can be cast, up to unimportant global factors, into a form that matches (2.18) and (2.20) in \cite{Samtleben:2019zrh} under the dictionary
\begin{equation} \label{eq: Gl4dictionary}
	\alpha^2_{\,\sst{\text{there}}}=\lambda^2\,,
	\qquad 
	e^{\varphi_{\,\sst{\text{there}}}}=e^{-\tilde\varphi}\,,
	\qquad 
	\tilde{m}_{AB\,\sst{\text{there}}}=(\det m)^{-1/2}
		\begin{pmatrix}	
			m	&	-m\vec{\phi}		\\
			-\left(m\vec{\phi}\right)^{T}	&	\det(m) + \vec{\phi}\cdot m\vec{\phi}	
		\end{pmatrix}\,.
\end{equation}
The solution \eqref{eq: etazetafamily} with $\zeta=0$ then recovers the one-parameter family for theory A in that reference upon identifying $\eta_{\sst{\rm\, there}}=\omega_{\sst{\rm\, here}}$.

\subsection{Supergravity spectrum at the $(\omega,\zeta)$-family}

At generic points of the critical locus~\eqref{eq: etazetafamily}, the bosonic spectrum is given by bringing the scalar values to \eqref{eq: bosonmasses}. This yields
\begin{equation} \label{eq: vectorspectrumatetazeta}
 \begin{aligned}	
	m_{\sst{(1)}}\ell_{\rm AdS}:\quad &
	0\ [2]\,,	\quad
	-2\ [5]\,,	\quad
	2\ [1]\,,	\\[5pt]
	&	-1\pm\sqrt{-1+2\,\zeta^2+e^{-2\omega}+e^{2\omega}\,\left(-1+\zeta^2\right)^2}\ [2+2]\,,	\\[5pt]
	&	1\pm\sqrt{-1+2\,\zeta^2+e^{-2\omega}+e^{2\omega}\,\left(1+\zeta^2\right)^2}\ [2+2]\,,
 \end{aligned}
\end{equation}
for the vectors fields. The integers between square brackets indicate the multiplicity of each eigenvalue. We include explicitly the two massless vectors corresponding to the unbroken ${\rm SO}(2)_{\rm L}\times {\rm SO}(2)_{\rm R}$ gauge symmetry, although in $D=3$ they and the massless graviton (together with possible massless gravitini) are non-propagating. 
For the scalars, the spectrum with Goldstone modes removed is 
\begin{equation} \label{eq: scalarspectrumatetazeta}
 \begin{aligned}	
	\left(m_{\sst{(0)}}\ell_{\rm AdS}\right)^2:\quad&
	0\ [5]\,,	\quad
	8\ [1]\,,	\quad
	-4+4\,e^{2\omega}\ [2]\,,	\quad
	-4+4\,e^{-2\omega}+8\,\zeta^{2}+4\,e^{2\omega}\,\zeta^{4}\ [2]\,,	\\[5pt]
	&	e^{-2\omega}+2\,\left(-1+\zeta^2\right)+e^{2\omega}\,\left(1+\zeta^2\right)^2\ [8]\,,
 \end{aligned}
\end{equation}
with the masses in the first line of \eqref{eq: scalarspectrumatetazeta} matching (2.33) of \cite{Samtleben:2019zrh} upon setting $\zeta=0$.
Regarding fermions, their masses are accordingly given by bringing \eqref{eq: etazetafamily} to \eqref{eq: fermionmasses}, resulting in
\begin{equation}	\label{eq: gravitinospectrumatetazeta}
	m_{{(\nicefrac32)}}\ell_{\rm AdS}:\quad 
	\quad \dfrac1{2}\bigg[-1\pm\sqrt{e^{-2\omega}+2\,(1+\zeta^2)+e^{2\omega}\left(\zeta^2-1\right)^2}\bigg]\ [4+4]\,,
\end{equation}
for the spin-$\nicefrac32$  gravitini, and
\begin{align}	\label{eq: s12spectrumatetazeta}
	m_{{(\nicefrac12)}}\ell_{\rm AdS}:	\quad & -\frac3{2}\pm\frac1{2}\sqrt{e^{-2\omega}+2\,(1+\zeta^2)+e^{2\omega}\,\left(-1+\zeta^2\right)^2}\ [4+4]\,,	 \nonumber\\[5pt]
	&\frac1{2}\pm\frac1{2}\sqrt{9\,e^{-2\omega}+6\,(-1+3\,\zeta^2)+e^{2\omega}\,\left(1+3\,\zeta^2\right)^2}\ [4+4]\,,	\nonumber\\[5pt]
	&\frac1{2}\pm\frac1{2}\sqrt{e^{-2\omega}+2\,(-3+\zeta^2)+e^{2\omega}\,\left(3+\zeta^2\right)^2}\ [4+4]\,,
\end{align}
for the spin-$\nicefrac12$ fields, with goldstino modes already subtracted.\footnote{The goldstino modes are identified using Eq.~(6.9) of Ref.~\cite{deWit:2003ja}.}

\begin{figure}[b!]
 \centering
 \centerline{\includegraphics[scale=1]{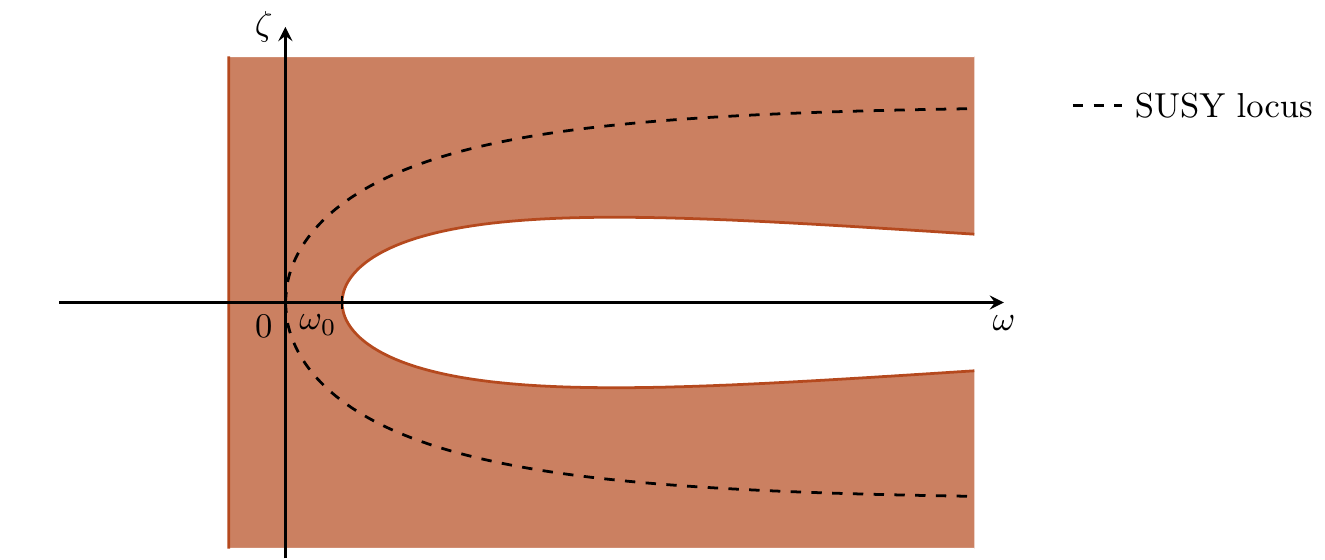}}
 \caption{Stability area at level 0, with $\omega_{0}=\ln\left(\sqrt{3}/2\right)$ saturating \eqref{eq: BFbound}.}
 \label{fig:stabarea0}
\end{figure}

\paragraph*{}
From the scalar spectrum \eqref{eq: scalarspectrumatetazeta}, stability against the Breitenlohner-Freedman bound \cite{Breitenlohner:1982jf}, 
in three dimensions given by $\left(m_{(0)}\ell_{\rm AdS}\right)^2\geq-1$, requires 
\begin{equation} \label{eq: BFbound}
	\begin{cases}
	e^{\omega} \geq \dfrac{\sqrt3}2, \smallskip\\
	\zeta^2 \geq \dfrac{\sqrt3}2\,e^{-\omega}-e^{-2\omega}\,,
	\end{cases}
\end{equation}
which  for non-zero $\zeta$ widens the stability region of equation (2.35) in \cite{Samtleben:2019zrh}. The corresponding stability region is drawn in figure~\ref{fig:stabarea0} in the $(\omega,\zeta)$ plane. The modes turning unstable on its boundary are found in the first line of \eqref{eq: scalarspectrumatetazeta} with masses $-4+4\,e^{2\omega}$, and $-4+4\,e^{-2\omega}+8\,\zeta^{2}+4\,e^{2\omega}\,\zeta^{4}$, respectively.

\paragraph*{}
At the scalar origin, where $\omega=\zeta=0$, \eqref{eq: gravitinospectrumatetazeta} shows supersymmetry enhancement to $\mathcal{N}=(0,4)$, and the bosonic symmetry becomes
\begin{equation}	\label{eq: SO4xSO4group}
	\text{SO(4)}_\text{gauge}\times\text{SO(4)}_{\text{global}}\,.
\end{equation}
For generic values of $\omega$ and $\zeta$, supersymmetry is completely broken, and the two massless vectors in~\eqref{eq: vectorspectrumatetazeta} signal that the symmetry group is broken down to
\begin{equation}	\label{eq: SO2xSO2xSO3group}
	\big[\SO(2)_{\text{L}}\times\SO(2)_{\text{R}}\big]_{\text{gauge}}\times\big[\SO(3)_{\text{diag}}\big]_{\text{global}}\,,
\end{equation}	
with $\SO(2)_{\text{L}}\times\SO(2)_{\text{R}}$ the Cartan subgroup of $\text{SO(4)}_\text{gauge}$ and $\SO(3)_{\text{diag}}$ being the diagonal subgroup of $\text{SO(4)}_{\text{global}}\simeq\text{SO}(3)_{\text{l}}\times\text{SO}(3)_{\text{r}}$ in the notation of appendix~\ref{sec: bosonicgrouptheory}.

At the particular locus 
\begin{equation}	\label{eq:susylocus}
	\zeta^2=1-e^{-2\omega}\,,
\end{equation}
the appearance of massless gravitini in \eqref{eq: gravitinospectrumatetazeta} shows that the $\mathcal{N}=(0, 4)$ supersymmetry of the scalar origin extends along a one-parameter family of vacua and four goldstini become part of the physical spectrum. Correspondingly, the residual gauge group enhances from $\SO(2)_{\text{L}}\times\SO(2)_{\text{R}}$ in \eqref{eq: SO2xSO2xSO3group} to $\SO(2)_{\text{L}}\times\SO(3)_{\text{R}}$ and two extra massless scalars arise in \eqref{eq: scalarspectrumatetazeta}. The full superalgebra controlling the spectrum is
\begin{equation}	\label{eq: susylocussuperalgebra}
	\text{SL}(2,\,\mathbb{R})_{\text{L}}\times \text{SO}(2)_{\text{L}}\times\big[(\text{SO}(3)_{\text{diag}})_{\text{global}}\ltimes \text{SU}(2\vert1,1)_{\text{R}}\big]\,, 
\end{equation}
with $\text{SL}(2,\,\mathbb{R})_{\text{R}}\times \text{SO}(3)_{\text{R}}$ the even part of $\text{SU}(2\vert1,1)_{\text{R}}$. Supermultiplets of \eqref{eq: susylocussuperalgebra} are thus labelled by two dimensions $\Delta_{\text{L}}$ and $\Delta_{\text{R}}$, two SO(3) half-integer spins and a charge normalised to be integer. They will be denoted by supplementing the $(\text{SO}(3)_{\text{diag}})_{\text{global}}\ltimes\text{SU}(2\vert1,1)_{\rm R}$ multiplets \eqref{eq: zerolongmult} and \eqref{eq: lowshortmult} with the extra SL$(2,\,\mathbb{R})_\text{L}\times\SO(2)_\text{L}$ flavours as subindices and superindices, respectively.
The SL$(2,\,\mathbb{R})$ dimensions can be determined out of the conformal dimension and spin of the different modes via
\begin{equation}
	\Delta=\Delta_{\text{L}}+\Delta_{\text{R}}\,,		\qquad\text{and}\qquad
	s=\Delta_{\text{R}}-\Delta_{\text{L}}\,,
\end{equation}
with conformal dimensions and masses related as
\begin{equation}
	\begin{aligned}
		\Delta_{\2}(\Delta_{\2}-2)=(m_{\2}\ell_{\rm AdS})^2\,,&	\qquad
		\Delta_{\1}=1+\vert m_\1\ell_{\rm AdS}\vert\,,		\qquad
		\Delta_{\0}(\Delta_{\0}-2)=(m_{\0}\ell_{\rm AdS})^2\,,	\\[7pt]
		\Delta_{{(\nicefrac32)}}=&1+\vert m_{{(\nicefrac32)}}\ell_{\rm AdS}\vert\,,		\qquad
		\Delta_{{(\nicefrac12)}}=1+\vert m_{{(\nicefrac12)}}\ell_{\rm AdS}\vert\,,
	\end{aligned}
\end{equation}
and the sign of the spin fixed by that of the mass. The spectrum of the three-dimensional supergravity around this family of supersymmetric vacua is given in table~\ref{tab:multsusyeta}.

\begin{table}[t!]
 \renewcommand{\arraystretch}{1.4}
 \centering
 \small
 \begin{tabular}{cccccc}
	&$\Delta_{\rm L}$ & $\Delta_{\rm R}$ & $\Delta$ & $s$ & $\Big(\SO(3)_{\rm diag}\times\SO(3)_{\rm R}\Big)^{\SO(2)_{\rm L}}$ \\\hline\hline
	$([0,0]_{\rm S})^0_2$ &	$2$ & $0$ & $2$ & $-2$ & $\big(0,0\big)^{0}$ \\\hline
	\multirow{2}{*}{$\left(\left[\tfrac12,\tfrac12\right]_{\rm S}\right)^0_2$} &\multirow{2}{*}{$2$} & $1$	& $3$ & $-1$ & $\big(0,0\big)^{0}+\big(1,0\big)^{0}$  \\
	&& $\nicefrac{1}{2}$ & $\nicefrac{5}{2}$ & $-\nicefrac{3}{2}$ & $\big(\nicefrac{1}{2},\nicefrac{1}{2}\big)^{0}$  \\\hline
	\multirow{3}{*}{$([0,1]_{\rm S})^0_2$} &\multirow{3}{*}{$2$} & $2$ & $4$ & $0$ & $\big(0,0\big)^{0}$ \\
	&& $\nicefrac{3}{2}$ & $\nicefrac{7}{2}$ & $-\nicefrac{1}{2}$ & $\big(\nicefrac{1}{2},\nicefrac{1}{2}\big)^{0}$  \\
	&& $1$ & $3$ & $-1$ & $\big(0,1\big)^{0}$  \\\hline
	$([0,0]_{\rm S})^0_1$ &$1$ & $0$ & $1$ & $-1$ & $\big(0,0\big)^{0}$ \\\hline
	\multirow{2}{*}{$\left(\left[\tfrac12,\tfrac12\right]_{\rm S}\right)^0_1$} &\multirow{2}{*}{$1$} & $1$ & $2$ & $0$ & $\big(0,0\big)^{0}+\big(1,0\big)^{0}$  \\
	&& $\nicefrac{1}{2}$ & $\nicefrac{3}{2}$ & $-\nicefrac{1}{2}$ & $\big(\nicefrac{1}{2},\nicefrac{1}{2}\big)^{0}$  \\\hline
	\multirow{3}{*}{$([0,1]_{\rm S})^0_1$} &\multirow{3}{*}{$1$} & $2$ & $3$ & $1$ & $\big(0,0\big)^{0}$ \\
	&& $\nicefrac{3}{2}$ & $\nicefrac{5}{2}$ & $\nicefrac{1}{2}$ & $\big(\nicefrac{1}{2},\nicefrac{1}{2}\big)^{0}$  \\
	&& $1$ & $2$ & $0$ & $\big(0,1\big)^{0}$  \\\hline
	\multirow{3}{*}{$([0,1]_{\rm S})^0_0$} &\multirow{3}{*}{$0$} & $2$ & $2$ & $2$ & $\big(0,0\big)^{0}$ \\
	&& $\nicefrac{3}{2}$ & $\nicefrac{3}{2}$ & $\nicefrac{3}{2}$ & $\big(\nicefrac{1}{2},\nicefrac{1}{2}\big)^{0}$  \\
	&& $1$ & $1$ & $1$ & $\big(0,1\big)^{0}$  \\\hline
	\multirow{4}{*}{$[0,0]^{\pm2}_{\tfrac{(1+\Gamma^{(0,\pm 2)})}{2}}$} &\multirow{5}{*}{$\nicefrac{(1+\Gamma^{(0,\pm 2)})}{2}$} & $\nicefrac{(3+\Gamma^{(0,\pm 2)})}{2}$ & $2+\Gamma^{(0,\pm 2)}$ & $1$ & $\big(0,0\big)^{\pm2}$ \\
	&& $\nicefrac{(2+\Gamma^{(0,\pm 2)})}{2}$ & $\dfrac{3}{2}+\Gamma^{(0,\pm 2)}$ & $\nicefrac{1}{2}$ & $\big(\nicefrac{1}{2},\nicefrac{1}{2}\big)^{\pm2}$  \\
	&& $\nicefrac{(1+\Gamma^{(0,\pm 2)})}{2}$ & $1+\Gamma^{(0,\pm 2)}$ & $0$ & $\big(1,0\big)^{\pm2}+\big(0,1\big)^{\pm2}$  \\
	&& $\nicefrac{\Gamma^{(0,\pm 2)}}{2}$ & $\dfrac{1}{2}+\Gamma^{(0,\pm 2)}$ & $-\nicefrac{1}{2}$ & $\big(\nicefrac{1}{2},\nicefrac{1}{2}\big)^{\pm2}$  \\
	&& $\nicefrac{(-1+\Gamma^{(0,\pm 2)})}{2}$ & $\Gamma^{(0,\pm 2)}$ & $-1$ & $\big(0,0\big)^{\pm2}$
  \end{tabular}
  \caption{Spectrum at KK level 0 of the family of supersymmetric vacua satisfying \eqref{eq:susylocus}, with $\Gamma^{(0,\pm 2)}=\sqrt{-3+4\,e^{2\omega}}$. Non-propagating massless modes in $([0,0]_{\rm S})^0_2$, $([0,0]_{\rm S})^0_1$ and $([0,1]_{\rm S})^0_0$ have been included for completeness.}
  \label{tab:multsusyeta}
\end{table}

\section{$D=6$ uplift}

 The field content of $\mathcal{N}_{\rm 6d}=(1,1)$ supergravity in $D=6$ comprises  the metric, the dilaton, four one-forms and a two-form~\cite{Giani:1984dw},
\begin{equation}	\label{eq: 6Dfields}
	\{g_{\hat\mu\hat\nu},\; \phi,\; A^\alpha{}_{\hat\mu},\;  B_{\hat\mu\hat\nu}\}\,,
\end{equation}
with indices $\hat\mu,\hat\nu\in\llbracket0,5\rrbracket$ and $\alpha\in\llbracket9,12\rrbracket$. The field strengths associated to these gauge potentials are
\begin{equation}	\label{eq: fieldstrenghts}
	F^\alpha=\d A^\alpha\qquad  {\rm and}\qquad   G=\d B-\dfrac{1}{2}\, A^{\alpha}\wedge F^{\alpha}\,, 
\end{equation}
 and the dynamics of these fields is described by the lagrangian
\begin{equation}	\label{eq: 6Dlagrangian}
	\mathcal{L}=\star_6\, R^{\6} - \d\phi\wedge\star_6\,\d\phi-\frac{1}{2}\,e^{-\phi}F^\alpha\wedge\star_6F^\alpha-\frac{1}{2}\,e^{-2\phi}G\wedge\star_6 G\,.
\end{equation}
From this lagrangian, the equations of motion for the dilaton and forms are
\begin{subequations}		\label{eq: 6Dequations}
	\begin{align}
		\nabla_{\hat\mu}\left(e^{-2\phi}\, G^{\hat\mu\hat\nu\hat\rho}\right)&=0\,,\\
		\nabla_{\hat\mu}\left(e^{-\phi}\,F^{\alpha\,\hat\mu\hat\rho}\right) +\frac{1}{2}\,e^{-2\phi}\, G^{\hat\mu\hat\nu\hat\rho}\,F^{\alpha}{}_{\hat\mu\hat\nu}&=0\,, \\
		\Box\phi+\frac{1}{8}\,e^{-\phi}\,F^{\alpha}{}_{\hat\mu\hat\nu}F^{\alpha\,\hat\mu\hat\nu}+\frac{1}{12}\,e^{-2\phi}\, G_{\hat\mu\hat\nu\hat\rho} G^{\hat\mu\hat\nu\hat\rho}&=0\,,
	\end{align}
\end{subequations}
and the Einstein equation reads
\begin{equation}	\label{eq: EinsteinEq}
	\begin{aligned}
	 R^{(6)}_{\hat\mu\hat\nu}-\frac{1}{2}\,R^{(6)}\,g_{\hat\mu\hat\nu} &= \partial_{\hat \mu}\phi\,\partial_{\hat\nu}\phi-\frac{1}{2}\,g_{\hat\mu\hat\nu}\,\partial_{\hat\rho}\phi\,\partial^{\hat\rho}\phi+ \frac{1}{2} \, e^{-\phi}\left(F^{\alpha}_{\hat\mu\hat\rho}F^{\alpha}{}_{\hat\nu}{}^{\hat\rho}-\frac{1}{4}\,g_{\hat\mu\hat\nu}\,F^{\alpha}_{\hat\rho\hat\sigma}F^{\alpha\,\hat\rho\hat\sigma}\right) \\ 
	 & \quad +\frac{1}{4}\,e^{-2\phi}\left( G_{\hat\mu\hat\rho\hat\sigma} G_{\hat\nu}{}^{\hat\rho\hat\sigma}-\frac{1}{6}\,g_{\hat\mu\hat\nu}\, G_{\hat\rho\hat\sigma\hat\lambda} G^{\hat\rho\hat\sigma\hat\lambda}\right)\,.
	\end{aligned}
\end{equation}

\subsection{ExFT recap}

We are interested here in the SO(8,4)-covariant formulation of $\mathcal{N}_{\rm 6d}=(1,1)$ supergravity in $D=6$, which is based on one of the solutions to the section constraint of SO(8,4) exceptional field theory~\cite{Hohm:2017wtr,Samtleben:2019zrh}. In this ExFT, we have the fields
\begin{equation}	\label{eq: ExFTfields}
	\{g_{\mu\nu},\ \mathcal{M}_{MN},\, \mathcal{A}_\mu^{MN},\, \mathcal{B}_{\mu MN}\}\,,
\end{equation}
with all them depending on both the external $x^\mu$ and the internal coordinates $Y^{MN}$, the latter in the adjoint of SO(8,4). The dependence of the fields on the internal coordinates is restricted to be on, at most, three of them by the section conditions
\begin{equation}	\label{eq: sectionconstr}
	\partial_{[MN}\otimes\partial_{KL]}=0=\eta^{NK}\partial_{MN}\otimes\partial_{KL}\,.
\end{equation}
The solution of \eqref{eq: sectionconstr} that describes $D=6$ $\mathcal{N}_{\rm 6d}=(1,1)$ supergravity is given by branching SO(8,4) as
\begin{equation}	\label{eq: gl3grading}
	\begin{aligned}
	 \text{SO}(8,4)	&\longrightarrow\text{GL}(3)\times\text{SO}(1,1)\times\text{SO}(4)\,,\\
	X^{M}		&\longrightarrow	\{X^{m},\; X_{m},\; X^0,\; X_0,\; X^\alpha\}\,,	
	\end{aligned}
\end{equation}
and keeping dependence on $y^m\equiv Y^{0m}$ only. Breaking the fields in \eqref{eq: ExFTfields} according to the branching~\eqref{eq: gl3grading} then recovers the fields of $D=6$ $\mathcal{N}_{\rm 6d}=(1,1)$ supergravity in \eqref{eq: 6Dfields}. 
In particular, the internal components of these $D=6$ fields are encoded in the generalised metric as \cite{Samtleben:2019zrh}\footnote{The r\^ole of $b$ and $\tilde{b}$ has been interchanged as compared with \cite{Samtleben:2019zrh}.}
\begin{equation}	\label{eq: ExFTdictionary}
	\begin{aligned}
	 \mathcal{M}^{00}&=g^{-1}e^{\phi}\,,	\\[4pt]
	\mathcal{M}^{0m}&=-\frac12\,\mathcal{M}^{00}\varepsilon^{mnp}\, \tilde{b}_{np}\,,	\\[4pt]
	\mathcal{M}^{00}\mathcal{M}^{mn}-\mathcal{M}^{0m}\mathcal{M}^{0n}&=g^{-1}g^{mn}\,,	\\[4pt]
	\mathcal{M}^{00}\mathcal{M}^{\alpha m}-\mathcal{M}^{0\alpha}\mathcal{M}^{0m}&=g^{-1}g^{mn}A^\alpha{}_n\,,		\\[4pt]
	\mathcal{M}^{00}\mathcal{M}^{m}{}_{n}-\mathcal{M}^{0m}\mathcal{M}^{0}{}_{n}&=g^{-1}g^{mp}\, b_{pn}+\frac12\,g^{-1}g^{mp}A^\alpha{}_p A^\alpha{}_n\,,
	\end{aligned}
\end{equation}
so that the generalised diffeomorphisms acting on the generalised metric reproduce standard diffeomorphisms in $D=6$, together with the gauge transformations\footnote{A typo has been corrected in $\delta  b$ as compared to \cite{Samtleben:2019zrh}.}
\begin{equation}
	\delta A^\alpha=\d\Lambda^\alpha\,,							\qquad
	\delta  b=\d\xi-\frac{1}{2}\,\d\Lambda^\alpha\wedge A^\alpha\,,		\qquad
	\delta \tilde{b}=\d\tilde{\xi}\,.										
\end{equation}
The $b$ and $\tilde{b}$ fields in \eqref{eq: ExFTdictionary} are not directly related to the two-form $ B$ in \eqref{eq: 6Dfields}, as they are taken to live only in the internal space. They do however reconstruct the full three-form field strength in \eqref{eq: fieldstrenghts} via
\begin{equation}	\label{eq: GisHHstar}
	 G=\d b-\frac{1}{2}A^{\alpha}\wedge F^{\alpha}+e^{2\phi}\,\star_6 \d \tilde{b}+ G_{\text{mixed}}\,.
\end{equation}
As such, $e^{2\phi}\,\d \tilde{b}$ describes the Freund-Rubin contribution on AdS$_3$, and $\d b-\frac{1}{2}A^{\alpha}\wedge F^{\alpha}$ the purely $S^3$ components. $ G_{\text{mixed}}$ in \eqref{eq: GisHHstar} stands for the mixed components of the $D=6$ three-form with legs both on AdS$_3$ and $S^3$, which are triggered by the ExFT vectors $\mathcal{A}_\mu^{MN}$ and $\mathcal{B}_{\mu MN}$. These will play no r\^ole in the following as they are absent in ${\rm AdS}_{3}$ solutions.

The consistent truncation of $D=6$ $\mathcal{N}_{\rm 6d}=(1,1)$ supergravity on $S^3$ (and more generally on suitable deformations $M^3$) down to a $D=3$ gauged supergravity can be described in terms of a generalised Scherk-Schwarz ansatz, where the dependence on external and internal coordinates factorises, with the former carried by the $D=3$ fields and the later by an SO(8,4) twist matrix $U_{M}{}^\bN(Y)$ and a scale factor $\rho(Y)$~\cite{Hohm:2017wtr}. The precise factorisation reads\footnote{We adopt here different conventions compared to the ones of \cite{Hohm:2017wtr}. We introduce $\sqrt{2}$ factors for the vector fields, and in the expression of the embedding tensor in~\eqref{eq: consistencyeqs}, to make sure that the three-dimensional theory that follows from the ExFT lagrangian (3.26) of \cite{Hohm:2017wtr} is given by~\eqref{eq: lagrangian}.}
\begin{equation}	\label{eq: SSansatz}
	\begin{aligned}
	g_{\mu\nu}(x,Y)&=\rho(Y)^{-2}g_{\mu\nu}(x)\,,\\[4pt]
	\mathcal{M}_{MN}(x,Y)&=U_{M}{}^\bM(Y)U_{N}{}^\bN(Y)M_{\bM\bN}(x)\,,\\[4pt]
	\mathcal{A}_\mu^{MN}(x,Y)&=\sqrt{2}\,\rho(Y)^{-1}U^{M}{}_\bM(Y)U^{N}{}_\bN(Y)A_\mu^{\bM\bN}(x)\,,			\\[4pt]
	\mathcal{B}_{\mu KL}(x,Y)&=-\frac{1}{2\sqrt{2}}\,\rho(Y)^{-1}U_{M \bN}(Y)\partial_{KL}U^{M}{}_\bM(Y)A_\mu^{\bM\bN}(x)\,,
	\end{aligned}
\end{equation}
with the scale factor and twist matrix subject to the consistency conditions
\begin{equation}	\label{eq: consistencyeqs}
	\begin{aligned}
		\theta_{\bK\bL\bM\bN}&= 3\sqrt{2}\,\rho^{-1}\,\partial_{LP}U_{N[\bK}U^N{}_\bL U^L{}_\bM U^P_{\bN]}\,,\\[4pt]
		\theta_{\bM\bN}&= 2\sqrt{2}\,\rho^{-1}\,U^K{}_\bM\partial_{KL}U^L{}_\bN-\frac{\rho^{-1}}{3\sqrt{2}}\,\eta_{\bM\bN}\,U^{K\bL}\partial_{KL}U^L{}_\bL-2\sqrt{2}\,\rho^{-2}U^K{}_\bM U^L{}_\bN\partial_{KL}\rho	\,,\\[4pt]
		\theta&=\frac{\rho^{-1}}{3\sqrt{2}}\,U^{K\bL}\partial_{KL}U^L{}_\bL\,.
	\end{aligned}
\end{equation}
%

\subsection{Scherk-Schwarz Reduction}

The SO(8,4) twist matrix $U_{M}{}^\bN(Y)$ relevant for the consistent truncation of $D=6$ $\mathcal{N}_{\rm 6d}=(1,1)$ supergravity on $S^3$,
corresponding to the three-dimensional theory \revA, can be constructed from the SO(3,3) result in DFT in \cite{Baguet:2015iou}. In this setting, the sphere can be defined as the locus $\delta_{ab}\mathcal{Y}^a\mathcal{Y}^b=1$ in terms of the $\mathbb{R}^4$ embedding coordinates $\mathcal{Y}^a$, $a,b=1,\dots,4$. The SO(4) Killing vectors of this sphere are then given by
\begin{equation}
	K_{ab\, m}=\mathcal{Y}_{[a\vert}\partial_m \mathcal{Y}_{\vert b]}\,,
\end{equation}
with $\partial_m$ being derivatives with respect to the three physical coordinates $y^m$, and the round metric on $S^3$ can be expressed in terms of them as
\begin{equation}	\label{eq:roundmetricS3}
	\mathring{g}_{m n}= 2\,K_{ab\,m}\,K_{cd}\,{}_{n}\, \delta^{ac}\delta^{bd}\,.
\end{equation}
The SO(4) Killing vectors can be split into SO(3)${}_{\text{L,R}}$ as
\begin{equation}
	\begin{cases}
	L_{\bar m}{}^{m} = \left(K_{4\bar m\,n} + \dfrac{1}{2}\,\varepsilon_{4\bar m \bar n \bar p}\,K_{\bar n\bar p\,n}\right)\,\mathring{g}^{nm}\,,\smallskip\\
	R_{\bar m}{}^{m} =  \left(K_{4\bar m\,n} - \dfrac{1}{2}\,\varepsilon_{4\bar m \bar n \bar p}\,K_{\bar n\bar p\,n}\right)\,\mathring{g}^{nm}\,,
	\end{cases}
\end{equation}
which satisfy the algebra
\begin{equation}
	{\cal L}_{L_{\bar m}}L_{\bar n} = \varepsilon_{\bar m\bar n\bar p}\,L_{\bar p}\,,\qquad 
	{\cal L}_{L_{\bar m}}R_{\bar n} = 0\,, \qquad 
	{\cal L}_{R_{\bar m}}R_{\bar n} = -\varepsilon_{\bar m\bar n\bar p}\,R_{\bar p}\,.
\end{equation}
These SO(3) vectors can be combined into an SO(3,3) doublet\footnote{${\cal K}_{\bar A}{}^{m}$ is defined such that $\delta^{\bar A\bar B}{\cal K}_{\bar A}{}^{m}{\cal K}_{\bar B}{}^{n}=\mathring{g}^{mn}$ and $\eta^{\bar A\bar B}{\cal K}_{\bar A}{}^{m}{\cal K}_{\bar B}{}^{n}=0$, with the ${\rm SO}(3,3)$-invariant matrix $\eta^{\bar A\bar B}$ in the off-diagonal form.}
\begin{equation}	\label{eq: SO33Killing}
	{\cal K}_{\bar A}{}^{m}=
		\begin{pmatrix}
			L_{\bar m}{}^{m}+R_{\bar m}{}^{m} \\
			(R_{\bar n}{}^{m}-L_{\bar n}{}^{m})\,\delta^{\bar n\bar m}
		\end{pmatrix}\,,
\end{equation}
with $\bar A\in\llbracket1,6\rrbracket$ and $X^{\bar A}=\{X^{\bar m}\,, X_{\bar m}\}$. Similarly, to account for the SO(3,3) duals of the coordinates $y^m$, we include six one-forms ${\cal Z}_{\bar A}$ defined as
\begin{equation}
	{\cal Z}_{\bar A\,m} = - \kappa_{\bar A}{}^{\bar B}{\cal K}_{\bar B}{}^{n}\mathring{g}_{nm}-2\,\sqrt{\mathring{g}}\,{\cal K}_{\bar A}{}^{n}\,\varepsilon_{mnp}\,\mathring{\xi}^{p}\,,
\end{equation}
with the vector $\mathring{\xi}$ satisfying $\mathring{\nabla}_m\mathring{\xi}^m=1$ with respect to the Levi-Civita connection associated to \eqref{eq:roundmetricS3}, and $\kappa_{\bar A}{}^{\bar B}$ the Cartan-Killing metric
\begin{equation}
	\kappa_{\bar A}{}^{\bar B}=
		-\begin{pmatrix}		
		0	&	\delta_{\bar m\, \bar n}	\\
		\delta^{\bar m\, \bar n}	&	0
		\end{pmatrix}\,.
\end{equation}
The parametrisation of the SO(8,4) twist matrix in terms of these ingredients follows from the embedding 
\begin{equation}
	\text{SO(3,3)}\subset \text{SO(4,4)}\subset \text{SO(8,4)}\,,
\end{equation}
with the DFT coordinates $Y^A\equiv\{y^m,\, y_m\}$ embedded into the SO(8,4) ExFT coordinates $Y^{MN}$ as $Y^{0A}=Y^{A}$. Accordingly, the twist matrix
\begin{equation}	\label{eq: twistmatrix}
	U^M{}_{\bar M}
		=
		\begin{pmatrix}
			\mathcal{K}_{\bar m}{}^m	& 	\mathcal{K}^{\bar mm}		&	-2\,\lambda\,\mathring{\xi}^{m}	&	0		&	0	\\
			\mathcal{Z}_{\bar m m}	& 	\mathcal{Z}^{\bar m}{}_{m}		&			0  			&	0		&	0	\\
			0				&	0					&			\rho			&	0		&	0	\\
			2\,\lambda\,\rho^{-1}\,{\cal Z}_{\bar m\,n}\,\mathring{\xi}^{n}		&	2\,\lambda\,\rho^{-1}\,{\cal Z}^{\bar m}{}_{n}\,\mathring{\xi}^{n}	&	0	&	\rho^{-1}	&	0	\\
			0				&	0					&			0			&	0		&	\delta^{\alpha}{}_{\bbeta}	\\
		\end{pmatrix}
\end{equation}
together with the scale factor $\rho=\mathring{g}^{-1/2}$ provide a solution to the consistency conditions \eqref{eq: consistencyeqs} with the embedding tensor in \eqref{eq:SO84embeddingsC}. The SO(3,3) contribution in the upper-left corner corresponds to the twist matrix in \cite{Baguet:2015iou}, and $\lambda$-dependent terms can be tracked to the integration of the non-dynamical two-form. The relevant embedding tensor \eqref{eq:SO84embeddingsC} is recovered upon choosing $\lambda=-1$, which we set in the following. Using the Scherk-Schwarz reduction \eqref{eq: SSansatz} for the inverse generalised metric,
\begin{equation}
	\mathcal{M}^{MN}=U^{M}{}_{\bar M}U^{N}{}_{\bar N}M^{\bar M\bar N}\,,
\end{equation}
and the $\mathcal{N}_{\rm 6d}=(1,1)$ dictionary \eqref{eq: ExFTdictionary}, we find the following $D=6$ fields.

\paragraph*{Metric} 
The six-dimensional metric is given by
\begin{equation}	\label{eq: 6dmetric}
	\d s^2_6=\Delta^{-2}g_{\mu\nu}(x)\, \d x^\mu \d x^\nu+g_{mn}(y)\,\d y^m\d y^n\,,
\end{equation}
with $g_{\mu\nu}$ being the AdS$_3$ metric with unit length, the inverse of the metric on $M^3$ given by
\begin{align}	\label{eq: metricgeneral}
	g^{mn}&=\Delta^2e^{-2\tilde\varphi}\Big\{(L^m+R^m)m^{-1}(L^n+R^n)+(R^m-L^m)[m+(\xi^2+\phi)m^{-1}(\xi^2-\phi)+2\xi^2](R^n-L^n)\nonumber\\
				&\quad+(L^m+R^m)m^{-1}(\xi^2-\phi)(R^n-L^n)+(R^m-L^m)(\xi^2+\phi)m^{-1}(L^n+R^n)\Big\}\,,
\end{align}
and the warp factor $\Delta^2=g/\mathring{g}$ computed from the determinant of the metrics $g_{mn}$ and $\mathring{g}_{m n}$ in \eqref{eq:roundmetricS3}.

\paragraph*{Dilaton} 
In terms of the warp factor, the dilaton reads
\begin{equation}	\label{eq: dilatongeneral}
	e^\phi=\Delta^2 e^{-2\tilde \varphi}\,.
\end{equation}

\paragraph*{Vectors} 
The vectors can be extracted from
\begin{equation} \label{eq: ggvecs}
	g^{-1}g^{mn}A_{n}^\alpha=\sqrt2\,\mathring{g}^{-1}e^{-2\tilde{\varphi}}\{(L^m+R^m)m^{-1}\xi^\balpha+(R^m-L^m)[1+(\xi^2+\phi)m^{-1}]\xi^\balpha\}\delta_{\balpha}{}^{\alpha}\,,
\end{equation}
upon using \eqref{eq: metricgeneral}. From the last equation in \eqref{eq: ExFTdictionary}, these vectors must satisfy
\begin{equation}	\label{eq: checkvecs}
	e^{\phi}\,g_{(m\vert p}\mathcal{M}^p{}_{\vert n)}=\delta_{\alpha\beta}\,A^\alpha_m\,A^\beta_n\,.
\end{equation}

\paragraph*{Two-form potentials} 
The two two-forms $ b$ and $\tilde{b}$ in \eqref{eq: GisHHstar} can be recovered via
{\setlength\arraycolsep{1.2pt}
\begin{align} \label{eq: dual2form}
	 b_{mn}&=\Delta^2\,e^{-2\tilde\varphi}\,g_{[m\vert p}\mathcal{M}{}^p{}_{\vert n]}\nonumber\\
		&=2\,\mathring{\omega}_{mnp}\,\mathring{\xi}^p+\Delta^2\,e^{-2\tilde\varphi}\,g_{p[m}\;\mathring{g}_ {n]q}\Big\{(L^p+R^p)m^{-1}(R^q-L^q)+(L^p+R^p)m^{-1}(\xi^2-\phi)(L^q+R^q)	\nonumber\\
		&\quad+(R^p-L^p)(\xi^2+\phi)m^{-1}(R^q-L^q)+(R^p-L^p)[m+(\xi^2+\phi)m^{-1}(\xi^2-\phi)+2\xi^2](L^q+R^q)\Big\}\,,	\nonumber\\
\end{align}
}%
and
\begin{equation} \label{eq: 2formgen}
	 \tilde{b}_{mn}=-2\, \sqrt{\mathring{g}}\,\varepsilon_{mnp}\,\mathring{\xi}^{p}=-2\, \mathring{\omega}_{mnp}\,\mathring{\xi}^{p}\,.
\end{equation}
Correspondingly, the derivative of $\tilde{b}$ is simply
\begin{equation} \label{eq: 2formgenFS}
	3\,\partial_{[m} \tilde{b}_{np]}=-2\, \mathring{\omega}_{mnp}\,.
\end{equation}
%

\subsection{Uplift of the $(\omega,\zeta)$-family}

For the bi-parametric family \eqref{eq: etazetafamily}, the inverse metric \eqref{eq: metricgeneral} becomes
\begin{equation}	\label{eq: inversemetricetazeta}
	g^{mn}=\Delta^2
	\begin{pmatrix}
		1-\mathcal{Y}_1^2-[1-e^{-2\omega}(1+e^{2\omega}\zeta^2)^2]\mathcal{Y}_2^2	
				&	-e^{-2\omega}(1+e^{2\omega}\zeta^2)^2\mathcal{Y}_1\mathcal{Y}_2	
				&	-\mathcal{Y}_1\mathcal{Y}_3-e^{2\omega}\zeta^2\,\mathcal{Y}_2\mathcal{Y}_4		\\
		-e^{-2\omega}(1+e^{2\omega}\zeta^2)^2\mathcal{Y}_1\mathcal{Y}_2	
				&	1-[1-e^{-2\omega}(1+e^{2\omega}\zeta^2)^2]\mathcal{Y}_1^2-\mathcal{Y}_2^2
				&	-\mathcal{Y}_2\mathcal{Y}_3+e^{2\omega}\zeta^2\,\mathcal{Y}_1\mathcal{Y}_4		\\
		-\mathcal{Y}_1\mathcal{Y}_3-e^{2\omega}\zeta^2\,\mathcal{Y}_2\mathcal{Y}_4	
				&	-\mathcal{Y}_2\mathcal{Y}_3+e^{2\omega}\zeta^2\,\mathcal{Y}_1\mathcal{Y}_4	
				&	\mathcal{Y}_1^2+\mathcal{Y}_2^2+e^{2\omega}\mathcal{Y}_4^2
	\end{pmatrix}
\end{equation}
in the basis in which $\{y^1\,,y^2\,,y^3\}=\{\mathcal{Y}^1\,,\mathcal{Y}^2\,,\mathcal{Y}^3\}$. Therefore, the warp factor is
\begin{equation}
	\Delta^2=\frac{e^{-\omega/2}}{\sqrt{1+(\zeta^2+e^{-2\omega}-1)(\mathcal{Y}_1^2+\mathcal{Y}_2^2)}}\,,
\end{equation}
which, at the supersymmetric locus \eqref{eq:susylocus}, reduces to the constant $\Delta^2=e^{-\omega/2}$. To write the direct metric it is useful to introduce the index splitting $\mathcal{Y}^a=\{\mathcal{Y}^{\hat{a}},\, \mathcal{Y}^{\check{a}}\}$, with $\hat{a}=1,2$ and $\check{a}=3,4$. Then, the internal manifold is the three-sphere with metric
\begin{align} \label{eq: etazetametric}
	\d s^2&(M_{\omega,\zeta}^3)=\Delta^6\Big\{e^{2\omega}\delta_{\hat{a}\hat{b}}\,\d\mathcal{Y}^{\hat{a}}\d\mathcal{Y}^{\hat{b}}+e^{-2\omega}(1+e^{2\omega}\zeta^2)^2\delta_{\check{a}\check{b}}\,\d\mathcal{Y}^{\check{a}}\d\mathcal{Y}^{\check{b}}-e^{2\omega}(e^{-2\omega}+\zeta^2-1)^2(\delta_{\hat{a}\hat{b}}\mathcal{Y}^{\hat{a}}\,\d\mathcal{Y}^{\hat{b}})^2	\nonumber\\
		&-(e^{2\omega}-1)(\varepsilon_{\hat{a}\hat{b}}\mathcal{Y}^{\hat{a}}\,\d\mathcal{Y}^{\hat{b}})^2-e^{-2\omega}[(1+e^{2\omega}\zeta^2)^2-e^{2\omega}](\varepsilon_{\check{a}\check{b}}\mathcal{Y}^{\check{a}}\,\d\mathcal{Y}^{\check{b}})^2+2e^{2\omega}\zeta^2(\varepsilon_{\hat{a}\hat{b}}\mathcal{Y}^{\hat{a}}\,\d\mathcal{Y}^{\hat{b}})(\varepsilon_{\check{a}\check{b}}\mathcal{Y}^{\check{a}}\,\d\mathcal{Y}^{\check{b}})
	\Big\}	\,,	\nonumber\\
\end{align}
and the gauge group in $D=3$ can be identified as rotations in the $\mathcal{Y}^1\mathcal{Y}^2$ and $\mathcal{Y}^3\mathcal{Y}^4$ planes, as more explicitly seen in the coordinates
\begin{equation}	\label{eq: u1u1coords}
	\mathcal{Y}^1=\cos\alpha\,\cos\beta\,,\qquad
	\mathcal{Y}^2=\cos\alpha\,\sin\beta\,,\qquad
	\mathcal{Y}^3=\sin\alpha\,\cos\gamma\,,\qquad
	\mathcal{Y}^4=\sin\alpha\,\sin\gamma\,.
\end{equation}
In these coordinates, the metric takes on the form\footnote{At $\zeta=0$, this metric recovers (6.6) in \cite{Samtleben:2019zrh} identifying $\alpha_{\rm here}=\theta_{\rm there}$, $\beta_{\rm here}=(\xi_1)_{\rm there}$ and $\gamma_{\rm here}=(\xi_2)_{\rm there}$.}
\begin{equation} \label{eq: etazetametricintr}
	\begin{aligned} 
	\d s^2(M_{\omega,\zeta}^3)&=\Delta^{-2}\Big\{\d\alpha^2+\Delta^{8}\big(\cos^2\!\alpha+e^{2\omega}\sin^2\!\alpha\big)\cos^2\!\alpha\, \d\beta^2+2\,\Delta^{8}\,e^{2\omega}\zeta^2\sin^2\!\alpha\cos^2\!\alpha\, \d\beta\d\gamma	\\
	&\quad\qquad+\Delta^{8}\big(\sin^2\!\alpha +e^{-2\omega}(1+e^{2\omega}\zeta^2)^2\cos^2\!\alpha\big)\sin^2\!\alpha\,\d\gamma^2\Big\}\,,
	\end{aligned}
\end{equation}
with the isometries realised as shifts in $\beta$ and $\gamma$ and the warp factor, which coincides with the dilaton by \eqref{eq: dilatongeneral}, given by
\begin{equation}	\label{eq: warpfactor}
	\Delta^2=\frac{e^{-\omega/2}}{\sqrt{1+(\zeta^2+e^{-2\omega}-1)\cos^2\!\alpha}}\,.
\end{equation}
The vectors in \eqref{eq: ggvecs} take the form
\begin{equation}	\label{eq: 6Dvectorsetazeta}
	A^\alpha=
	\begin{cases}
		0\,,	 &\alpha\in\{9,10,11\}\,,	\\
		\sqrt{2}\, e^{\omega}\zeta\,\Delta^4\left(\cos^{2}\!\alpha\,\d\beta-\sin^{2}\!\alpha\,\d\gamma\right)\,,	 &\alpha=12\,,
	\end{cases}
\end{equation}
which can be checked to satisfy \eqref{eq: checkvecs}. The corresponding field strengths are therefore
\begin{equation}	\label{eq: 6DdAetazeta}
	F^{12}=\d A^{12}= -2\,\sqrt{2}\,\zeta\,\sin\alpha\cos\alpha\,\Delta^{8}\Big(e^{2\omega}\,\d\alpha\wedge\d\beta+\left(1+e^{2\omega}\zeta^{2}\right)\,\d\alpha\wedge\d\gamma\Big)
\end{equation}
and $F^{\alpha}=0$ for $\alpha\in\{9,10,11\}$.
Finally, the three-form field strength has purely AdS$_3$ and $S^3$ contributions, with the possible mixed components in \eqref{eq: GisHHstar} vanishing. For this family of solutions, the two-form $b$ on the two parameter family is
\begin{equation}
	 b_{mn}= 2\,\mathring{\omega}_{mnp}\,\big(\mathring{\xi}^p - \Delta^{-3}\,g^{pq}\partial_q\Delta\big)\,,
\end{equation}
and the gauge invariant combination in \eqref{eq: GisHHstar} reads
\begin{equation}	\label{eq: Htildeetazeta}
	\d b-\frac{1}{2}\,A^\alpha\wedge F^{\alpha}=2\,\Delta^{7}\, \vol(M_{\omega,\zeta}^3)\,.
\end{equation}
From \eqref{eq: 2formgenFS}, the three-form $\d \tilde{b}$ is simply 
\begin{equation}	\label{eq: Hetazeta}
	\d \tilde{b}=-2\,\Delta^{-1}\,\vol(M_{\omega,\zeta}^3)\,,
\end{equation}
and the total three-form field strength in $D=6$ then results from combining \eqref{eq: Htildeetazeta} and the dual of~\eqref{eq: Hetazeta} into \eqref{eq: GisHHstar} with $G_{\rm mixed}=0$,
\begin{equation}	\label{eq: total3formetazeta}
	G=2\,\vol({\rm AdS}_3)+2\,\Delta^{7}\, \vol(M_{\omega,\zeta}^3)\,.
\end{equation}

\paragraph*{}
At the supersymmetric locus~\eqref{eq:susylocus}, the three-sphere $M_\omega^3$ comes equipped with the Sasaki-Einstein structure
\begin{equation} \label{eq: SE3str}
	\bm{\eta}=J_{ab}\,\mathcal{Y}^a\, \d\mathcal{Y}^b\,,	\qquad
	\bm{J}=\frac12J_{ab}\,\d\mathcal{Y}^a\wedge \d\mathcal{Y}^b\,,	\qquad
	\bm{\Omega}=\Omega_{ab}\,\mathcal{Y}^a\, \d\mathcal{Y}^b\,,
\end{equation}
inherited from the Calabi-Yau forms $J$ and $\Omega$ on the $\mathbb{R}^4$ in which the $S^3$ is embedded:
\begin{equation}
	J=\d\mathcal{Y}^1\wedge \d\mathcal{Y}^2-\d\mathcal{Y}^3\wedge \d\mathcal{Y}^4\,,\qquad
	\Omega=(\d\mathcal{Y}^1+i \d\mathcal{Y}^2)\wedge(\d\mathcal{Y}^3-i \d\mathcal{Y}^4)\,.
\end{equation}
These forms therefore satisfy
\begin{equation}	\label{eq: SEalg}
	\bm{J}\wedge\bm{\Omega}=0\,,	\qquad
	\bm{\eta}\wedge \bm{J}=\frac{i}2\,\bm{\eta}\wedge \bm{\Omega}\wedge \bar{\bm{\Omega}}=\vol(S^3)\,,
\end{equation}
and
\begin{equation}	\label{eq: SEdiff}
	\d\bm{\eta}=2\bm{J}\,,	\qquad
	\d\bm{\Omega}=2i\,\bm{\eta}\wedge \bm{\Omega}\,.
\end{equation}

In terms of these forms, the metric \eqref{eq: etazetametric} on \eqref{eq:susylocus} reads
\begin{equation}
	\d s^2=e^{\omega/2}\delta_{ab}\d\mathcal{Y}^a\d\mathcal{Y}^b-2\,e^{-\omega/2}\sinh\omega\; \bm{\eta}^2\,,
\end{equation}
corresponding to a stretching along the Hopf fibre. This fibred $S^1\hookrightarrow M_\omega^3\to\mathbb{CP}^1$  structure can also be seen in terms of the coordinates
\begin{equation}	\label{eq: changeofcoords}
	\mathcal{Y}_1+i\mathcal{Y}_2=\frac1{\sqrt{1+\vert z\vert^2}}\,e^{i\theta}\,,	\qquad\quad
	\mathcal{Y}_3+i\mathcal{Y}_4=\frac{z}{\sqrt{1+\vert z\vert^2}}\,e^{-i\theta}\,,
\end{equation}
with $\theta\in(0,2\pi)$ and $z=z^R+iz^I$ an inhomogeneous $\mathbb{CP}^1$ coordinate. In terms of these coordinates, the line element takes on the form 
\begin{equation} \label{eq: susymetric}
	\d s^2(M_\omega^3)=e^{\omega/2}\d s^2(\mathbb{CP}^1)+e^{-3\omega/2}\,\bm{\eta}^2\,,
\end{equation}
with the Fubini-Study metric and connection $\bm{\eta}=\d\theta+\sigma$ given by
\begin{equation}
	\d s^2(\mathbb{CP}^1)=\frac{(\d z^R)^2+(\d z^I)^2}{(1+\vert z\vert^2)^2}\,,	\qquad\quad
	\sigma=\frac{z^I\d z^R-z^R\d z^I}{1+\vert z\vert^2}\,.
 \end{equation}
This stretched fibration naturally accounts for the $\SO(2)\times\SO(3)$ gauge symmetry preserved by the $D=3$ solution, with the $\SO(3)\simeq\SU(2)$ realised as the isometries of the Fubini-Study metric, and SO(2) being shifts of the angle along the Hopf fibre.

Remarkably, the vectors in \eqref{eq: 6Dvectorsetazeta} are governed by the Sasaki-Einstein structure \eqref{eq: SE3str} on the entire two-parameter family, and not only at the supersymmetric locus, as in the coordinates \eqref{eq: u1u1coords} the contact one-form becomes
\begin{equation}	\label{eq: contactform}
	\bm{\eta}=\cos^{2}\!\alpha\,\d\beta-\sin^{2}\!\alpha\,\d\gamma\,,
\end{equation}
and the non-vanishing components in \eqref{eq: 6Dvectorsetazeta} and \eqref{eq: 6DdAetazeta} read
\begin{equation}		\label{eq: vectorsSEstructure}
	\begin{aligned}
		A^{12}&=\sqrt{2}\, e^{\omega}\zeta\,\Delta^4\; \bm{\eta}\,,		\\[5pt]
		F^{12}&= 2\sqrt{2}\, e^{\omega}\zeta\,\Delta^4\,\big(2\,\Delta^{-1}\,\d\Delta\wedge \bm{\eta}+\bm{J}\big)\,.
	\end{aligned}
\end{equation}
On  the supersymmetric solutions \eqref{eq:susylocus}, the prefactors in \eqref{eq: vectorsSEstructure} simplify and the warp factor becomes coordinate-independent, with the forms reducing to
\begin{equation}		\label{eq: vectorsSEstructure_susy}
	\begin{aligned}
		A^{12}&=\sqrt{2}\, \sqrt{1-e^{-2\omega}}\; \bm{\eta}\,,		\\[5pt]
		F^{12}&= 2\sqrt{2}\, \sqrt{1-e^{-2\omega}}\; \bm{J}\,.
	\end{aligned}
\end{equation}
Similarly, the three-form in \eqref{eq: total3formetazeta} becomes
\begin{equation}	\label{eq: total3formsusy}
	G=2\,\vol({\rm AdS}_3)+2\,e^{-7\omega/4}\, \vol(M_{\omega}^3)\,.
\end{equation}

\paragraph*{}

We have verified that the metric \eqref{eq: etazetametricintr}, vector field strengths \eqref{eq: 6DdAetazeta}, and two-form field strengths \eqref{eq: total3formetazeta} do satisfy all the $D=6$ equations of motion \eqref{eq: 6Dequations} and \eqref{eq: EinsteinEq} for any value of the marginal scalars. Some details about the Einstein equation can be found in appendix~\ref{app: EinsteinEq}.

\section{Kaluza-Klein Spectra}


Given a $D=6$ background uplifting from a gauged supergravity solution via the Scherk-Schwarz ansatz~\eqref{eq: SSansatz}, and described by the $D=3$ fields 
\begin{equation}	\label{eq: exftbkg}
	\{g_{\mu\nu},\ M_{\bM\bN},\, A_\mu^{\bM\bN}\}=\{\bar{g}_{\mu\nu},\ \Delta_{\bM\bN},\, 0\}\,,
\end{equation}
its associated Kaluza-Klein spectrum can be obtained by making the ExFT fields depend on the linearised perturbations as an extension to the Scherk-Schwarz ansatz:
\begin{equation}
	\begin{aligned}
	g_{\mu\nu}(x,Y)&=\rho(Y)^{-2}\big(\bar{g}_{\mu\nu}(x)+h_{\mu\nu}(x,Y)\big)\,,\\[4pt]
	\mathcal{M}_{MN}(x,Y)&=U_{M}{}^\bM(Y)U_{N}{}^\bN(Y)\big(\Delta_{\bM\bN}+j_{\bM\bN}(x,Y)\big)\,,	\\[4pt]
	\mathcal{A}_\mu^{MN}(x,Y)&=\sqrt{2}\,\rho(Y)^{-1}U^{M}{}_\bM(Y)U^{N}{}_\bN(Y)A_\mu^{\bM\bN}(x,Y)\,,	\\[4pt]
	\mathcal{B}_{\mu KL}(x,Y)&=-\frac1{2\sqrt{2}}\,\rho(Y)^{-1}U_{M \bN}(Y)\partial_{KL}U^{M}{}_\bM(Y)A_\mu^{\bM\bN}(x,Y)\,.
	\end{aligned}	
\end{equation}
These linear perturbations have a natural tower structure when expanded in terms of the harmonics of the background solution. In fact, the expansion only requires the harmonics corresponding to the maximally symmetric case, and the fluctuation ansatz is simply~\cite{Malek:2019eaz,Malek:2020yue,Eloy:2020uix}\footnote{This ansatz differs from that of Ref.~\cite{Eloy:2020uix} by some factors, in agreement with the generalized Scherk-Schwarz ansatz~\eqref{eq: SSansatz}.}
\begin{equation}	
	\begin{aligned}
	g_{\mu\nu}(x,Y)&=\rho(Y)^{-2}\big(\bar{g}_{\mu\nu}(x)+h_{\mu\nu}{}^{\Lambda}(x)\mathcal{Y}^{\Lambda}\big)\,,	\\[4pt]
	\mathcal{M}_{MN}(x,Y)&=U_{M}{}^\bM(Y)U_{N}{}^\bN(Y)\big(\Delta_{\bM\bN}+j_{\bM\bN}{}^{\Lambda}(x)\mathcal{Y}^{\Lambda}\big)\,, \\[4pt]
	\mathcal{A}_\mu^{MN}(x,Y)&=\sqrt{2}\,\rho(Y)^{-1}U^{M}{}_\bM(Y)U^{N}{}_\bN(Y)A_\mu^{\bM\bN\, \Lambda}(x)\mathcal{Y}^{\Lambda}\,,\\[4pt]
	\mathcal{B}_{\mu KL}(x,Y)&=-\frac1{2\sqrt{2}}\,\rho(Y)^{-1}U_{M \bN}(Y)\partial_{KL}U^{M}{}_\bM(Y)A_\mu^{\bM\bN\, \Lambda}(x)\mathcal{Y}^{\Lambda}\,,
	\end{aligned}
\end{equation}
where $\Lambda$ denotes Kaluza-Klein indices in the tower of symmetric traceless representations of the maximal isometry group SO(4),
\begin{equation}
	\bigoplus_{n=0}^{\infty}\Big(\frac n2,\frac n2\Big)
	\;.
	\label{eq:tower_harm}
\end{equation}
The choice of $\mathcal{Y}^{\Lambda}$ as the harmonics corresponding to the configuration with maximal symmetry translates into the fact that they furnish ${\rm SO}(4)$ representations under the action of the relevant Killing vector fields. This leads to the definition of $\mathring{\mathcal{T}}_{\bM\bN}{}^{\Lambda\Sigma}$ as the $\big(\nicefrac n2,\nicefrac n2\big)$ representation matrix encoded in the twist matrix as
\begin{equation}
	\rho^{-1}U^{M}{}_{\bM}U^{N}{}_{\bN}\partial_{MN}\mathcal{Y}^{\Lambda}=
	-\sqrt{2}\,\mathring{\mathcal{T}}_{\bM\bN}{}^{\Lambda\Sigma}\mathcal{Y}^{\Sigma}\,.
\end{equation}
The properties of the twist matrix (\ref{eq: consistencyeqs}) guarantee that the $\mathring{\mathcal{T}}_{\bM\bN}{}^{\Lambda\Sigma}$ 
represent the gauge algebra, with the commutator normalised as \cite{Eloy:2020uix}
\begin{equation}
	\left[\mathring{\mathcal{T}}_{\bM\bN},\mathring{\mathcal{T}}_{\bP\bQ}\right]=-\Theta_{\bM\bN,[\bP}{}^{\bK}\, \mathring{\mathcal{T}}_{\bQ]\bK}+\Theta_{\bP\bQ,[\bM}{}^{\bK}\, \mathring{\mathcal{T}}_{\bN]\bK}\,.
\end{equation}
For the $S^3$ background in \eqref{eq: twistmatrix}, the matrix $\mathring{\mathcal{T}}_{\bM\bN}{}^{ab}$ has non-vanishing components
\begin{equation}	\label{eq: curlyTkk1}
	\mathring{\mathcal{T}}\,_{\bar m\bar0}\,{}^{ab}=\sqrt{2}\,\delta^{[a}_{4}\delta^{b]}_{\bar m}\,,		\qquad\qquad
	\mathring{\mathcal{T}}\,^{\bar m}{}_{\bar0}\,{}^{ab}=-\dfrac{1}{\sqrt{2}}\,\varepsilon^{4\bar{m}ab }\,,
\end{equation}
when acting on the lowest non-trivial level $n=1$ within the tower of harmonics~(\ref{eq:tower_harm}).
At higher levels, these tensors can be constructed recursively from \eqref{eq: curlyTkk1}
\begin{equation}
	(\mathring{\mathcal{T}}_{\bM\bN})_{a_1\dots a_n}{}^{ \tilde{b}_1\dots  \tilde{b}_n}=n(\mathring{\mathcal{T}}_{\bM\bN})_{\{a_1}{}^{\{ \tilde{b}_1}\delta_{a_2}^{ \tilde{b}_2}\dots\delta_{a_n\}}^{ \tilde{b}_n\}}\,.
\end{equation}

To describe backgrounds corresponding to other points of the scalar manifold, it is convenient to dress this tensor analogously to \eqref{eq: dressedembtens},
\begin{equation} \label{eq:dressedKKT}
	\mathcal{T}_{\bM\bN}=(\mathcal{V}^{-1})_{\bM}{}^{\bK}(\mathcal{V}^{-1})_{\bN}{}^{\bL}\mathring{\mathcal{T}}_{\bK\bL}\,.
\end{equation}
Then, the Kaluza-Klein mass matrices for the bosonic fields are those presented in \cite{Eloy:2020uix}, 
\begin{subequations}{\label{eq: KKbosonmasses}}
	\begin{align} 
	 	M_{(2)}^{2}{}^{\Sigma\Omega} &= -\,2\,\delta^{\bar M\bar P}\delta^{\bar N\bar Q}{\cal T}_{\bar M\bar N}{}^{\Sigma \Gamma}{\cal T}_{\bar P\bar Q}{}^{\Gamma\Omega}\,, \label{eq: KKbosonmassesG}\\[5pt]
	 	M_{(1)}{}^{\bar M\bar N\,\Sigma}{}^{}_{\bar P\bar Q}{}^{\Omega} &= \Big(\eta^{\bar K[\bar M}\eta^{\bar N]\bar L}-\delta^{\bar K[\bar M}\delta^{\bar N]\bar L}\Big)\Big(T_{\bar K\bar L\vert\bar P\bar Q}\delta^{\Sigma\Omega}+ 4\,{\cal T}_{\bar K[\bar P}{}^{\Sigma\Omega}\eta_{\bar Q]\bar L}\Big)\,,\\[5pt]
		M^{2}_{(0)\,\bar M\bar N}{}^{\Sigma}{}_{\bar P\bar Q}{}^{\Omega} \,j^{\bar M\bar N,\Sigma}j^{\bar P\bar Q,\Omega} &= \left(m_{\bar M\bar N, \bar P\bar Q}\,\delta^{\Sigma\Omega} + m'_{\bar M\bar N}{}^{\Sigma}{}_{\bar P\bar Q}{}^{\Omega}  \right)j^{\bar M\bar N,\Sigma}j^{\bar P\bar Q,\Omega}\,,
		\label{eq: KKbosonmassesS}
	\end{align}
\end{subequations}
where
\begin{align}
&\quad \begin{aligned}
 m_{\bar M\bar N, \bar P\bar Q} =&\  4\,T_{\bar M\bar P\bar K\bar L}\,T_{\bar N\bar Q\bar R\bar S}\,\delta^{\bar K\bar R}\delta^{\bar L\bar S}+\frac{4}{3}\,T_{\bar M\bar U\bar K\bar L}\,T_{\bar P\bar V\bar R\bar S}\,\delta_{\bar N\bar Q}\,\delta^{\bar U\bar V}\delta^{\bar K\bar R}\delta^{\bar L\bar S} \\
 & - 4\,T_{\bar M\bar P\bar K\bar L}\,T_{\bar N\bar Q}{}^{\bar K\bar L}-4\,T_{\bar M\bar U\bar K\bar L}\,T_{\bar P\bar V}{}^{\bar K\bar L}\delta_{\bar N\bar Q}\,\delta^{\bar U\bar V}+\frac{8}{3}\,T_{\bar M\bar U\bar K\bar L}\,T_{\bar P}{}^{\bar U\bar K\bar L}\delta_{\bar N\bar Q}\\
 & + 2\,T_{\bar M \bar P}\,T_{\bar N \bar Q} - T_{\bar M \bar N}\,T_{\bar P \bar Q}  + 2\,T_{\bar M \bar K}\,T_{\bar P \bar L}\,\delta_{\bar N \bar Q}\,\delta^{\bar K\bar L}\\
 &-\,T_{\bar M \bar P}\,T_{\bar K \bar L}\,\delta_{\bar N \bar Q}\,\delta^{\bar K\bar L} + 16\,T\,T_{\bar M\bar P}\,\delta_{\bar N \bar Q}\,,
 \end{aligned}\\[7pt]
 &\begin{aligned}
 m'_{\bar M\bar N}{}^{\Sigma}{}_{\bar P\bar Q}{}^{\Omega}  =&\ 8\,T_{\bar M\bar P\bar R\bar K}\, \delta_{\bar N}{}^{\bar R}\delta^{\bar K\bar L}\,{\cal T}_{\bar Q\bar L}{}^{\Sigma\Omega}+8\,T_{\bar M\bar P\bar R\bar K} \,\delta_{\bar Q}{}^{\bar R}\delta^{\bar K\bar L}\,{\cal T}_{\bar N\bar L}{}^{\Sigma\Omega}  \\
 &-8\,\eta_{\bar M\bar P}\,T_{\bar N\bar Q\bar K\bar L} \,\delta^{\bar K\bar R}\delta^{\bar L\bar S}\,{\cal T}_{\bar R\bar S}{}^{\Sigma\Omega}  +8\,\eta_{\bar M\bar P}\,T_{\bar N\bar Q\bar K\bar L} \, {\cal T}^{\bar K\bar L\,\Sigma\Omega}\\
 &+8\,\left(T_{\bar M\bar P}+ T\,\eta_{\bar M\bar P}\right)\,{\cal T}_{\bar N\bar Q}{}^{\Sigma\Omega} + 2 \,\eta_{\bar M\bar P}\,\eta_{\bar N\bar Q} \,\delta^{\bar K\bar R}\delta^{\bar L\bar S}\, {\cal T}_{\bar K\bar L}{}^{\Sigma\Lambda}{\cal T}_{\bar R\bar S}{}^{\Lambda\Omega}\\
 & + 16\,\delta_{\bar M\bar P}\, \delta^{\bar K\bar L}\,{\cal T}_{\bar Q\bar L}{}^{\Sigma\Lambda}{\cal T}_{\bar N\bar K}{}^{\Lambda\Omega}-4\,\delta_{\bar M}{}^{\bar K}\delta_{\bar P}{}^{\bar L}\,{\cal T}_{\bar Q\bar L}{}^{\Sigma\Lambda}{\cal T}_{\bar N\bar K}{}^{\Lambda\Omega}+16\,{\cal T}_{\bar M\bar P}{}^{\Sigma\Lambda}{\cal T}_{\bar N\bar Q}{}^{\Lambda\Omega}\,.
\end{aligned}
\end{align}
The fermionic counterparts can be computed in analogy to~\cite{Cesaro:2020soq} and read 
\begin{subequations}		\label{eq: KKfermionmasses}
\begin{align}	
	M_{(\nicefrac{3}{2})}{}^{A\Lambda,\, B\Sigma}&=- A_{1}^{AB}\, \delta^{\Lambda\Sigma}-2\,\Gamma^{IJ}_{AB}\,\mathcal{T}_{IJ}{}^{\Lambda\Sigma}\,,	\\[5pt]
	M_{(\nicefrac{1}{2})}{}^{\Dot{A}r\Lambda,\, \Dot{B}s\Sigma}&=-A_{3}^{\Dot{A}r\, \Dot{B}s}\, \delta^{\Lambda\Sigma}-2\,\Gamma^{IJ}_{\Dot{A}\Dot{B}}\,\delta_{rs}\,\mathcal{T}_{IJ}{}^{\Lambda\Sigma}+8\,\delta_{\dA\dB}\,\mathcal{T}_{rs}{}^{\Lambda\Sigma}\,.
\end{align}
\end{subequations}

All the eigenvalues of the graviton and gravitino mass matrices in \eqref{eq: KKbosonmasses} and \eqref{eq: KKfermionmasses} correspond to physical modes in the spectrum. For the gravitons, it must be taken into account that each of the eigenvalues of \eqref{eq: KKbosonmassesG} in fact corresponds to two states of oposite spin. The eigenvalues of the remaining matrices include the Goldstone modes which are absorbed by massive gravitons, gravitini and vectors in the super-BEH mechanism upon taking into account the off-diagonal couplings between modes of different spin. For the explicit computations, it is useful to observe that the naive eigenvalues of the above mass matrices corresponding to Goldstone modes  (ignoring their off-diagonal couplings) are related to the masses of the corresponding gravitons and gravitini modes~\cite{Cesaro:2021haf}. In $D=3$, we observe the following relations:
\begin{equation}	\label{eq: goldstonerelations}
	\left(m_\1\ell_{\rm AdS}\right)_{\rm goldstone}=\pm\,2\,\sqrt{1+\left(m_\2\ell_{\rm AdS}\right)^{2}}\,,		\qquad
	\left(m_{{(\nicefrac12)}}\ell_{\rm AdS}\right)_{\rm goldstino}=3\, m_{{(\nicefrac32)}}\ell_{\rm AdS}\,.
\end{equation}
Similarly, for each eigenvalue of the spin-$2$ matrix in \eqref{eq: KKbosonmassesG}, two eigenvalues of the scalar mass matrix
\eqref{eq: KKbosonmassesS} are disregarded, one always being zero and the other having value
\begin{equation}
	\left(m_{{(0)}}\ell_{\rm AdS}\right)^{2}_{\rm goldstone}=-3\, \left(m_\2\ell_{\rm AdS}\right)^{2}\,.
\end{equation}

\subsection{Kaluza-Klein spectrum around the $(\omega,\zeta)$-backgrounds}

For generic values of the parameters $(\omega,\zeta)$, we can evaluate the above mass formulae at Kaluza-Klein level $n=1$. In the following, we summarize the results. The graviton masses are 
\begin{equation}	\label{eq: etazetagravkk1}
	\left(m_\2\ell_{\rm AdS}\right)^{2}:\quad  2+e^{2\omega}\ [2+2],\ \ 2+e^{-2\omega}+2\,\zeta^{2}+e^{2\omega}\,\zeta^{4}\ [2+2]\,,
\end{equation}
with the eight modes having a definite spin projection. In turn, the vectors masses read
\begin{equation}	\label{eq: etazetaveckk1}
	\begin{aligned}
		m_\1\ell_{\rm AdS}:\quad -1&\pm\sqrt{3+e^{2\omega}}\ [10+10],\ \ 1\pm\sqrt{3+e^{2\omega}}\ [2+2],\\
		1&\pm\sqrt{3+e^{-2\omega}+2\,\zeta^{2}+e^{2\omega}\,\zeta^{4}}\ [2+2],\\
		-1&\pm\sqrt{3+e^{-2\omega}+2\,\zeta^{2}+e^{2\omega}\,\zeta^{4}}\ [10+10],\\
		1&\pm\sqrt{-1+e^{-2\omega}+2\,\zeta^{2}+e^{2\omega}\,\left(2+\zeta^{2}\right) ^{2}}\ [2+2],\\ 
		-1&\pm\sqrt{-1+e^{-2\omega}+2\,\zeta^{2}+e^{2\omega}\,\left(-2+\zeta^{2}\right) ^{2}}\ [2+2],\\
		1&\pm\sqrt{-1+4\,e^{-2\omega}+8\,\zeta^{2}+e^{2\omega}\,\left(1+2\,\zeta^{2}\right) ^{2}}\ [2+2],\\ 
		-1&\pm\sqrt{-1+4\,e^{-2\omega}+8\,\zeta^{2}+e^{2\omega}\,\left(1-2\,\zeta^{2}\right) ^{2}}\ [2+2]\,.
	\end{aligned}
\end{equation}
Finally, the scalar masses are given by
\begin{equation}	\label{eq: etazetascalkk1}
	\begin{aligned}
		\left(m_\0\ell_{\rm AdS}\right)^{2}:\quad & 2+e^{2\omega}\ [10],\quad -6+9\,e^{2\omega}\ [2],\\
		&6+e^{2\omega}\pm4\,\sqrt{3+e^{2\omega}}\ [2+2],\\
		-&6+9\,e^{-2\omega}+18\,\zeta^{2}+9\,e^{2\omega}\,\zeta^{4}\ [2],\\ 
		2&+e^{-2\omega}+2\,\zeta^{2}+e^{2\omega}\,\zeta^{4}\ [10], \\
		-&2+4\,e^{-2\,\omega}+8\,\zeta^{2}+e^{2\omega}\,\left(1-2\,\zeta^{2}\right)^{2}\ [2],\\
		-&2+4\,e^{-2\,\omega}+8\,\zeta^{2}+e^{2\omega}\,\left(1+2\,\zeta^{2}\right)^{2}\ [10], \\
		-&2+2\,\zeta^{2}+e^{-2\omega}+e^{2\omega}\,\left(-2+\zeta^{2}\right)^{2}\ [2],\\ 
		-&2+2\,\zeta^{2}+e^{-2\omega}+e^{2\omega}\,\left(2+\zeta^{2}\right)^{2}\ [10],\\
		-&1+\left(2\pm\sqrt{3+e^{-2\omega}+2\,\zeta^{2}+e^{2\omega}\,\zeta^{4}}\right)^{2}\ [2+2]\,,
	\end{aligned}
\end{equation}
with Goldstone modes already removed. Regarding fermionic states, the gravitino masses are
\begin{equation}	\label{eq: etazetagravitinikk1}
	\begin{aligned}
		m_{{(\nicefrac32)}}\ell_{\rm AdS}:\quad & -\frac{1}{2}\pm\frac{1}{2}\sqrt{\left(\zeta^{2}-3\right)^{2}e^{2\omega}+2(\zeta^{2}+3)+e^{-2\omega}}\ [4+4]\,, \\
		&-\frac{1}{2}\pm\frac{1}{2}\sqrt{\left(\zeta^{2}+1\right)^{2}e^{2\omega}+2(\zeta^{2}+7)+e^{-2\omega}}\ [8+8]\,, \\
		&-\frac{1}{2}\pm\frac{1}{2}\sqrt{\left(1-3\,\zeta^{2}\right)^{2}e^{2\omega}+6(1+3\,\zeta^{2})+9\,e^{-2\omega}}\ [4+4]\,,
	\end{aligned}
\end{equation}
and those of the spin-$1/2$ modes with goldstini already removed read
\begin{equation}	\label{eq: etazetafermkk1}
	\begin{aligned}
		m_{{(\nicefrac12)}}\ell_{\rm AdS}: 
		-&\frac{3}{2}\pm\frac{1}{2}\sqrt{\left(\zeta ^2-3\right)^2 e^{2 \omega }+2 \left(\zeta ^2+3\right)+e^{-2 \omega }}\ [4+4]\,,	\\
		& \frac{1}{2} \pm \frac{1}{2} \sqrt{\left(\zeta ^2-3\right)^2 e^{2 \omega }+2 \left(\zeta ^2+3\right)+e^{-2 \omega }}\ [4+4]\,, \\
		-&\frac{3}{2} \pm\frac{1}{2}\sqrt{\left(1-3 \zeta ^2\right)^2 e^{2 \omega }+6 \left(3 \zeta ^2+1\right)+9 e^{-2 \omega }}\ [4+4]\,, \\
		&\frac{1}{2}\pm\frac{1}{2}\sqrt{\left(1-3 \zeta ^2\right)^2 e^{2 \omega }+6 \left(3 \zeta ^2+1\right)+9 e^{-2 \omega }}\ [4+4] \,, \\
 		&\frac{1}{2}\pm\frac{1}{2}\sqrt{\left(\zeta ^2+5\right)^2 e^{2 \omega }+2 \left(\zeta ^2-5\right)+e^{-2 \omega }}\ [4+4] \,, \\
		&\frac{1}{2}\pm \frac{1}{2}\sqrt{\left(5 \zeta ^2+1\right)^2 e^{2 \omega }+10 \left(5 \zeta ^2-1\right)+25 e^{-2 \omega }}\ [4+4]\,, \\
		-&\frac{3}{2}\pm\frac{1}{2}\sqrt{\left(\zeta ^2+1\right)^2 e^{2 \omega }+2 \left(\zeta ^2+7\right)+e^{-2 \omega }}\ [8+8]\,, \\
		&\frac{1}{2}\pm\frac{1}{2} \sqrt{\left(\zeta ^2+1\right)^2 e^{2 \omega }+2 \left(\zeta ^2+7\right)+e^{-2 \omega }}\ [8+8]\,, \\
		&\frac{1}{2}\pm\frac{1}{2}\sqrt{9 \left(\zeta ^2+1\right)^2 e^{2 \omega }+2 \left(9 \zeta ^2-1\right)+9 e^{-2 \omega }}\ [8+8]\,.
	\end{aligned}
\end{equation}
At this level, there is no choice of parameters for which any of the vectors or gravitini become massless and (super-)symmetry is enhanced. 

\subsection{Stability}
\begin{figure}[b!]
 \centering
 \centerline{\includegraphics[scale=1]{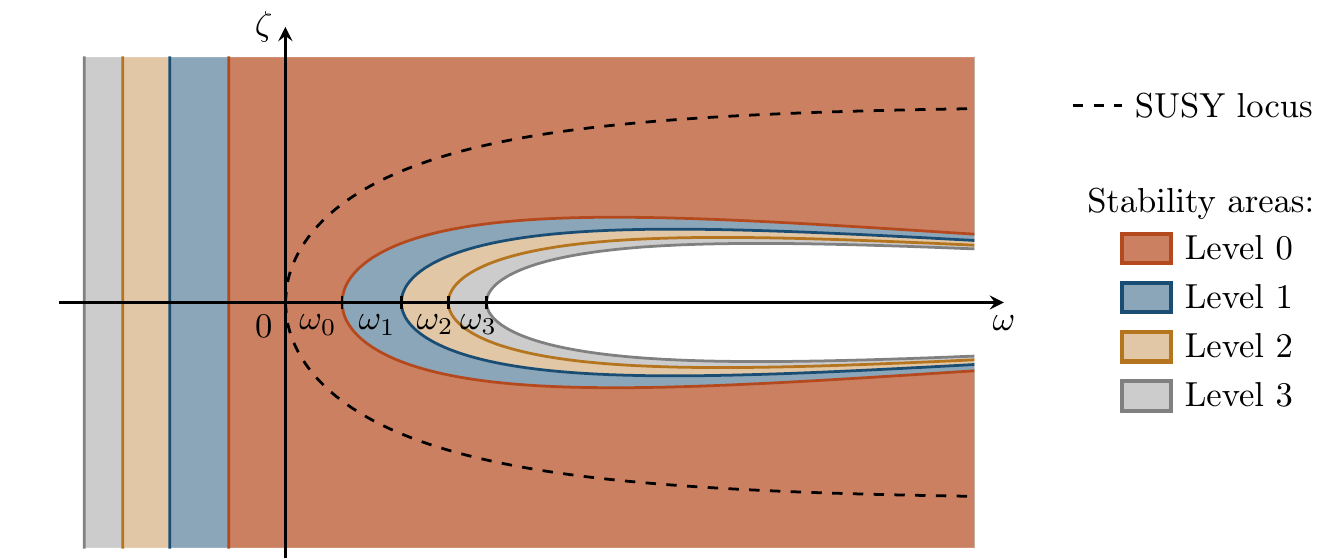}}
 \caption{Stability area at levels 0, 1, 2 and 3, with $\omega_{n}=\ln\Big((2+n)/\sqrt{3+2\,n}\Big)$ saturating~\eqref{eq: BFboundkkn}. The area at level $n$ include the ones at levels $m<n$.}
 \label{fig:stabarea}
\end{figure}

From \eqref{eq: etazetascalkk1}, we observe that BF stability requirements at Kaluza-Klein level $n=1$ are less severe than \eqref{eq: BFbound} at level zero. All scalar modes at the former level are stable within the larger region
\begin{equation} \label{eq: BFboundkk1}
	\begin{cases}
	e^{\omega} \geq \dfrac{\sqrt5}3\,, \smallskip\\
	\zeta^2 \geq \dfrac{\sqrt5}3\,e^{-\omega}-e^{-2\omega}\,,
	\end{cases}
\end{equation}
with the eventually unstable modes having squared masses $-6+9\,e^{2\omega}$ and $-6+9\,e^{-2\omega}+18\,\zeta^{2}+9\,e^{2\omega}\,\zeta^{4}$, respectively. Extrapolating these results to the higher Kaluza-Klein levels, 
we expect that for each level $n$ there are only two pairs of unstable modes, whose masses are given by
\begin{equation}
	-(4+2\,n)+\left(2+n\right)^{2}\,e^{2\omega}\ [2]\,,\quad -(4+2\,n)+\left(2+n\right)^{2}\,e^{-2\omega}\,\left(1+e^{2\omega}\,\zeta^{2}\right)^{2}\ [2]\,,
\end{equation}
which lead to the stability conditions
\begin{equation} \label{eq: BFboundkkn}
	\begin{cases}
	e^{\omega} \geq \dfrac{\sqrt{3+2\,n}}{2+n}\,, \smallskip\\
	\zeta^2 \geq \dfrac{\sqrt{3+2\,n}}{2+n}\,e^{-\omega}-e^{-2\omega}\,.
	\end{cases}
\end{equation}
This has been tested up to and including level 3. We have represented the regions of stability in figure~\ref{fig:stabarea} for increasing values of the KK level.
We expect this pattern to persist to all higher levels. In this case, the vacua satisfying~\eqref{eq: BFbound} would constitute, to our knowledge, the first family of non-supersymmetric ${\rm AdS}_{3}$ vacua enjoying pertubative stability of the full Kaluza-Klein tower.\footnote{Other pertubatively stable non-supersymmetric AdS vacua have already been found in other dimensions, for example in $D=4$ and $D=7$~\cite{Duff:1986hr,Page:1984ae,Page:1984ad,Gaiotto:2009mv,Basile:2018irz,Guarino:2020flh,Marchesano:2021ycx}.} 

\subsection{Spectrum around the supersymmetric backgrounds}
As discussed at the supergravity level, for the family of supersymmetric vacua satisfying \eqref{eq:susylocus} the spectrum organises in representations of the superalgebra \eqref{eq: susylocussuperalgebra}. At level $n$, the representations of the $(\text{SO}(3)_{\text{diag}})_{\text{global}}\ltimes \text{SU}(2\vert1,1)_{\text{R}}$ factor are  $[h,0,\frac{n}{2}]$ in the notation of appendix~\ref{sec: superalgebra}, and are long at generic points in the parameter space. These supermultiplets carry definite SO(2)$_\text{L}$ charge $q$, and SL$(2,\,\mathbb{R})_\text{L}$ dimension $\Delta_{\rm L}$ and will be denoted
\begin{equation}	\label{eq: susylocusmultiplet}
	\big[h,0,\tfrac{n}{2}\big]^{q}_{\Delta_{\rm L}}\,,
\end{equation}
with the content of long supermultiplets following from \eqref{eq: zerolongmult}.

As a result of our spectrum computation, we find that at level $n$ the superconformal primary of the multiplet~\eqref{eq: susylocusmultiplet} with charge $q$ 
carries the $\omega$-dependent conformal dimension
\begin{equation} \label{eq:lowestdelta0}
	h = \frac{-1+\Gamma^{(n,q)}}{2}\,, 		\quad \text{with}\quad
	\Gamma^{(n,q)}=\sqrt{\left(n+1\right)^{2}-q^{2}+q^{2}\,e^{2\omega}}\,,
\end{equation}
and the complete spectrum can be given as
\begin{equation}	\label{eq: spectrumsusyfamily}
	\begin{aligned}
		{\cal S}^{(n)}_{\omega} =& \sum_{q\,\in\,{\cal P}_{n}} \Big( \big[0,\tfrac{n}{2}\big]^{q}_{\nicefrac{\left(3+\Gamma^{(n,q)}\right)}{2}} + \big[0,\tfrac{n}{2}\big]^{q}_{\nicefrac{\left(-1+\Gamma^{(n,q)}\right)}{2}} \Big)\\
		&+\sum_{q\,\in\,{\cal P}_{n+2}} \big[0,\tfrac{n}{2}\big]^{q}_{\nicefrac{\left(1+\Gamma^{(n,q)}\right)}{2}}+\sum_{q\,\in\,{\cal P}_{n-2}} \big[0,\tfrac{n}{2}\big]^{q}_{\nicefrac{\left(1+\Gamma^{(n,q)}\right)}{2}}\,,
	\end{aligned}
\end{equation}
where we denote as ${\cal P}_{k}$ the set of integers ranging from $-k$ to $k$ in steps of two and suppressed the $h$ given by \eqref{eq:lowestdelta0} for notational simplicity. This spectrum has been checked at generic $n$ for the multiplets that contain gravitons following \cite{Bachas:2011xa} and the explicit metric \eqref{eq: susymetric}. For lower-spin multiplets, it has been checked for $n\leq3$.
The explicit allocation of the modes into supermultiplets with this structure at Kaluza-Klein levels one and two can be found in tables~\ref{tab:multsusyeta1} and~\ref{tab:multsusyeta2} of appendix~\ref{app:tablessusy}, respectively.

It is worth noting that the spectrum in table~\ref{tab:multsusyeta} also follows from \eqref{eq: spectrumsusyfamily} upon breaking multiplets which saturate the unitarity bound according to \eqref{eq: breakingrule}. Additionally, we also find shortening at higher Kaluza-Klein levels whenever $n$ is even and $q$ vanishes. These multiplets however are not protected, as they can recombine into long multiplets away from the BPS bound.

\subsection{Spectrum around AdS$_3 \times S^3$}

At the scalar origin, corresponding to the background AdS$_3 \times S^3$, the spectrum \eqref{eq: spectrumsusyfamily} recombines into representations of
\begin{equation}	\label{eq: originsuperalgebra}
	\text{SL}(2,\,\mathbb{R})_{\text{L}}\times \text{SO}(3)_{\text{L}}\times \text{SO}(3)_{\text{l}}\times\big[\text{SO}(3)_{\rm r}\ltimes \text{SU}(2\vert1,1)_{\text{R}}\big]\,. 
\end{equation}
All $\text{SO}(3)_{\text{r}}\ltimes \text{SU}(2\vert1,1)_{\text{R}}$ supermultiplets appearing in the spectrum saturate the unitarity bound \eqref{eq: BPSbound} and are thus short. Their content is summarised in \eqref{eq: lowshortmult}, and the full multiplet, including flavour charges, will be denoted as
\begin{equation}	\label{eq: originmultiplet}
	(\bm{2k+1})^{(j_{\rm gl},\, j_{\rm ga})}_{\Delta_{\rm L}}\,,
\end{equation}
following \cite{deBoer:1998kjm}. Here, $k$ labels the $\SO(3)_{\rm R}\subset\text{SU}(2\vert1,1)_{\text{R}}$ representation, and  $j_{\rm gl}$ and $j_{\rm ga}$ are spins of $\text{SO}(3)_{\text{l}}$ and $\SO(3)_{\rm L}$, respectively. The superconformal primary of these multiplets always transforms trivially under $\text{SO}(3)_{\text{r}}$.

We find that, at levels $0,1$ and $n\geq2$, the perturbations around the uplift of the scalar origin arrange themselves as
\begin{equation}
  \begin{cases}
  \S^{(0)}_{({\rotatebox[origin=c]{180}{\rm \scriptsize A}})} =\bm{2}^{(\nicefrac{1}{2},0)}_{2}+\bm{2}^{(\nicefrac{1}{2},1)}_{1}+\bm{3}^{(0,0)}_{2}+\bm{3}^{(0,1)}_{1}\,, \\[10pt]
  \S^{(1)}_{(\rotatebox[origin=c]{180}{\rm \scriptsize A})} = \bm{2}^{(0,\nicefrac{1}{2})}_{\nicefrac{5}{2}}+\bm{2}^{(0,\nicefrac{3}{2})}_{\nicefrac{3}{2}}+\bm{2}^{(0,\nicefrac{1}{2})}_{\nicefrac{1}{2}}+\bm{3}^{(\nicefrac{1}{2},\nicefrac{1}{2})}_{\nicefrac{5}{2}}+\bm{3}^{(\nicefrac{1}{2},\nicefrac{3}{2})}_{\nicefrac{3}{2}} + \bm{3}^{(\nicefrac{1}{2},\nicefrac{1}{2})}_{\nicefrac{1}{2}} + \bm{4}^{(0,\nicefrac{1}{2})}_{\nicefrac{5}{2}} + \bm{4}^{(0,\nicefrac{3}{2})}_{\nicefrac{3}{2}} + \bm{4}^{(0,\nicefrac{1}{2})}_{\nicefrac{1}{2}}\,, \\[10pt]
  \displaystyle \S^{(n)}_{(\rotatebox[origin=c]{180}{\rm \scriptsize A})} =\!(\bm{n+1})^{(0,\nicefrac{n}{2})}_{\nicefrac{(n+4)}{2}}\!+(\bm{n+1})^{(0,\nicefrac{(n+2)}{2})}_{\nicefrac{(n+2)}{2}}\!+(\bm{n+1})^{(0,\nicefrac{(n-2)}{2})}_{\nicefrac{(n+2)}{2}}\!+(\bm{n+1})^{(0,\nicefrac{n}{2})}_{\nicefrac{n}{2}}+(\bm{n+2})^{(\nicefrac{1}{2},\nicefrac{n}{2})}_{\nicefrac{(n+4)}{2}}\!\!\medskip\\
  \qquad\quad +(\bm{n+2})^{(\nicefrac{1}{2},\nicefrac{(n+2)}{2})}_{\nicefrac{(n+2)}{2}}+(\bm{n+2})^{(\nicefrac{1}{2},\nicefrac{(n-2)}{2})}_{\nicefrac{(n+2)}{2}}\!\!+\!(\bm{n+2})^{(\nicefrac{1}{2},\nicefrac{n}{2})}_{\nicefrac{n}{2}}+(\bm{n+3})^{(0,\nicefrac{n}{2})}_{\nicefrac{(n+4)}{2}}+(\bm{n+3})^{(0,\nicefrac{(n+2)}{2})}_{\nicefrac{(n+2)}{2}}\medskip\\
 \qquad\quad +(\bm{n+3})^{(0,\nicefrac{(n-2)}{2})}_{\nicefrac{(n+2)}{2}}+(\bm{n+3})^{(0,\nicefrac{n}{2})}_{\nicefrac{n}{2}}\,.
  \end{cases}
\end{equation}
This agrees with the breaking of \eqref{eq: spectrumsusyfamily} under \eqref{eq: breakingruleorigin} after realising the embedding $\SO(3)_{\rm diag}\subset\text{SO}(3)_{\text{l}}\times\text{SO}(3)_{\text{r}}$. The detailed field content of these multiplets can be found in tables~\ref{tab:specrevA0}, \ref{tab:specrevA1} and~\ref{tab:specrevAn} in appendix~\ref{sec: detailedspectra}.

The $D=6$ background corresponding to the scalar origin of our consistent truncation \revA is exactly the same as that obtained from an uplift of theory A: the unique supersymmetric AdS$_3\times S^3$ solution of $D=6$, $\mathcal{N}_{\rm 6d}=(1,1)$ supergravity. Accordingly, the full Kaluza-Klein spectrum obtained here must coincide with the spectrum as reported in equation (4.18) of \cite{Eloy:2020uix}, and indeed it does. However, even though the multiplet content in both approaches is the same, the organisation in terms of KK levels differs, and levels can be matched only through
\begin{figure}[t!]
 \centering
 \centerline{\includegraphics[scale=0.85]{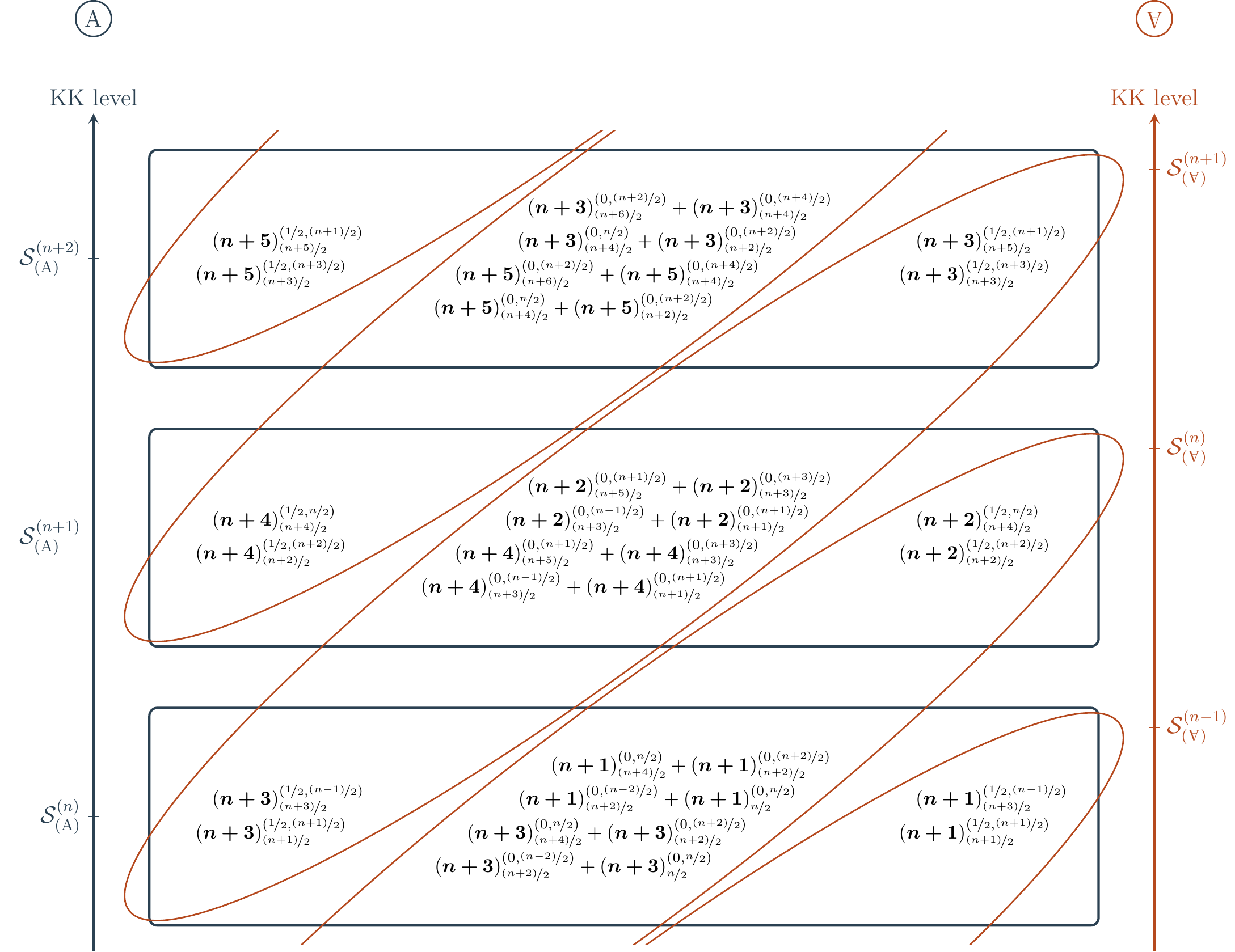}}
 \caption{Matching of the spectra.}
 \label{fig:matching}
\end{figure}
\begin{equation}	\label{eq: matchingspec}
	\begin{aligned}
		\S = \sum_{k\geq2}&\ \bigg[\ \tikz[remember picture] \coordinate(figtop);\ \bm{k}^{(0,\nicefrac{(k-1)}{2})}_{\nicefrac{(k+3)}{2}}+\bm{k}^{(0,\nicefrac{(k+1)}{2})}_{\nicefrac{(k+1)}{2}}+\bm{k}^{(0,\nicefrac{(k-3)}{2})}_{\nicefrac{(k+1)}{2}}\,+\,\bm{k}^{(0,\nicefrac{(k-1)}{2})}_{\nicefrac{(k-1)}{2}} \tikz[remember picture] \coordinate(figtopr);\, \\
		& +\tikz[remember picture] \coordinate(figmid);\ \bm{k}^{(\nicefrac{1}{2},\nicefrac{(k-2)}{2})}_{\nicefrac{(k+2)}{2}}+\bm{k}^{(\nicefrac{1}{2},\nicefrac{k}{2})}_{\nicefrac{k}{2}}+\bm{k}^{(\nicefrac{1}{2},\nicefrac{(k-4)}{2})}_{\nicefrac{k}{2}}\!+\!\bm{k}^{(\nicefrac{1}{2},\nicefrac{(k-2)}{2})}_{\nicefrac{(k-2)}{2}} \phantom{\sum_{k\geq2}\bigg[}\\
		& +\tikz[remember picture] \coordinate(figbot);\ \bm{k}^{(0,\nicefrac{(k-3)}{2})}_{\nicefrac{(k+1)}{2}}\,+\,\bm{k}^{(0,\nicefrac{(k-1)}{2})}_{\nicefrac{(k-1)}{2}}+\bm{k}^{(0,\nicefrac{(k-5)}{2})}_{\nicefrac{(k-1)}{2}}+\bm{k}^{(0,\nicefrac{(k-3)}{2})}_{\nicefrac{(k-3)}{2}}\ \ \bigg]\,,
	\end{aligned}
\end{equation}
\begin{tikzpicture}[overlay,remember picture]
	\draw [dashed,myblue,thick,rounded corners=3pt] (figtop) -- +(0,-1.7) -- +(3.8,-1.7)-- +(3.8,-0.45) -- +(8.32,-0.45) -- +(8.32,0.75) -- +(0,0.75) -- cycle;
	\draw [dotted,myblue,very thick,rounded corners=3pt] (figbot) -- +(0,-0.45) -- +(8.32,-0.45) -- +(8.32,1.95) -- +(4.15,1.95) -- +(4.15,0.7) -- +(0,0.7) -- cycle;
	\draw [dashed,myred,thick,rounded corners=3pt] ($(figtop) + (0.1,0)$) -- +(0,-0.35) -- +(8.10,-0.35) -- +(8.10,0.65) -- +(0,0.65) -- cycle;
	\draw [dash dot dot,myred,thick,rounded corners=3pt] ($(figmid) + (0.1,0)$) -- +(0,-0.35) -- +(8.10,-0.35) -- +(8.10,0.6) -- +(0,0.6) -- cycle;
	\draw [dotted,myred,very thick,rounded corners=3pt] ($(figbot) + (0.1,0)$) -- +(0,-0.35) -- +(8.10,-0.35) -- +(8.10,0.6) -- +(0,0.6) -- cycle;
\end{tikzpicture}
\!\!disregarding the multiplets with negative $\SU(2)$ spins for $k<5$. Multiplets that appear at level $n = k-1$ in theory A (resp. theory \revA) have been framed with blue (resp. red) dashed lines \tikz[baseline=0.4ex]{\draw[dashed, thick,rounded corners=1pt] (0,0) rectangle (0.5,0.4);}\,, those at level $k-2$ in dashed dotted lines \tikz[baseline=0.4ex]{\draw[dash dot dot, thick,rounded corners=1pt] (0,0) rectangle (0.5,0.4);}\,, and those at $k-3$ in dotted lines \tikz[baseline=0.4ex]{\draw[dotted, thick,rounded corners=1pt] (0,0) rectangle (0.5,0.4);}\,. This reassembling has been pictorially represented in figure~\ref{fig:matching} which provides a more detailed
version of the mechanism discussed with figure~\ref{fig:matching_schematic} in the introduction.

\subsection{Coupling to vector multiplets in $D=6$}
The minimal ${\cal N}_{6{\rm d}}=(1,1)$ six-dimensional theory that we considered so far can be further coupled to vector multiplets. The three-dimensional supergravity resulting from compactification is then based on an enhanced coset space 
\begin{equation}
	\SO(8,4+m)/\left(\SO(8)\times\SO(4+m)\right)\;,
\end{equation}
with $m$ the number of vector multiplets added in $D=6$.
The gauged supergravity can be described by embedding the $\SO(8,4)$ tensors $\eta_{\bar M\bar N}$, $\Theta_{\bar M\bar N\vert\bar P\bar Q}$ and $M_{\bar M\bar N}$ (see~\eqref{eq: Paulieta}, \eqref{eq:SO84embeddingsC} and~\eqref{eq: scalarmatrix}) into $\SO(8,4+m)$. It inherits the vacua~\eqref{eq: etazetafamily} and their Kaluza-Klein spectra can be computed upon furthermore embedding ${\cal T}_{\bar M\bar N}$ from~\eqref{eq:dressedKKT} into $\SO(8,4+m)$. At the supersymmetric locus~\eqref{eq:susylocus}, the spectrum at level $n$ in~\eqref{eq: spectrumsusyfamily} is supplemented by $m$ multiplets $[0,\tfrac{n}{2}]_{\nicefrac{\left(1+\Gamma^{(n,q)}\right)}{2}}^{q}$ transforming as vectors under $\SO(m)$:
\begin{equation}
	{\cal S}_{\omega}^{(n),\,\SO(8,4+m)}={\cal S}_{\omega}^{(n)} + \sum_{q\,\in\,{\cal P}_{n}} \big[0,\tfrac{n}{2}\big]^{q,\,m}_{\nicefrac{\left(1+\Gamma^{(n,q)}\right)}{2}},
\end{equation}
where we added the superscript $m$ to indicate the behaviour under $\SO(m)$. It can then be checked that the full Kaluza-Klein spectrum still agrees with the result obtained from an expansion around theory A, as computed in~\cite{Eloy:2020uix}.

Concerning the full non-supersymmetric $(\omega,\zeta)$ family, the additional vector fields induce new massive scalars to the spectrum, and we have checked up to level 3 that the stability condition~\eqref{eq: BFboundkkn} is not affected by these additional modes.

\section{Further discussion}


In this paper we have constructed a new gauged supergravity in $D=3$ arising from consistent truncation of $\mathcal{N}_{\rm 6d}=(1,1), D=6$ supergravity on a three-sphere. This gauged supergravity has a two-parameter family of AdS$_3$ solutions that uplifts to a six-dimensional background with a squashed sphere, non-vanishing fluxes for the vectors and two-form, and a non-trivial dilaton profile \eqref{eq: etazetametricintr}--\eqref{eq: total3formetazeta}. Remarkably, this family of solutions is endowed with a one-parameter subfamily preserving $\mathcal{N}=(0,4)$ supersymmetry. Building on recent ExFT techniques \cite{Malek:2019eaz,Malek:2020yue,Eloy:2020uix}, it has been possible to obtain the KK spectrum on the squashed sphere for arbitrary values of the marginal deformations in the entire two-dimensional conformal manifold. The results indicate that, for large regions in parameter space, the non-supersymmetric solutions are still perturbatively stable. 

To the best of our knowledge, this is the first instance of non-supersymmetric but perturbatively stable AdS$_3$ vacua continuously connected to supersymmetric solutions. This makes them very interesting cases of study in the context of the AdS swampland conjecture \cite{Ooguri:2016pdq}, which states the instability of all non-supersymmetric AdS vacua within string theory. Any possible decay of our vacua must therefore be necessarily through a non-pertubative channel, such as brane-jet instabilities~\cite{Bena:2020xxb,Guarino:2020jwv} or bubble nucleations (see \emph{e.g.} \cite{Witten:1981gj,Coleman:1980aw,Brown:1988kg,Bomans:2021ara}). These possible decay mechanisms will have to deal with the presence of the $\mathcal{N}=(0,4)$ vacua continuously connected to the non-supersymmetric family~\eqref{eq: BFboundkkn}. 

Concerning the supersymmetric family at the locus \eqref{eq:susylocus}, its associated $D=6$ geometry, already presented in \eqref{eq: susyconfiguration} in the introduction, is controlled by the Sasaki-Einstein structure \eqref{eq: SEalg}--\eqref{eq: SEdiff}. This allows a straightforward generalisation to orbifold solutions where either the $\mathbb{CP}^1$ base is replaced by a discrete quotient preserving the K\"ahler-Einstein structure, the Hopf fibre of the $S^3$ is quotiented, or both.

When the marginal deformations are turned off and the solution includes the round $S^3$, the configuration obtained from our uplift matches the ${\rm AdS}_3\times S^3$ solution from theory A~\cite{Samtleben:2019zrh}, and the Kaluza-Klein spectra obtained through both routes match accordingly. However, as discussed around~\eqref{eq: matchingspec} and pictorially represented in figure~\ref{fig:matching}, the notion of KK level differs from both perspectives. Interestingly, for graviton multiplets, whose dimensions can also be computed via \cite{Bachas:2011xa}, both notions agree and they also coincide with the KK level associated to the eigenvalues of the laplacian on the $S^3$.

The absence of an absolute definition for KK level was fundamental to obtain the two-dimensional conformal manifold in \eqref{eq: etazetafamily} from gauged supergravity. This can be traced back to the fact that, while table~\ref{tab:specrevA0} shows that for theory \revA the multiplets containing the marginal scalars, $\bm{2}_1^{(\nicefrac12,1)}$ and $\bm{3}_1^{(0,1)}$, sit at the supergravity level, in theory A the former is not seen in the supergravity truncation but is located in the first level, as shown in tables~\ref{tab:specA0} and~\ref{tab:specA1}.
This prompts the question of whether similar phenomena could take place in other solutions. For instance, if the $\beta$ deformation of $\mathcal{N}=4$ SYM described holographically by the Lunin-Maldacena solution \cite{Lunin:2005jy} could be captured at lowest level by a non-maximal consistent truncation of type IIB on the $S^5$. 

Our ungauged six-dimensional theory \eqref{eq: 6Dlagrangian} arises as a $\mathbb{T}^{4}$ truncation of half-maximal $D=10$ supergravity. Interestingly, the consistent truncation to theory \revA can also be obtained from a different 10D solution~\cite{Eloy:2020uix} with an $S^{3}\times S^{3}\times S^{1}$ factor in the limit where the quotient of the two three-sphere radii degenerates.
This ${\rm AdS}_3\times S^3\times S^3\times S^1$ solution is conjectured \cite{deBoer:1999gea,Eberhardt:2017pty} to be holographically dual to a non-linear sigma model on a symmetric U(2) orbifold, while the ${\rm AdS}_3\times S^3\times \mathbb{T}^{4}$ one has recently been argued \cite{Eberhardt:2019ywk,Gaberdiel:2021kkp} to be dual to a symmetric $\mathbb{T}^{4}$ orbifold. The spectrum on the former solution is controlled by the large $\mathcal{N}=4$ superalgebra $D^1(2,1;\gamma/(1-\gamma))_{\rm R}$, which reduces to SU(2$\vert$1,1)$_{\rm R}\rtimes\SO(3)_{\rm r}$ in \eqref{eq: originsuperalgebra} for $\gamma=1$, as discussed in appendix~\ref{sec: superalgebra}. This corresponds precisely to the limit in which the ratio of the radii of the two three-spheres is zero, and this decompactification limit clarifies the presence of the global factors in the symmetry group controlling the spectrum from the ten-dimensonal perspective. 
For the dual CFTs, this limit relates the sigma model on ${\rm Sym}^N\big({\rm U}(2)\big)$ with a single unit of two-form flux and vanishing central charge to the ${\rm Sym}^N\big(\mathbb{T}^{4}\big)$ theory.\footnote{Non-trivial relations between these CFTs have already been noticed in \cite{Gaberdiel:2018rqv}.} It would be very interesting to understand the conformal manifold from the field theory point of view and extend our results to finite-$N$. From our analysis, we observe that the supersymmetry-preserving marginal operator belongs to the $\bm{2}_1^{(\nicefrac12,1)}$ multiplet, whereas triggering the marginal deformation in $\bm{3}_1^{(0,1)}$ necessarily breaks supersymmetry.

Finally, $\SO(4,4)$ triality has been a key ingredient to generate the new theory \revA. Similarly, one may try to employ triality in the search for new theories within supergravities whose global symmetry groups contain such subgroups. A prominent candidate is half-maximal supergravity in $D=6$, with symmetry group $\SO(4,n)$. Inequivalent embeddings of  $\SO(4,4)\subset \SO(4,4+n)$ might be used to relate different gaugings \cite{Karndumri:2016ruc} and construct new higher-dimensional uplifts. The same idea may be exploited based on the different embeddings into the ${\rm E}_{5(5)}\simeq\SO(5,5)$ global symmetry group of maximal six-dimensional supergravity.

\section*{Acknowledgements}

We would like to thank Mattia Ces\`aro, Olaf Hohm, Emanuel Malek, Alessandro Tomasiello, and Oscar Varela, for useful discussions. GL wants to thank ENS de Lyon for hospitality at the early stages of this work. CE is supported by the FWO-Vlaanderen through the project G006119N and by the Vrije Universiteit Brussel through the Strategic Research Program “High-Energy Physics”. GL is supported by an FPI-UAM predoctoral fellowship and a Spain-US Fulbright scholarship, and partially supported by grants SEV-2016-0597 and PGC2018-095976-B-C21 from MCIU/AEI/FEDER, UE.


\appendix

\addtocontents{toc}{\setcounter{tocdepth}{1}}

\section{Some group theory} \label{sec: grouptheory}

\subsection{The SO(8,4) algebra}	\label{sec: bosonicgrouptheory}
Given the SO(8,4) invariant tensor $\eta^{\bM\bN}$, the algebra generators can be taken as
\begin{equation}
	\big(T^{\bK\bL}\big)_{\bP}{}^{\bQ}=\big(\delta^{\bK}_{\bP}\delta^{\bL}_{\bM}-\delta^{\bK}_{\bM}\delta^{\bL}_{\bP}\big)\eta^{\bM\bQ}\,,
\end{equation}
satisfying commutation relations
\begin{equation}
	[T^{\bK\bL},\,T^{\bM\bN}]=2\,\big(\eta^{\bK[\bM}T^{\bN]\bL}-\eta^{\bL[\bM}T^{\bN]\bK}\big)\,.
\end{equation}
Given the embedding
\begin{equation}	\label{eq: groupbranching}
	\text{SO}(8,4) \supset \text{SO}(7,3)\times \text{SO}(1,1) \supset\text{GL}(3,\mathbb{R})\times\text{SO}(1,1)\times\text{SO}(4)_\text{global}\,,
\end{equation}
it is useful to split indices following \eqref{eq: gl3gradingbar} with invariant metric \eqref{eq: Paulieta}. Then, in the first step in~\eqref{eq: groupbranching}, the adjoint of SO(8,4) breaks as
\begin{equation}
	\text{SO}(8,4) \to \bm{10}_++[\text{SO}(7,3)\times \text{SO}(1,1)]+\bm{10}_-\,,
\end{equation}
with SO(1,1) generated by $T\,{}^{\bzero}{}_{\bzero}$, and $\bm{10}_\pm$ given by
\begin{equation}	\label{eq: 10ofSO73}
	\bm{10}_+=\{T\,{}^{\bar{m}\bzero},\, T\,{}_{\bar{m}}{}^{\bzero},\, T\,{}^{\balpha\bzero}\}\,,	\qquad\quad
	\bm{10}_-=\{T\,{}^{\bar{m}}{}_{\bzero},\, T\,{}_{\bar{m}\bzero},\, T\,{}^{\balpha}{}_{\bzero}\}\,.
\end{equation}
The SO(7,3) subalgebra can itself be decomposed under GL(3$,\mathbb{R})\,\times\,$SO(4)$_\text{global}$ as
\begin{equation}	\label{eq: so73intoGl3xSO4}
	\text{SO}(7,3) \to (\bm{3'},\bm{1})_{+2}+(\bm{3},\bm{4})_{+1}+[\text{GL}(3,\mathbb{R})\times\text{SO(4)}_\text{global}]+(\bm{3'},\bm{4})_{-1}+(\bm{3},\bm{1})_{-2}\,.
\end{equation}
The GL(1$,\mathbb{R}$) factor in GL(3$,\mathbb{R}$) is generated by $T\,{}^{\bar{1}}{}_{\bar{1}}+T\,{}^{\bar{2}}{}_{\bar{2}}+T\,{}^{\bar{3}}{}_{\bar{3}}$, and the SO(4)$_\text{global}\simeq {\rm SO(3)}_{\rm l}\times{\rm SO(3)}_{\rm r}$ factor by $T\,{}^{\balpha\bbeta}$, as
\begin{align}
	\text{SO(3)}_{\rm l}&=\{T\, {}^{\bar 9\,\bar{12}}+T\,{}^{\bar{10}\,\bar{11}},\; T\,{}^{\bar{10}\,\bar{12}}-T\,{}^{\bar{9}\,\bar{11}},\; T\,{}^{\bar{11}\,\bar{12}}+T\,{}^{\bar9\,\bar{10}}\}\,, \notag\\[5pt]
	\text{SO(3)}_{\rm r}&=\{-T\, {}^{\bar 9\,\bar{12}}+T\,{}^{\bar{10}\,\bar{11}},\; -T\,{}^{\bar{10}\,\bar{12}}-T\,{}^{\bar{9}\,\bar{11}},\; -T\,{}^{\bar{11}\,\bar{12}}+T\,{}^{\bar9\,\bar{10}}\}\,.
\end{align}
In turn, the ${\rm SL}(3,\mathbb{R})$ factor has a Cartan subalgebra generated by 
\begin{equation}
	\mathfrak{h}_{\text{SL(3)}}=\big\{\tfrac12\big(T\,{}^{\bar1}{}_{\bar1}-T\,{}^{\bar2}{}_{\bar2}\big),\ \tfrac1{\sqrt{12}}\big(T\,{}^{\bar1}{}_{\bar1}+T\,{}^{\bar2}{}_{\bar2}-2T\,{}^{\bar3}{}_{\bar3}\big)\big\}\,,
\end{equation}
with the corresponding set of positive and negative roots respectively being
\begin{equation}
	\Delta^+_{\text{SL(3)}}=\{\tfrac1{\sqrt2}\,T\,{}^{\bar1}{}_{\bar2},\; \tfrac1{\sqrt2}\,T\,{}^{\bar1}{}_{\bar3},\; \tfrac1{\sqrt2}\,T\,{}^{\bar2}{}_{\bar3}\}\,, \qquad\quad	
	\Delta^-_{\text{SL(3)}}=\{\tfrac1{\sqrt2}\,T\,{}^{\bar2}{}_{\bar1},\; \tfrac1{\sqrt2}\,T\,{}^{\bar3}{}_{\bar1},\; \tfrac1{\sqrt2}\,T\,{}^{\bar3}{}_{\bar2}\}\,.
\end{equation}
The representations in \eqref{eq: so73intoGl3xSO4} with non-zero GL(1$,\mathbb{R}) \subset$ GL(3$,\mathbb{R}$) weight are finally given by
\begin{align}
	(\bm{3'},\bm{1})_{(0,+2)}	&=	\big\{T\,{}^{\bar1\bar2},\ T\,{}^{\bar1\bar3},\ T\,{}^{\bar2\bar3}\big\}\,,	\nonumber\\[5pt]
	(\bm{3},\bm{1})_{(0,-2)}	&=	\big\{T\,{}_{\bar1\bar2},\ T\,{}_{\bar1\bar3},\ T\,{}_{\bar2\bar3}\big\}\,,	\nonumber\\[5pt]
	(\bm{3},\bm{4})_{(0,+1)}	&=	\big\{T\,{}^{\bar1\balpha},\ T\,{}^{\bar2\balpha},\ T\,{}^{\bar3\balpha}\big\}\,,		\nonumber\\[5pt]
	(\bm{3'},\bm{4})_{(0,-1)}	&=	\big\{T\,{}_{\bar1}{}^{\balpha},\ T\,{}_{\bar2}{}^{\balpha},\ T\,{}_{\bar3}{}^{\balpha}\big\}\,,
\end{align}
with the $\bm{10}_\pm$ of SO(7,3)$\times$SO(1,1) in \eqref{eq: 10ofSO73} analogously breaking into
\begin{align}
	&(\bm{3},\bm{1})_{(+1,+1)}=\{T\,{}^{\bar{m}\bzero}\}\,,		\qquad
	&&(\bm{3'},\bm{1})_{(+1,-1)}=\{T\,{}_{\bar{m}}{}^{\bzero}\}\,,	\qquad
	&&&(\bm{1},\bm{4})_{(+1,0)}=\{T\,{}^{\balpha\bzero}\}\,,		\nonumber\\
	&(\bm{3},\bm{1})_{(-1,+1)}=\{T\,{}^{\bar{m}}{}_{\bzero}\}\,,		\qquad
	&&(\bm{3'},\bm{1})_{(-1,-1)}=\{T\,{}_{\bar{m}\bzero}\}\,,		\qquad
	&&&(\bm{1},\bm{4})_{(-1,0)}=\{T\,{}^{\balpha}{}_{\bzero}\}\,.
\end{align}
In these equations, the subindices refer to the $\SO(1,1)\times\GL(1,\mathbb{R})$ charges in \eqref{eq: groupbranching}.
From the embedding tensor \eqref{eq:SO84embeddingsC}, the gauge group in \eqref{eq: ggauge} has translations
\begin{equation}
	(T^1)^4	=	\big\{T\,{}^{\bzero\balpha}\big\}\,,	\qquad
	T^3_{\text{left}}		=	\big\{T\,{}^{\bar{m}\bzero}+T\,{}_{\bar{m}}{}^{\bzero}\big\}\,,	\qquad
	T^3_{\text{right}}	=	\big\{T\,{}^{\bar{m}\bzero}-T\,{}_{\bar{m}}{}^{\bzero}\big\}\,,
\end{equation}
with $T^3_{\text{left}}$ and $T^3_{\text{right}}$ in the (1,0) and (0,1) of SO(4)$_{\text{gauge}}\simeq$ SO(3)$_{\text{L}}\times$SO(3)$_{\text{R}}$, respectively. The later subgroups are in turn generated by 
\begin{align}
	\text{SO(3)}_{\text{L}}&=\big\{T\,{}^{\bar1\bar2}+T\,{}_{\bar1\bar2}+T\,{}^{\bar1}{}_{\bar2}-T\,{}^{\bar2}{}_{\bar1},\ T\,{}^{\bar1\bar3}+T\,{}_{\bar1\bar3}+T\,{}^{\bar1}{}_{\bar3}-T\,{}^{\bar3}{}_{\bar1},\ T\,{}^{\bar2\bar3}+T\,{}_{\bar2\bar3}+T\,{}^{\bar2}{}_{\bar3}-T\,{}^{\bar3}{}_{\bar2}\big\}\,,	\notag\\[5pt]
	\text{SO(3)}_{\text{R}}&=\big\{T\,{}^{\bar1\bar2}+T\,{}_{\bar1\bar2}-T\,{}^{\bar1}{}_{\bar2}+T\,{}^{\bar2}{}_{\bar1},\ T\,{}^{\bar1\bar3}+T\,{}_{\bar1\bar3}-T\,{}^{\bar1}{}_{\bar3}+T\,{}^{\bar3}{}_{\bar1},\ T\,{}^{\bar2\bar3}+T\,{}_{\bar2\bar3}-T\,{}^{\bar2}{}_{\bar3}+T\,{}^{\bar3}{}_{\bar2}\big\}\,,\notag\\
\end{align}
which makes manifest that the compact part of \eqref{eq: ggauge} does not sit at level zero of the grading \eqref{eq: branching}, but involves the $(\bm{3'},\bm{1})_{(0,+2)}$ and $(\bm{3},\bm{1})_{(0,-2)}$ factors.


\subsection{The SU(2$\vert$1,1)$\rtimes$SU(2) algebra} \label{sec: superalgebra}

Like the spectra discussed in \cite{Eloy:2020uix, deBoer:1998kjm}, the Kaluza-Klein tower on the ${\cal N}=(0,4)$ supersymmetric solutions organises in terms of the non-semisimple $\SU(2\vert1,1)_+\rtimes\SU(2)_-$. This superalgebra is generated by $L_{m}$ with $m=-1,0,1$, which constitutes the SL(2,$\mathbb{R}$) subgroup and $A_i^{\pm}$ with $i=1,2,3$, which generate SU(2)$_{\pm}$. The fermionic generators are $G_{r}^a$ with $r=\pm\frac12$ and $a=1,2,3,4$, which furnish a~$(\bm{2},\bm{2})$ representation of SU(2)$_+\times$ SU(2)$_-$. The (anti-)commutation rules associated to this algebra are
\begin{align}
	[L_m,\, L_n]&=(m-n)L_{m+n}\,,							\qquad\quad
	[A^{\pm}_i,\, A^{\pm}_j]=i\epsilon_{ijk}A^{\pm}_k\,,			\qquad\quad
	[L_m,\, A^{\pm}_i]=0\,,								\nonumber\\[5pt]
	[L_m,\, G_r^a]&=\Big(\frac m2-r\Big)G_r^a\,,				\qquad\qquad\!\!
	[A^{\pm}_i,\, G_r^a]=i\alpha^{\pm i}_{ab}G_r^b\,,			\nonumber\\[5pt]
	\{G_r^a,\, G_s^b\}&=2\delta^{ab}L_{r+s}+4i(r-s)\alpha_{ab}^{+i}A^+_i\,,
\end{align}
with
\begin{equation}
	\alpha^{\pm i}_{ab}=\pm\delta_{i+1,[a}\delta_{b]1}+\tfrac12 \epsilon_{i,a-1,b-1,4}\,.
\end{equation}
These rules can be understood as the limit of the exceptional $\mathcal{N}=4$ superalgebra \cite{Sevrin:1988ew,Eberhardt:2017fsi}
\begin{equation}	\label{eq: superalgebralimit}
	\lim_{\alpha\rightarrow\infty} D(2,1;\alpha)=\text{SU}(2\vert1,1)_+\rtimes\text{SU}(2)_-\,.
\end{equation}
This limit does not affect the matter content of long multiplets, whose states can be given in terms of the weights under the bosonic subalgebra as $(h,\, j^-,\, j^+)$, with $h$ denoting the SL(2,$\mathbb{R}$) dimension and $j^\pm$ being half-integer spins for SU(2)$_\pm$. Supermultiplets are then determined by a primary state which is annihilated by all $G_{\frac12}^a$ and $L_1$. A supermultiplet with superconformal primary $(h,\, j^-,\, j^+)$ will be denoted $[h, j^-, j^+]$, and its full content is given by
\begin{align} \label{eq: genericlongmult}
	\vert\psi\rangle &:\quad (j^-,\, j^+),	\nonumber\\[5pt]
	G_{-\frac12}^a\vert\psi\rangle &:\quad (j^-+\tfrac12,\, j^++\tfrac12),\; (j^--\tfrac12,\, j^++\tfrac12),\; (j^-+\tfrac12,\, j^+-\tfrac12),\; (j^--\tfrac12,\, j^+-\tfrac12)\,,	\nonumber\\[5pt]
	G_{-\frac12}^{[a}G_{-\frac12}^{b]}\vert\psi\rangle &:\quad (j^-,\, j^++1),\; (j^-+1,\, j^+),\; 2\cdot(j^-,\, j^+),\; (j^-,\, j^+-1),\; (j^--1,\, j^+)\,,	\nonumber\\[5pt]
	G_{-\frac12}^{[a}G_{-\frac12}^{b}G_{-\frac12}^{c]}\vert\psi\rangle &:\quad(j^-+\tfrac12,\, j^++\tfrac12),\; (j^--\tfrac12,\, j^++\tfrac12),\; (j^-+\tfrac12,\, j^+-\tfrac12),\; (j^--\tfrac12,\, j^+-\tfrac12)\,,	\nonumber\\[5pt]
	G_{-\frac12}^{1}G_{-\frac12}^{2}G_{-\frac12}^{3}G_{-\frac12}^{4}\vert\psi\rangle &:\quad (j^-,\, j^+)\,,
\end{align}
with the scaling dimensions, suppressed here for clarity, increasing by $\nicefrac12$ in every line. Long multiplets do not saturate the BPS bound%
\begin{equation}	\label{eq: BPSbound}
	h\geq j^+\,.
\end{equation}
When $h=j^+$, multiplet shortening occurs and the state content of $[j^+, j^-, j^+]_{\rm S}$ reduces to
\begin{align} \label{eq: genericshortmult}
	\vert\psi\rangle &:\quad (j^+,\, j^-,\, j^+)\,,	\nonumber\\[5pt]
	G_{-\frac12}^a\vert\psi\rangle &:\quad (j^++\tfrac12,\, j^-+\tfrac12,\, j^+-\tfrac12),\; (j^++\tfrac12,\, j^--\tfrac12,\, j^+-\tfrac12)\,,	\nonumber\\[5pt]
	G_{-\frac12}^{[a}G_{-\frac12}^{b]}\vert\psi\rangle &:\quad (j^++1,\, j^-,\, j^+-1)\,.
\end{align}
Compared to the representation theory of $D(2,1;\alpha)$ \cite{Eberhardt:2017fsi}, we observe that the limit \eqref{eq: superalgebralimit} implies that all states for which the SU(2)$_+$ weight rises become null at the BPS bound.
A shortening of different type can also take place for low values of $j^{\pm}$. In particular, for $[h,0, j^+]$ with $h>j^+\geq1$, the result is
\begin{align} \label{eq: zerolongmult}
	\vert\psi\rangle &:\quad (h,\, 0,\, j^+),	\nonumber\\[5pt]
	G_{-\frac12}^a\vert\psi\rangle &:\quad (h+\tfrac12,\, \tfrac12,\, j^++\tfrac12),\; \underdashed{$(h+\tfrac12,\, \tfrac12,\, j^+-\tfrac12)$}\,,	\nonumber\\[5pt]
	G_{-\frac12}^{[a}G_{-\frac12}^{b]}\vert\psi\rangle &:\quad (h+1,\, 1,\, j^+),\; (h+1,\, 0,\, j^++1),\; \underdashed{$(h+1,\, 0,\, j^+)$},\; \underline{(h+1,\, 0,\, j^+-1)}\,,	\nonumber\\[5pt]
	G_{-\frac12}^{[a}G_{-\frac12}^{b}G_{-\frac12}^{c]}\vert\psi\rangle &:\quad(h+\tfrac32,\, \tfrac12,\, j^++\tfrac12),\; \underdashed{$(h+\tfrac32,\, \tfrac12,\, j^+-\tfrac12)$}\,,	\nonumber\\[5pt]
	G_{-\frac12}^{1}G_{-\frac12}^{2}G_{-\frac12}^{3}G_{-\frac12}^{4}\vert\psi\rangle &:\quad (h+2,\, 0,\, j^+)\,,
\end{align}
with underlined states missing for $j^+\leq\nicefrac12$, and those with dashed underlining also missing for $j^+=0$. When \eqref{eq: BPSbound} is saturated, similar considerations apply to \eqref{eq: genericshortmult} and the state contents of $[j^+,0,j^+]_{\rm S}$ and $[j^+,\frac12,j^+]_{\rm S}$ are
\begin{align} \label{eq: lowshortmult}
	[j^+,0,j^+]_{\rm S}&\quad:\quad	
		(j^+,\, 0,\, j^+)\,,\quad	
		(j^++\tfrac12,\, \tfrac12,\, j^+-\tfrac12)\,,\quad
		(j^++1,\, 0,\, j^+-1)\,,	
	\\[5pt]
	[j^+,\tfrac12,j^+]_{\rm S}&\quad	:\quad
		(j^+,\, \tfrac12,\, j^+)\,,\quad	
		(j^++\tfrac12,\, 0,\, j^+-\tfrac12)\,,\quad
		(j^++\tfrac12,\, 1,\, j^+-\tfrac12)\,,\quad
		(j^++1,\, \tfrac12,\, j^+-1)\,.
	\nonumber
\end{align}
The breaking rule at the unitarity bound is therefore
\begin{equation} \label{eq: generalbreakingrule}
	[j^++\epsilon,j^-,j^+]\xrightarrow[\epsilon\rightarrow0]{} [j^+,j^-,j^+]_{\rm S}+[j^++\tfrac12,j^-+\tfrac12,j^++\tfrac12]_{\rm S}+[j^++\tfrac12,j^--\tfrac12,j^++\tfrac12]_{\rm S}+[j^++1,j^-,j^++1]_{\rm S}\,,
\end{equation}
which for $j^-=0$ particularises to
\begin{equation} \label{eq: breakingrule}
	[j^++\epsilon,0,j^+]\xrightarrow[\epsilon\rightarrow0]{} [j^+,0,j^+]_{\rm S}+[j^++\tfrac12,\tfrac12,j^++\tfrac12]_{\rm S}+[j^++1,0,j^++1]_{\rm S}\,.
\end{equation}

At the supersymmetric family \eqref{eq:susylocus}, the SU(2)$_+\times$ SU(2)$_-$ factors can be identified with SO(3)$_\text{diag}\times$ SO(3)$_\text{R}$, with SO(2)$_\text{L}$ being outside the superalgebra. The relevant supermultiplets in this case are $[h,0, j^+]$ in \eqref{eq: zerolongmult}. At the scalar origin, the factor within the superalgebra becomes SO(3)$_{\rm r}\times$ SO(3)$_\text{R}$ and the flavour symmetry enhances to SO(3)$_{\rm l}\times$ SO(3)$_\text{L}$, while all multiplets saturate the BPS bound and decompose according to \eqref{eq: breakingrule}. Taking into account the superalgebra specifications in \eqref{eq: susylocusmultiplet} and \eqref{eq: originmultiplet} (except for the SL$(2,\mathbb{R})_{\rm L}$ factor), the breaking rule becomes
\begin{align} \label{eq: breakingruleorigin}
	\sum_{q\,\in\,{\cal P}_{k}} [j^++\epsilon,0,j^+]^{q}
	&\xrightarrow[\epsilon\rightarrow0]{} [j^+,0,j^+]_{\rm S}^{(0,\nicefrac k2)}+[j^++\tfrac12,0,j^++\tfrac12]_{\rm S}^{(\nicefrac12, \nicefrac k2)}+[j^++1,0,j^++1]_{\rm S}^{(0, \nicefrac k2)}	\nonumber\\[5pt]
	&\simeq \bm{(2j^++1)}^{(0,\nicefrac k2)}+\bm{(2j^++2)}^{(\nicefrac12, \nicefrac k2)}+\bm{(2j^++3)}^{(0, \nicefrac k2)}\,,
\end{align}
employing the notation in (4.14) of \cite{deBoer:1998kjm}.

\section{Details on Einstein equations}	\label{app: EinsteinEq}

The Ricci tensor associated to \eqref{eq: etazetametricintr} is given by
\begin{align}	\label{eq: RicciTensor}
	R^{\6}_{\hat\mu\hat\nu}\, \d \hat{x}^{\hat\mu}\d \hat{x}^{\hat\nu}
	&=-\big(1+ (1+e^{2\omega}\zeta^2)\Delta^8\big)\d s^2(\text{AdS}_3)
		+\big[e^{-\omega}\big(1+e^{2\omega}(1+\zeta^2)\big)\Delta^4+2\,e^{2\omega}\zeta^2\Delta^8\big]\d\alpha^2					\nonumber\\[5pt]
	&+\Big[\big(1- (1-e^{2\omega})\sin^2\!\alpha\big)\Delta^8+
		\big((1-e^{2\omega}\zeta^2)\cos^2\!\alpha+e^{2\omega}(1+3\, e^{2\omega}\zeta^2)\sin^2\!\alpha\big)\Delta^{16}\Big]\cos^2\!\alpha\,\d\beta^2
	\nonumber\\[5pt]
	&+2\,e^{2\omega}\zeta^2\big(1+(5+3e^{2\omega}\zeta^2)\Delta^8\big)\Delta^8\,\cos^2\!\alpha\,\sin^2\!\alpha\,\d\beta\,\d\gamma		\nonumber\\[5pt]
	&+\Big[\big[1-\big(1-e^{-2\omega}(1+e^{2\omega}\zeta^2)^2\big)\cos^2\!\alpha\big]\Delta^8									\nonumber\\[5pt]
	&\qquad+\big(e^{-2\omega}(1+e^{2\omega}\zeta^2)^2(1+3e^{2\omega}\zeta^2)\cos^2\!\alpha+(1-e^{2\omega}\zeta^2)\sin^2\!\alpha\big)\Delta^{16}\Big]\sin^2\!\alpha\,\d\gamma^2	\,.
\end{align}
with $\text{AdS}_3$ of unit length and $\Delta$ in \eqref{eq: warpfactor}. Correspondingly, the contributions of the stress-energy tensor in \eqref{eq: EinsteinEq} are given by:
\begin{subequations} \label{eq: StressTensorContribs}
\begin{equation}
	\partial_{\hat \mu}\phi\,\partial_{\hat\nu}\phi\, \d \hat{x}^{\hat\mu}\d \hat{x}^{\hat\nu}
	=\Delta^8 e^{2\omega} \left(\zeta^2+e^{-2\omega}-1\right)^2 \sin^2\!\alpha \cos^2\!\alpha\, \d \alpha^2\,,
\end{equation}	
\begin{equation}	
	e^{-\phi}\, F^{\alpha}_{\hat\mu\hat\rho}F^{\alpha}{}_{\hat\nu}{}^{\hat\rho}\, \d \hat{x}^{\hat\mu}\d \hat{x}^{\hat\nu}
	=8\, \Delta^8 \zeta ^2 \Big[e^{2\omega}\, \d\alpha^2+\Delta^8 \left(e^{2\omega}\d\beta+\left(\zeta^2 e^{2\omega}+1\right)\d\gamma\right)^2\sin^2\!\alpha \cos^2\!\alpha\Big]\,,
\end{equation}	
\begin{equation}	
	e^{-2\phi}\,  G_{\hat\mu\hat\rho\hat\sigma} G_{\hat\nu}{}^{\hat\rho\hat\sigma}\, \d \hat{x}^{\hat\mu}\d \hat{x}^{\hat\nu}
	=-8\,  \d s^2({\rm AdS}_3)+8\, \Delta^{10}\,\d s^2(S^3)\,,
\end{equation}
\end{subequations}
with the internal metric in \eqref{eq: etazetametricintr}. Then, it can be checked that the terms in \eqref{eq: StressTensorContribs} together with their traces add up to a stress-energy tensor that matches the Einstein tensor constructed out of \eqref{eq: RicciTensor}.

At the locus \eqref{eq:susylocus}, the Ricci tensor simplifies into
\begin{align}
	R^{\6}_{\hat\mu\hat\nu}\, \d \hat{x}^{\hat\mu}\d \hat{x}^{\hat\nu}
	&=-2\,\d s^2(\text{AdS}_3)+2\,e^{-\omega/2}(2-e^{-2\omega})\,\d s^2(S^3)-8\, e^{-3\omega}\sinh{\omega}\,\bm{\eta}^2\,.
\end{align}
with the internal metric in \eqref{eq: susymetric} and the contact form \eqref{eq: contactform}. Similarly, \eqref{eq: StressTensorContribs} reduces to
\begin{subequations} 
\begin{equation}
	\partial_{\hat \mu}\phi\,\partial_{\hat\nu}\phi\, \d \hat{x}^{\hat\mu}\d \hat{x}^{\hat\nu}
	=0\,,
\end{equation}	
\begin{equation}	
	e^{-\phi}\, F^{\alpha}_{\hat\mu\hat\rho}F^{\alpha}{}_{\hat\nu}{}^{\hat\rho}\, \d \hat{x}^{\hat\mu}\d \hat{x}^{\hat\nu}
	=8\left(1-e^{-2\omega}\right) \left(\d\alpha^2+(\text{d$\beta $}+\text{d$\gamma $})^2\sin^2\!\alpha \cos^2\!\alpha\right)\,,
\end{equation}	
\begin{equation}	
	e^{-2\phi}\,  G_{\hat\mu\hat\rho\hat\sigma} G_{\hat\nu}{}^{\hat\rho\hat\sigma}\, \d \hat{x}^{\hat\mu}\d \hat{x}^{\hat\nu}
	=-8\,  \d s^2({\rm AdS}_3)+8\, e^{-5\omega/2}\,\d s^2(S^3)\,.
\end{equation}
\end{subequations}
%

\section{Spectra for theory A and \revA at scalar origin}
\label{sec: detailedspectra}

In this section we collect the spectrum at the scalar origin for theories A and \revA in \eqref{eq:SO84embeddings}. The results for theory A were already presented in \cite{Eloy:2020uix} and are brought here to facilitate comparison. Both spectra are organised in representations of the superalgebra \eqref{eq: originsuperalgebra}, in the notation of  \eqref{eq: originmultiplet}, and with $\SO(4)_{\rm L}\simeq \text{SO}(3)_{\text{l}}\times\SO(3)_{\rm L}$ and $\SO(4)_{\rm R}\simeq \text{SO}(3)_{\text{r}}\times\SO(3)_{\rm R}$.

Tables~\ref{tab:specA0} and \ref{tab:specrevA0} contain the spectra at level zero of theory A and \revA respectively. Similarly, level $n=1$ is presented in tables \ref{tab:specA1} and \ref{tab:specrevA1}, and levels $n>1$ in tables \ref{tab:specAn} and \ref{tab:specrevAn}.


\begin{table}[p]
\centering
  \begin{tabular}{ccccccl}
	$\Delta_{\rm L}$ & $\Delta_{\rm R}$ & $\Delta$ & $s$ & $\SO(4)_{\rm L}$ & $\SO(4)_{\rm R}$ &  \\\cmidrule{1-7}\morecmidrules\cmidrule{1-7}
	$2$  &  0  &  2  &  $-2$  &  $(0,0)$  &  $(0,0)$  &  $\bm{1}_2^{(0,0)}$ \\\cmidrule{1-7}
	$1$  &  0  &  1  &  $-1$  &  $(0,1)$  &  $(0,0)$  &  $\bm{1}_1^{(0,1)}$ \\\cmidrule{1-7}
	\multirow{3}{*}{$2$} & $2$ & $4$ & $0$ & \multirow{3}{*}{$\big(0,0\big)$} & $\big(0,0\big)$ &  \multirow{3}{*}{$\bm{3}^{(0,0)}_{2}$} \\
	& $\nicefrac{3}{2}$ & $\nicefrac{7}{2}$ & $-\nicefrac{1}{2}$ & & $\big(\nicefrac{1}{2},\nicefrac{1}{2}\big)$ & \\
	& $1$ & $3$ & $-1$ & & $\big(0,1\big)$ \\\cmidrule{1-7}
	\multirow{3}{*}{$1$} & $2$ & $3$ & $1$ & \multirow{3}{*}{$\big(0,1\big)$} & $\big(0,0\big)$ &  \multirow{3}{*}{$\bm{3}^{(0,1)}_{1}$} \\
	& $\nicefrac{3}{2}$ & $\nicefrac{5}{2}$ & $\nicefrac{1}{2}$ & & $\big(\nicefrac{1}{2},\nicefrac{1}{2}\big)$ & \\
	& $1$ & $2$ & $0$ & & $\big(0,1\big)$ \\\cmidrule{1-7}
	\multirow{3}{*}{$\nicefrac{1}{2}$} & $2$ & $\nicefrac{5}{2}$ & $\nicefrac{3}{2}$ & \multirow{3}{*}{$\big(\nicefrac{1}{2},\nicefrac{1}{2}\big)$} & $\big(0,0\big)$ &  \multirow{3}{*}{$\bm{3}^{(\nicefrac{1}{2},\nicefrac{1}{2})}_{\nicefrac{1}{2}}$} \\
	& $\nicefrac{3}{2}$ & $2$ & $1$ & & $\big(\nicefrac{1}{2},\nicefrac{1}{2}\big)$ & \\
	& $1$ & $\nicefrac{3}{2}$ & $\nicefrac{1}{2}$ & & $\big(0,1\big)$ \\\cmidrule{1-7}
	\multirow{3}{*}{$0$} & $2$ & $2$ & $2$ & \multirow{3}{*}{$\big(0,0\big)$} & $\big(0,0\big)$ &  \multirow{3}{*}{$\bm{3}^{(0,0)}_{0}$} \\
	& $\nicefrac{3}{2}$ & $\nicefrac{3}{2}$ & $\nicefrac{3}{2}$ & & $\big(\nicefrac{1}{2},\nicefrac{1}{2}\big)$ & \\
	& $1$ & $1$ & $1$ & & $\big(0,1\big)$
  \end{tabular}
  \caption{Spectrum at the scalar origin for theory A at level 0.}
  \label{tab:specA0}
\end{table}


\begin{table}[p]
\centering
  \begin{tabular}{cccccccl}
	$\Delta_{\rm L}$ & $\Delta_{\rm R}$ & $\Delta$ & $s$ & $\SO(4)_{\rm L}$ & $\SO(4)_{\rm R}$ &  \\\cmidrule{1-7}\morecmidrules\cmidrule{1-7}
	$2$  &  0  &  2  &  $-2$  &  $(0,0)$  &  $(0,0)$  &  $\bm{1}_2^{(0,0)}$ \\\cmidrule{1-7}
	$1$  &  0  &  1  &  $-1$  &  $(0,1)$  &  $(0,0)$  &  $\bm{1}_1^{(0,1)}$ \\\cmidrule{1-7}
	\multirow{2}{*}{$2$} & $1$ & $3$ & $-1$ & \multirow{2}{*}{$\big(\nicefrac{1}{2},0\big)$} & $\big(\nicefrac{1}{2},0\big)$ &  \multirow{2}{*}{$\bm{2}^{(\nicefrac{1}{2},0)}_{2}$} \\
	& $\nicefrac{1}{2}$ & $\nicefrac{5}{2}$ & $-\nicefrac{3}{2}$ & & $\big(0,\nicefrac{1}{2}\big)$ & \\\cmidrule{1-7}
	\multirow{2}{*}{$1$} & $1$ & $2$ & $0$ & \multirow{2}{*}{$\big(\nicefrac{1}{2},1\big)$} & $\big(\nicefrac{1}{2},0\big)$ &  \multirow{2}{*}{$\bm{2}^{(\nicefrac{1}{2},1)}_{1}$} \\
	& $\nicefrac{1}{2}$ & $\nicefrac{3}{2}$ & $-\nicefrac{1}{2}$ & & $\big(0,\nicefrac{1}{2}\big)$ & \\\cmidrule{1-7}
	\multirow{3}{*}{$2$} & $2$ & $4$ & $0$ & \multirow{3}{*}{$\big(0,0\big)$} & $\big(0,0\big)$ &  \multirow{3}{*}{$\bm{3}^{(0,0)}_{2}$} \\
	& $\nicefrac{3}{2}$ & $\nicefrac{7}{2}$ & $-\nicefrac{1}{2}$ & & $\big(\nicefrac{1}{2},\nicefrac{1}{2}\big)$ & \\
	& $1$ & $3$ & $-1$ & & $\big(0,1\big)$ \\\cmidrule{1-7}
	\multirow{3}{*}{$1$} & $2$ & $3$ & $1$ & \multirow{3}{*}{$\big(0,1\big)$} & $\big(0,0\big)$ &  \multirow{3}{*}{$\bm{3}^{(0,1)}_{1}$} \\
	& $\nicefrac{3}{2}$ & $\nicefrac{5}{2}$ & $\nicefrac{1}{2}$ & & $\big(\nicefrac{1}{2},\nicefrac{1}{2}\big)$ & \\
	& $1$ & $2$ & $0$ & & $\big(0,1\big)$ \\\cmidrule{1-7}
	\multirow{3}{*}{$0$} & $2$ & $2$ & $2$ & \multirow{3}{*}{$\big(0,0\big)$} & $\big(0,0\big)$ &  \multirow{3}{*}{$\bm{3}^{(0,0)}_{0}$} \\
	& $\nicefrac{3}{2}$ & $\nicefrac{3}{2}$ & $\nicefrac{3}{2}$ & & $\big(\nicefrac{1}{2},\nicefrac{1}{2}\big)$ & \\
	& $1$ & $1$ & $1$ & & $\big(0,1\big)$
  \end{tabular}
  \caption{Spectrum at the scalar origin for theory \revA at level 0.}
  \label{tab:specrevA0}
\end{table}


\begin{table}
\centering
  \begin{tabular}{ccccccl}
	$\Delta_{\rm L}$ & $\Delta_{\rm R}$ & $\Delta$ & $s$ & $\SO(4)_{\rm L}$ & $\SO(4)_{\rm R}$ &  \\\cmidrule{1-7}\morecmidrules\cmidrule{1-7}
	\multirow{2}{*}{$\nicefrac{5}{2}$} & $1$ & $\nicefrac{7}{2}$ & $-\nicefrac{3}{2}$ & \multirow{2}{*}{$\big(0,\nicefrac{1}{2}\big)$} & $\big(\nicefrac{1}{2},0\big)$ &  \multirow{2}{*}{$\bm{2}^{(0,\nicefrac{1}{2})}_{\nicefrac{5}{2}}$} \\
	& $\nicefrac{1}{2}$ & $3$ & $-2$ & & $\big(0,\nicefrac{1}{2}\big)$ & \\\cmidrule{1-7}
	\multirow{2}{*}{$2$} & $1$ & $3$ & $-1$ & \multirow{2}{*}{$\big(\nicefrac{1}{2},0\big)$} & $\big(\nicefrac{1}{2},0\big)$ &  \multirow{2}{*}{$\bm{2}^{(\nicefrac{1}{2},0)}_{2}$} \\
	& $\nicefrac{1}{2}$ & $\nicefrac{5}{2}$ & $-\nicefrac{3}{2}$ & & $\big(0,\nicefrac{1}{2}\big)$ & \\\cmidrule{1-7}
	\multirow{2}{*}{$\nicefrac{3}{2}$} & $1$ & $\nicefrac{5}{2}$ & $-\nicefrac{1}{2}$ & \multirow{2}{*}{$\big(0,\nicefrac{3}{2}\big)$} & $\big(\nicefrac{1}{2},0\big)$ &  \multirow{2}{*}{$\bm{2}^{(0,\nicefrac{3}{2})}_{\nicefrac{3}{2}}$} \\
	& $\nicefrac{1}{2}$ & $2$ & $-1$ & & $\big(0,\nicefrac{1}{2}\big)$ & \\\cmidrule{1-7}
	\multirow{2}{*}{$1$} & $1$ & $2$ & $0$ & \multirow{2}{*}{$\big(\nicefrac{1}{2},1\big)$} & $\big(\nicefrac{1}{2},0\big)$ &  \multirow{2}{*}{$\bm{2}^{(\nicefrac{1}{2},1)}_{1}$} \\
	& $\nicefrac{1}{2}$ & $\nicefrac{3}{2}$ & $-\nicefrac{1}{2}$ & & $\big(0,\nicefrac{1}{2}\big)$ & \\\cmidrule{1-7}
	\multirow{2}{*}{$\nicefrac{1}{2}$} & $1$ & $\nicefrac{3}{2}$ & $\nicefrac{1}{2}$ & \multirow{2}{*}{$\big(0,\nicefrac{1}{2}\big)$} & $\big(\nicefrac{1}{2},0\big)$ &  \multirow{2}{*}{$\bm{2}^{(0,\nicefrac{1}{2})}_{\nicefrac{1}{2}}$} \\
	& $\nicefrac{1}{2}$ & $1$ & $0$ & & $\big(0,\nicefrac{1}{2}\big)$ & \\\cmidrule{1-7}
	\multirow{3}{*}{$\nicefrac{5}{2}$} & $\nicefrac{5}{2}$ & $5$ & $0$ & \multirow{3}{*}{$\big(0,\nicefrac{1}{2}\big)$} & $\big(0,\nicefrac{1}{2}\big)$ &  \multirow{3}{*}{$\bm{4}^{(0,\nicefrac{1}{2})}_{\nicefrac{5}{2}}$} \\
	& $2$ & $\nicefrac{9}{2}$ & $-\nicefrac{1}{2}$ & & $\big(\nicefrac{1}{2},1\big)$ & \\
	& $\nicefrac{3}{2}$ & $4$ & $-1$ & & $\big(0,\nicefrac{3}{2}\big)$ \\\cmidrule{1-7}
	\multirow{3}{*}{$2$} & $\nicefrac{5}{2}$ & $\nicefrac{9}{2}$ & $\nicefrac{1}{2}$ & \multirow{3}{*}{$\big(\nicefrac{1}{2},0\big)$} & $\big(0,\nicefrac{1}{2}\big)$ &  \multirow{3}{*}{$\bm{4}^{(\nicefrac{1}{2},0)}_{2}$} \\
	& $2$ & $4$ & $0$ & & $\big(\nicefrac{1}{2},1\big)$ & \\
	& $\nicefrac{3}{2}$ & $\nicefrac{7}{2}$ & $-\nicefrac{1}{2}$ & & $\big(0,\nicefrac{3}{2}\big)$ \\\cmidrule{1-7}
	\multirow{3}{*}{$\nicefrac{3}{2}$} & $\nicefrac{5}{2}$ & $4$ & $1$ & \multirow{3}{*}{$\big(0,\nicefrac{3}{2}\big)$} & $\big(0,\nicefrac{1}{2}\big)$ &  \multirow{3}{*}{$\bm{4}^{(0,\nicefrac{3}{2})}_{\nicefrac{3}{2}}$} \\
	& $2$ & $\nicefrac{7}{2}$ & $\nicefrac{1}{2}$ & & $\big(\nicefrac{1}{2},1\big)$ & \\
	& $\nicefrac{3}{2}$ & $3$ & $0$ & & $\big(0,\nicefrac{3}{2}\big)$ \\\cmidrule{1-7}
	\multirow{3}{*}{$1$} & $\nicefrac{5}{2}$ & $\nicefrac{7}{2}$ & $\nicefrac{3}{2}$ & \multirow{3}{*}{$\big(\nicefrac{1}{2},1\big)$} & $\big(0,\nicefrac{1}{2}\big)$ &  \multirow{3}{*}{$\bm{4}^{(\nicefrac{1}{2},1)}_{1}$} \\
	& $2$ & $3$ & $1$ & & $\big(\nicefrac{1}{2},1\big)$ & \\
	& $\nicefrac{3}{2}$ & $\nicefrac{5}{2}$ & $\nicefrac{1}{2}$ & & $\big(0,\nicefrac{3}{2}\big)$ \\\cmidrule{1-7}
	\multirow{3}{*}{$\nicefrac{1}{2}$} & $\nicefrac{5}{2}$ & $3$ & $2$ & \multirow{3}{*}{$\big(0,\nicefrac{1}{2}\big)$} & $\big(0,\nicefrac{1}{2}\big)$ &  \multirow{3}{*}{$\bm{4}^{(0,\nicefrac{1}{2})}_{\nicefrac{1}{2}}$} \\
	& $2$ & $\nicefrac{5}{2}$ & $\nicefrac{3}{2}$ & & $\big(\nicefrac{1}{2},1\big)$ & \\
	& $\nicefrac{3}{2}$ & $2$ & $1$ & & $\big(0,\nicefrac{3}{2}\big)$
  \end{tabular}
  \caption{Spectrum at the scalar origin for theory A at level 1.}
  \label{tab:specA1}
\end{table}


\begin{table}[p]
\centering
  \begin{tabular}{ccccccl}
	$\Delta_{\rm L}$ & $\Delta_{\rm R}$ & $\Delta$ & $s$ & $\SO(4)_{\rm L}$ & $\SO(4)_{\rm R}$ &  \\\cmidrule{1-7}\morecmidrules\cmidrule{1-7}
	\multirow{2}{*}{$\nicefrac{5}{2}$} & $1$ & $\nicefrac{7}{2}$ & $-\nicefrac{3}{2}$ & \multirow{2}{*}{$\big(0,\nicefrac{1}{2}\big)$} & $\big(\nicefrac{1}{2},0\big)$ &  \multirow{2}{*}{$\bm{2}^{(0,\nicefrac{1}{2})}_{\nicefrac{5}{2}}$} \\
	& $\nicefrac{1}{2}$ & $3$ & $-2$ & & $\big(0,\nicefrac{1}{2}\big)$ & \\\cmidrule{1-7}
	\multirow{2}{*}{$\nicefrac{3}{2}$} & $1$ & $\nicefrac{5}{2}$ & $-\nicefrac{1}{2}$ & \multirow{2}{*}{$\big(0,\nicefrac{3}{2}\big)$} & $\big(\nicefrac{1}{2},0\big)$ &  \multirow{2}{*}{$\bm{2}^{(0,\nicefrac{3}{2})}_{\nicefrac{3}{2}}$} \\
	& $\nicefrac{1}{2}$ & $2$ & $-1$ & & $\big(0,\nicefrac{1}{2}\big)$ & \\\cmidrule{1-7}
	\multirow{2}{*}{$\nicefrac{1}{2}$} & $1$ & $\nicefrac{3}{2}$ & $\nicefrac{1}{2}$ & \multirow{2}{*}{$\big(0,\nicefrac{1}{2}\big)$} & $\big(\nicefrac{1}{2},0\big)$ &  \multirow{2}{*}{$\bm{2}^{(0,\nicefrac{1}{2})}_{\nicefrac{1}{2}}$} \\
	& $\nicefrac{1}{2}$ & $1$ & $0$ & & $\big(0,\nicefrac{1}{2}\big)$ & \\\cmidrule{1-7}
	\multirow{3}{*}{$\nicefrac{5}{2}$} & $2$ & $\nicefrac{9}{2}$ & $-\nicefrac{1}{2}$ & \multirow{3}{*}{$\big(\nicefrac{1}{2},\nicefrac{1}{2}\big)$} & $\big(0,0\big)$ &  \multirow{3}{*}{$\bm{3}^{(\nicefrac{1}{2},\nicefrac{1}{2})}_{\nicefrac{5}{2}}$} \\
	& $\nicefrac{3}{2}$ & $4$ & $-1$ & & $\big(\nicefrac{1}{2},\nicefrac{1}{2}\big)$ & \\
	& $1$ & $\nicefrac{7}{2}$ & $-\nicefrac{3}{2}$ & & $\big(0,1\big)$\\\cmidrule{1-7}
	\multirow{3}{*}{$\nicefrac{3}{2}$} & $2$ & $\nicefrac{7}{2}$ & $\nicefrac{1}{2}$ & \multirow{3}{*}{$\big(\nicefrac{1}{2},\nicefrac{3}{2}\big)$} & $\big(0,0\big)$ &  \multirow{3}{*}{$\bm{3}^{(\nicefrac{1}{2},\nicefrac{3}{2})}_{\nicefrac{3}{2}}$} \\
	& $\nicefrac{3}{2}$ & $3$ & $0$ & & $\big(\nicefrac{1}{2},\nicefrac{1}{2}\big)$ & \\
	& $1$ & $\nicefrac{5}{2}$ & $-\nicefrac{1}{2}$ & & $\big(0,1\big)$\\\cmidrule{1-7}
	\multirow{3}{*}{$\nicefrac{1}{2}$} & $2$ & $\nicefrac{5}{2}$ & $\nicefrac{3}{2}$ & \multirow{3}{*}{$\big(\nicefrac{1}{2},\nicefrac{1}{2}\big)$} & $\big(0,0\big)$ &  \multirow{3}{*}{$\bm{3}^{(\nicefrac{1}{2},\nicefrac{1}{2})}_{\nicefrac{1}{2}}$} \\
	& $\nicefrac{3}{2}$ & $2$ & $1$ & & $\big(\nicefrac{1}{2},\nicefrac{1}{2}\big)$ & \\
	& $1$ & $\nicefrac{3}{2}$ & $\nicefrac{1}{2}$ & & $\big(0,1\big)$\\\cmidrule{1-7}
	\multirow{3}{*}{$\nicefrac{5}{2}$} & $\nicefrac{5}{2}$ & $5$ & $0$ & \multirow{3}{*}{$\big(0,\nicefrac{1}{2}\big)$} & $\big(0,\nicefrac{1}{2}\big)$ &  \multirow{3}{*}{$\bm{4}^{(0,\nicefrac{1}{2})}_{\nicefrac{5}{2}}$} \\
	& $2$ & $\nicefrac{9}{2}$ & $-\nicefrac{1}{2}$ & & $\big(\nicefrac{1}{2},1\big)$ & \\
	& $\nicefrac{3}{2}$ & $4$ & $-1$ & & $\big(0,\nicefrac{3}{2}\big)$ \\\cmidrule{1-7}
	\multirow{3}{*}{$\nicefrac{3}{2}$} & $\nicefrac{5}{2}$ & $4$ & $1$ & \multirow{3}{*}{$\big(0,\nicefrac{3}{2}\big)$} & $\big(0,\nicefrac{1}{2}\big)$ &  \multirow{3}{*}{$\bm{4}^{(0,\nicefrac{3}{2})}_{\nicefrac{3}{2}}$} \\
	& $2$ & $\nicefrac{7}{2}$ & $\nicefrac{1}{2}$ & & $\big(\nicefrac{1}{2},1\big)$ & \\
	& $\nicefrac{3}{2}$ & $3$ & $0$ & & $\big(0,\nicefrac{3}{2}\big)$ \\\cmidrule{1-7}
	\multirow{3}{*}{$\nicefrac{1}{2}$} & $\nicefrac{5}{2}$ & $3$ & $2$ & \multirow{3}{*}{$\big(0,\nicefrac{1}{2}\big)$} & $\big(0,\nicefrac{1}{2}\big)$ &  \multirow{3}{*}{$\bm{4}^{(0,\nicefrac{1}{2})}_{\nicefrac{1}{2}}$} \\
	& $2$ & $\nicefrac{5}{2}$ & $\nicefrac{3}{2}$ & & $\big(\nicefrac{1}{2},1\big)$ & \\
	& $\nicefrac{3}{2}$ & $2$ & $1$ & & $\big(0,\nicefrac{3}{2}\big)$
  \end{tabular}
  \caption{Spectrum at the scalar origin for theory \revA at level 1.}
  \label{tab:specrevA1}
\end{table}


\begin{table}[p]
\centering
\small
  \begin{tabular}{ccccccl}
	$\Delta_{\rm L}$ & $\Delta_{\rm R}$ & $\Delta$ & $s$ & $\SO(4)_{\rm L}$ & $\SO(4)_{\rm R}$ &  \\\cmidrule{1-7}\morecmidrules\cmidrule{1-7}
	\multirow{3}{*}{$\nicefrac{(n+4)}2$} & $\nicefrac{(n+2)}2$ & $n+3$ & $-1$ & \multirow{3}{*}{$(0,\nicefrac{n}2)$} & $(0,\nicefrac{(n-2)}2)$ & \multirow{3}{*}{$\bm{(n+1)}_{\nicefrac{(n+4)}2}^{(0,\nicefrac n2)}$} \\
	& $\nicefrac{(n+1)}2$ & $n+\nicefrac52$ & $-\nicefrac32$ & & $(\nicefrac{1}2,\nicefrac{(n-1)}2)$ & \\
	& $\nicefrac{n}2$ & $n+2$ & $-2$ & & $(0,\nicefrac{n}2)$\\\cmidrule{1-7}
	\multirow{3}{*}{$\nicefrac{(n+3)}2$} & $\nicefrac{(n+2)}2$ & $n+\nicefrac52$ & $-\nicefrac12$ & \multirow{3}{*}{$(\nicefrac12,\nicefrac{(n-1)}2)$} & $(0,\nicefrac{(n-2)}2)$ & \multirow{3}{*}{$\bm{(n+1)}_{\nicefrac{(n+3)}2}^{(\nicefrac12,\nicefrac {(n-1)}2)}$} \\
	& $\nicefrac{(n+1)}2$ & $n+2$ & $-1$ & & $(\nicefrac{1}2,\nicefrac{(n-1)}2)$ & \\
	& $\nicefrac{n}2$ & $n+\nicefrac32$ & $-\nicefrac32$ & & $(0,\nicefrac{n}2)$\\\cmidrule{1-7}
	\multirow{3}{*}{$\nicefrac{(n+2)}2$} & $\nicefrac{(n+2)}2$ & $n+2$ & $0$ & \multirow{3}{*}{$(0,\nicefrac{(n+2)}2)$} & $(0,\nicefrac{(n-2)}2)$ & \multirow{3}{*}{$\bm{(n+1)}_{\nicefrac{(n+2)}2}^{(0,\nicefrac {(n+2)}2)}$} \\
	& $\nicefrac{(n+1)}2$ & $n+\nicefrac32$ & $-\nicefrac12$ & & $(\nicefrac{1}2,\nicefrac{(n-1)}2)$ & \\
	& $\nicefrac{n}2$ & $n+1$ & $-1$ & & $(0,\nicefrac{n}2)$\\\cmidrule{1-7}
	\multirow{3}{*}{$\nicefrac{(n+2)}2$} & $\nicefrac{(n+2)}2$ & $n+2$ & $0$ & \multirow{3}{*}{$(0,\nicefrac{(n-2)}2)$} & $(0,\nicefrac{(n-2)}2)$ & \multirow{3}{*}{$\bm{(n+1)}_{\nicefrac{(n+2)}2}^{(0,\nicefrac {(n-2)}2)}$} \\
	& $\nicefrac{(n+1)}2$ & $n+\nicefrac32$ & $-\nicefrac12$ & & $(\nicefrac12,\nicefrac{(n-1)}2)$ & \\
	& $\nicefrac{n}2$ & $n+1$ & $-1$ & & $(0,\nicefrac{n}2)$\\\cmidrule{1-7}
	\multirow{3}{*}{$\nicefrac{(n+1)}2$} & $\nicefrac{(n+2)}2$ & $n+\nicefrac32$ & $\nicefrac12$ & \multirow{3}{*}{$(\nicefrac12,\nicefrac{(n+1)}2)$} & $(0,\nicefrac{(n-2)}2)$ & \multirow{3}{*}{$\bm{(n+1)}_{\nicefrac{(n+1)}2}^{(\nicefrac12,\nicefrac {(n+1)}2)}$} \\
	& $\nicefrac{(n+1)}2$ & $n+1$ & $0$ & & $(\nicefrac12,\nicefrac{(n-1)}2)$ & \\
	& $\nicefrac{n}2$ & $n+\nicefrac12$ & $-\nicefrac12$ & & $(0,\nicefrac{n}2)$\\\cmidrule{1-7}
	\multirow{3}{*}{$\nicefrac{n}2$} & $\nicefrac{(n+2)}2$ & $n+1$ & $1$ & \multirow{3}{*}{$(0,\nicefrac{n}2)$} & $(0,\nicefrac{(n-2)}2)$ & \multirow{3}{*}{$\bm{(n+1)}_{\nicefrac{n}2}^{(0,\nicefrac {n}2)}$} \\
	& $\nicefrac{(n+1)}2$ & $n+\nicefrac12$ & $\nicefrac12$ & & $(\nicefrac12,\nicefrac{(n-1)}2)$ & \\
	& $\nicefrac{n}2$ & $n$ & $0$ & & $(0,\nicefrac{n}2)$\\\cmidrule{1-7}
	\multirow{3}{*}{$\nicefrac{(n+4)}2$} & $\nicefrac{(n+4)}2$ & $n+4$ & $0$ & \multirow{3}{*}{$(0,\nicefrac{n}2)$} & $(0,\nicefrac{n}2)$ & \multirow{3}{*}{$\bm{(n+3)}_{\nicefrac{(n+4)}2}^{(0,\nicefrac n2)}$} \\
	& $\nicefrac{(n+3)}2$ & $n+\nicefrac72$ & $-\nicefrac12$ & & $(\nicefrac{1}2,\nicefrac{(n+1)}2)$ & \\
	& $\nicefrac{(n+2)}2$ & $n+3$ & $-1$ & & $(0,\nicefrac{(n+2)}2)$\\\cmidrule{1-7}
	\multirow{3}{*}{$\nicefrac{(n+3)}2$} & $\nicefrac{(n+4)}2$ & $n+\nicefrac72$ & $\nicefrac12$ & \multirow{3}{*}{$(\nicefrac12,\nicefrac{(n-1)}2)$} & $(0,\nicefrac n2)$ & \multirow{3}{*}{$\bm{(n+3)}_{\nicefrac{(n+3)}2}^{(\nicefrac12,\nicefrac {(n-1)}2)}$} \\
	& $\nicefrac{(n+3)}2$ & $n+3$ & $0$ & & $(\nicefrac{1}2,\nicefrac{(n+1)}2)$ & \\
	& $\nicefrac{(n+2)}2$ & $n+\nicefrac52$ & $-\nicefrac12$ & & $(0,\nicefrac{(n+2)}2)$\\\cmidrule{1-7}
	\multirow{3}{*}{$\nicefrac{(n+2)}2$} & $\nicefrac{(n+4)}2$ & $n+3$ & $1$ & \multirow{3}{*}{$(0,\nicefrac{(n+2)}2)$} & $(0,\nicefrac n2)$ & \multirow{3}{*}{$\bm{(n+3)}_{\nicefrac{(n+2)}2}^{(0,\nicefrac {(n+2)}2)}$} \\
	& $\nicefrac{(n+3)}2$ & $n+\nicefrac52$ & $\nicefrac12$ & & $(\nicefrac{1}2,\nicefrac{(n+1)}2)$ & \\
	& $\nicefrac{(n+2)}2$ & $n+2$ & $0$ & & $(0,\nicefrac{(n+2)}2)$\\\cmidrule{1-7}
	\multirow{3}{*}{$\nicefrac{(n+2)}2$} & $\nicefrac{(n+4)}2$ & $n+3$ & $1$ & \multirow{3}{*}{$(0,\nicefrac {(n-2)}2)$} & $(0,\nicefrac n2)$ & \multirow{3}{*}{$\bm{(n+3)}_{\nicefrac{(n+2)}2}^{(0,\nicefrac {(n-2)}2)}$} \\
	& $\nicefrac{(n+3)}2$ & $n+\nicefrac52$ & $\nicefrac12$ & & $(\nicefrac12,\nicefrac{(n+1)}2)$ & \\
	& $\nicefrac{(n+2)}2$ & $n+2$ & $0$ & & $(0,\nicefrac{(n+2)}2)$\\\cmidrule{1-7}
	\multirow{3}{*}{$\nicefrac{(n+1)}2$} & $\nicefrac{(n+4)}2$ & $n+\nicefrac52$ & $\nicefrac32$ & \multirow{3}{*}{$(\nicefrac12,\nicefrac{(n+1)}2)$} & $(0,\nicefrac n2)$ & \multirow{3}{*}{$\bm{(n+3)}_{\nicefrac{(n+1)}2}^{(\nicefrac12,\nicefrac {(n+1)}2)}$} \\
	& $\nicefrac{(n+3)}2$ & $n+2$ & $1$ & & $(\nicefrac12,\nicefrac{(n+1)}2)$ & \\
	& $\nicefrac{(n+2)}2$ & $n+\nicefrac32$ & $\nicefrac12$ & & $(0,\nicefrac{(n+2)}2)$\\\cmidrule{1-7}
	\multirow{3}{*}{$\nicefrac{n}2$} & $\nicefrac{(n+4)}2$ & $n+2$ & $2$ & \multirow{3}{*}{$(0,\nicefrac{n}2)$} & $(0,\nicefrac n2)$ & \multirow{3}{*}{$\bm{(n+3)}_{\nicefrac{n}2}^{(0,\nicefrac {n}2)}$} \\
	& $\nicefrac{(n+3)}2$ & $n+\nicefrac32$ & $\nicefrac32$ & & $(\nicefrac12,\nicefrac{(n+1)}2)$ & \\
	& $\nicefrac{(n+2)}2$ & $n+1$ & $1$ & & $(0,\nicefrac{(n+2)}2)$\\
  \end{tabular}
  \caption{Spectrum at the scalar origin for theory A at level $n\geq2$.}
  \label{tab:specAn}
\end{table}


\begin{table}[p]
\centering
\small
  \begin{tabular}{ccccccl}
	$\Delta_{\rm L}$ & $\Delta_{\rm R}$ & $\Delta$ & $s$ & $\SO(4)_{\rm L}$ & $\SO(4)_{\rm R}$ &\\\cmidrule{1-7}\morecmidrules\cmidrule{1-7}
	\multirow{3}{*}{$\nicefrac{(n+4)}{2}$} & $\nicefrac{(n+2)}{2}$ & $n+3$ & $-1$ & \multirow{3}{*}{$\big(0,\nicefrac{n}{2}\big)$} & $\big(0,\nicefrac{(n-2)}{2}\big)$ &  \multirow{3}{*}{$\bm{(n+1)}^{(0,\nicefrac{n}{2})}_{\nicefrac{(n+4)}{2}}$} \\
	& $\nicefrac{(n+1)}{2}$ & $n+\nicefrac{5}{2}$ & $-\nicefrac{3}{2}$ & & $\big(\nicefrac{1}{2},\nicefrac{(n-1)}{2}\big)$ & \\
	& $\nicefrac{n}{2}$ & $n+2$ & $-2$ & & $\big(0,\nicefrac{n}{2}\big)$\\\cmidrule{1-7}
	\multirow{3}{*}{$\nicefrac{(n+2)}{2}$} & $\nicefrac{(n+2)}{2}$ & $n+2$ & $0$ & \multirow{3}{*}{$\big(0,\nicefrac{(n-2)}{2}\big)$} & $\big(0,\nicefrac{(n-2)}{2}\big)$ &  \multirow{3}{*}{$\bm{(n+1)}^{(0,\nicefrac{(n-2)}{2})}_{\nicefrac{(n+2)}{2}}$} \\
	& $\nicefrac{(n+1)}{2}$ & $n+\nicefrac{3}{2}$ & $-\nicefrac{1}{2}$ & & $\big(\nicefrac{1}{2},\nicefrac{(n-1)}{2}\big)$ & \\
	& $\nicefrac{n}{2}$ & $n+1$ & $-1$ & & $\big(0,\nicefrac{n}{2}\big)$\\\cmidrule{1-7}
	\multirow{3}{*}{$\nicefrac{(n+2)}{2}$} & $\nicefrac{(n+2)}{2}$ & $n+2$ & $0$ & \multirow{3}{*}{$\big(0,\nicefrac{(n+2)}{2}\big)$} & $\big(0,\nicefrac{(n-2)}{2}\big)$ &  \multirow{3}{*}{$\bm{(n+1)}^{(0,\nicefrac{(n+2)}{2})}_{\nicefrac{(n+2)}{2}}$} \\
	& $\nicefrac{(n+1)}{2}$ & $n+\nicefrac{3}{2}$ & $-\nicefrac{1}{2}$ & & $\big(\nicefrac{1}{2},\nicefrac{(n-1)}{2}\big)$ & \\
	& $\nicefrac{n}{2}$ & $n+1$ & $-1$ & & $\big(0,\nicefrac{n}{2}\big)$\\\cmidrule{1-7}
	\multirow{3}{*}{$\nicefrac{n}{2}$} & $\nicefrac{(n+2)}{2}$ & $n+1$ & $1$ & \multirow{3}{*}{$\big(0,\nicefrac{n}{2}\big)$} & $\big(0,\nicefrac{(n-2)}{2}\big)$ &  \multirow{3}{*}{$\bm{(n+1)}^{(0,\nicefrac{n}{2})}_{\nicefrac{n}{2}}$} \\
	& $\nicefrac{(n+1)}{2}$ & $n+\nicefrac{1}{2}$ & $\nicefrac{1}{2}$ & & $\big(\nicefrac{1}{2},\nicefrac{(n-1)}{2}\big)$ & \\
	& $\nicefrac{n}{2}$ & $n$ & $0$ & & $\big(0,\nicefrac{n}{2}\big)$\\\cmidrule{1-7}
	\multirow{3}{*}{$\nicefrac{(n+4)}{2}$} & $\nicefrac{(n+3)}{2}$ & $n+\nicefrac{7}{2}$ & $-\nicefrac{1}{2}$ & \multirow{3}{*}{$\big(\nicefrac{1}{2},\nicefrac{n}{2}\big)$} & $\big(0,\nicefrac{(n-1)}{2}\big)$ &  \multirow{3}{*}{$\bm{(n+2)}^{(\nicefrac{1}{2},\nicefrac{n}{2})}_{\nicefrac{(n+4)}{2}}$} \\
	& $\nicefrac{(n+2)}{2}$ & $n+3$ & $-1$ & & $\big(\nicefrac{1}{2},\nicefrac{n}{2}\big)$ & \\
	& $\nicefrac{(n+1)}{2}$ & $n+\nicefrac{5}{2}$ & $-\nicefrac{3}{2}$ & & $\big(0,\nicefrac{(n+1)}{2}\big)$\\\cmidrule{1-7}
	\multirow{3}{*}{$\nicefrac{(n+2)}{2}$} & $\nicefrac{(n+3)}{2}$ & $n+\nicefrac{5}{2}$ & $\nicefrac{1}{2}$ & \multirow{3}{*}{$\big(\nicefrac{1}{2},\nicefrac{(n-2)}{2}\big)$} & $\big(0,\nicefrac{(n-1)}{2}\big)$ &  \multirow{3}{*}{$\bm{(n+2)}^{(\nicefrac{1}{2},\nicefrac{(n-2)}{2})}_{\nicefrac{(n+2)}{2}}$} \\
	& $\nicefrac{(n+2)}{2}$ & $n+2$ & $0$ & & $\big(\nicefrac{1}{2},\nicefrac{n}{2}\big)$ & \\
	& $\nicefrac{(n+1)}{2}$ & $n+\nicefrac{3}{2}$ & $-\nicefrac{1}{2}$ & & $\big(0,\nicefrac{(n+1)}{2}\big)$\\\cmidrule{1-7}
	\multirow{3}{*}{$\nicefrac{(n+2)}{2}$} & $\nicefrac{(n+3)}{2}$ & $n+\nicefrac{5}{2}$ & $\nicefrac{1}{2}$ & \multirow{3}{*}{$\big(\nicefrac{1}{2},\nicefrac{(n+2)}{2}\big)$} & $\big(0,\nicefrac{(n-1)}{2}\big)$ &  \multirow{3}{*}{$\bm{(n+2)}^{(\nicefrac{1}{2},\nicefrac{(n+2)}{2})}_{\nicefrac{(n+2)}{2}}$} \\
	& $\nicefrac{(n+2)}{2}$ & $n+2$ & $0$ & & $\big(\nicefrac{1}{2},\nicefrac{n}{2}\big)$ & \\
	& $\nicefrac{(n+1)}{2}$ & $n+\nicefrac{3}{2}$ & $-\nicefrac{1}{2}$ & & $\big(0,\nicefrac{(n+1)}{2}\big)$\\\cmidrule{1-7}
	\multirow{3}{*}{$\nicefrac{n}{2}$} & $\nicefrac{(n+3)}{2}$ & $n+\nicefrac{3}{2}$ & $\nicefrac{3}{2}$ & \multirow{3}{*}{$\big(\nicefrac{1}{2},\nicefrac{n}{2}\big)$} & $\big(0,\nicefrac{(n-1)}{2}\big)$ &  \multirow{3}{*}{$\bm{(n+2)}^{(\nicefrac{1}{2},\nicefrac{n}{2})}_{\nicefrac{n}{2}}$} \\
	& $\nicefrac{(n+2)}{2}$ & $n+1$ & $1$ & & $\big(\nicefrac{1}{2},\nicefrac{n}{2}\big)$ & \\
	& $\nicefrac{(n+1)}{2}$ & $n+\nicefrac{1}{2}$ & $\nicefrac{1}{2}$ & & $\big(0,\nicefrac{(n+1)}{2}\big)$\\\cmidrule{1-7}
	\multirow{3}{*}{$\nicefrac{(n+4)}{2}$} & $\nicefrac{(n+4)}{2}$ & $n+4$ & $0$ & \multirow{3}{*}{$\big(0,\nicefrac{n}{2}\big)$} & $\big(0,\nicefrac{n}{2}\big)$ &  \multirow{3}{*}{$\bm{(n+3)}^{(0,\nicefrac{n}{2})}_{\nicefrac{(n+4)}{2}}$} \\
	& $\nicefrac{(n+3)}{2}$ & $n+\nicefrac{7}{2}$ & $-\nicefrac{1}{2}$ & & $\big(\nicefrac{1}{2},\nicefrac{(n+1)}{2}\big)$ & \\
	& $\nicefrac{(n+2)}{2}$ & $n+3$ & $-1$ & & $\big(0,\nicefrac{(n+2)}{2}\big)$\\\cmidrule{1-7}
	\multirow{3}{*}{$\nicefrac{(n+2)}{2}$} & $\nicefrac{(n+4)}{2}$ & $n+3$ & $1$ & \multirow{3}{*}{$\big(0,\nicefrac{(n-2)}{2}\big)$} & $\big(0,\nicefrac{n}{2}\big)$ &  \multirow{3}{*}{$\bm{(n+3)}^{(0,\nicefrac{(n-2)}{2})}_{\nicefrac{(n+2)}{2}}$} \\
	& $\nicefrac{(n+3)}{2}$ & $n+\nicefrac{5}{2}$ & $\nicefrac{1}{2}$ & & $\big(\nicefrac{1}{2},\nicefrac{(n+1)}{2}\big)$ & \\
	& $\nicefrac{(n+2)}{2}$ & $n+2$ & $0$ & & $\big(0,\nicefrac{(n+2)}{2}\big)$\\\cmidrule{1-7}
	\multirow{3}{*}{$\nicefrac{(n+2)}{2}$} & $\nicefrac{(n+4)}{2}$ & $n+3$ & $1$ & \multirow{3}{*}{$\big(0,\nicefrac{(n+2)}{2}\big)$} & $\big(0,\nicefrac{n}{2}\big)$ &  \multirow{3}{*}{$\bm{(n+3)}^{(0,\nicefrac{(n+2)}{2})}_{\nicefrac{(n+2)}{2}}$} \\
	& $\nicefrac{(n+3)}{2}$ & $n+\nicefrac{5}{2}$ & $\nicefrac{1}{2}$ & & $\big(\nicefrac{1}{2},\nicefrac{(n+1)}{2}\big)$ & \\
	& $\nicefrac{(n+2)}{2}$ & $n+2$ & $0$ & & $\big(0,\nicefrac{(n+2)}{2}\big)$\\\cmidrule{1-7}
	\multirow{3}{*}{$\nicefrac{n}{2}$} & $\nicefrac{(n+4)}{2}$ & $n+2$ & $2$ & \multirow{3}{*}{$\big(0,\nicefrac{n}{2}\big)$} & $\big(0,\nicefrac{n}{2}\big)$ &  \multirow{3}{*}{$\bm{(n+3)}^{(0,\nicefrac{n}{2})}_{\nicefrac{n}{2}}$} \\
	& $\nicefrac{(n+3)}{2}$ & $n+\nicefrac{3}{2}$ & $\nicefrac{3}{2}$ & & $\big(\nicefrac{1}{2},\nicefrac{(n+1)}{2}\big)$ & \\
	& $\nicefrac{(n+2)}{2}$ & $n+1$ & $1$ & & $\big(0,\nicefrac{(n+2)}{2}\big)$
  \end{tabular}
  \caption{Spectrum at the scalar origin for theory \revA at level $n\geq2$.}
  \label{tab:specrevAn}
\end{table}

\clearpage

\section{Lowest KK levels at supersymmetric locus}
\label{app:tablessusy}

The tables in this appendix show the  content of the spectrum in \eqref{eq: spectrumsusyfamily} for levels $n=1$ in table~\ref{tab:multsusyeta1} and $n=2$ in table~\ref{tab:multsusyeta2}.

\begin{table}[H]
 \renewcommand{\arraystretch}{1.4}
 \centering
 \centerline{\begin{tabular}{cccccc}
	&$\Delta_{\rm L}$ & $\Delta_{\rm R}$ & $\Delta$ & $s$ & $\Big(\SO(3)_{\rm diag}\times\SO(3)_{\rm R}\Big)^{\SO(2)_{\rm L}}$ \\\cmidrule{1-6}\morecmidrules\cmidrule{1-6}
	\multirow{5}{*}{$\big[0,\tfrac12\big]^{\pm1}_{(3+\Gamma^{(1,\pm 1)})/2}$} & \multirow{5}{*}{$\nicefrac{(3+\Gamma^{(1,\pm 1)})}{2}$} & $\nicefrac{(3+\Gamma^{(1,\pm 1)})}{2}$ & $3+\Gamma^{(1,\pm 1)}$ & $0$ & $\big(0,\nicefrac{1}{2}\big)^{\pm1}$ \\
	&& $\nicefrac{(2+\Gamma^{(1,\pm 1)})}{2}$ & $\dfrac{5}{2}+\Gamma^{(1,\pm 1)}$ & $-\nicefrac{1}{2}$ & $\big(\nicefrac{1}{2},0\big)^{\pm1}+\big(\nicefrac{1}{2},1\big)^{\pm1}$  \\
	&& $\nicefrac{(1+\Gamma^{(1,\pm 1)})}{2}$ & $2+\Gamma^{(1,\pm 1)}$ & $-1$ & $\big(0,\nicefrac{3}{2}\big)^{\pm1}+\big(0,\nicefrac{1}{2}\big)^{\pm1}+\big(1,\nicefrac{1}{2}\big)^{\pm1}$  \\
	&& $\nicefrac{\Gamma^{(1,\pm 1)}}{2}$ & $\dfrac{3}{2}+\Gamma^{(1,\pm 1)}$ & $-\nicefrac{3}{2}$ & $\big(\nicefrac{1}{2},0\big)^{\pm1}+\big(\nicefrac{1}{2},1\big)^{\pm1}$  \\
	&& $\nicefrac{(-1+\Gamma^{(1,\pm 1)})}{2}$ & $1+\Gamma^{(1,\pm 1)}$ & $-2$ & $\big(0,\nicefrac{1}{2}\big)^{\pm1}$\\\cmidrule{1-6}
	\multirow{5}{*}{$\big[0,\tfrac12\big]^{\pm1}_{\tfrac{-1+\Gamma^{(1,\pm 1)}}{2}}$}&\multirow{5}{*}{$\nicefrac{(-1+\Gamma^{(1,\pm 1)})}{2}$} & $\nicefrac{(3+\Gamma^{(1,\pm 1)})}{2}$ & $1+\Gamma^{(1,\pm 1)}$ & $2$ & $\big(0,\nicefrac{1}{2}\big)^{\pm1}$ \\
	&& $\nicefrac{(2+\Gamma^{(1,\pm 1)})}{2}$ & $\dfrac{1}{2}+\Gamma^{(1,\pm 1)}$ & $\nicefrac{3}{2}$ & $\big(\nicefrac{1}{2},0\big)^{\pm1}+\big(\nicefrac{1}{2},1\big)^{\pm1}$  \\
	&& $\nicefrac{(1+\Gamma^{(1,\pm 1)})}{2}$ & $\Gamma^{(1,\pm 1)}$ & $1$ & $\big(0,\nicefrac{3}{2}\big)^{\pm1}+\big(0,\nicefrac{1}{2}\big)^{\pm1}+\big(1,\nicefrac{1}{2}\big)^{\pm1}$  \\
	&& $\nicefrac{\Gamma^{(1,\pm 1)}}{2}$ & $-\dfrac{1}{2}+\Gamma^{(1,\pm 1)}$ & $\nicefrac{1}{2}$ & $\big(\nicefrac{1}{2},0\big)^{\pm1}+\big(\nicefrac{1}{2},1\big)^{\pm1}$  \\
	&& $\nicefrac{(-1+\Gamma^{(1,\pm 1)})}{2}$ & $-1+\Gamma^{(1,\pm 1)}$ & $0$ & $\big(0,\nicefrac{1}{2}\big)^{\pm1}$\\\cmidrule{1-6}
	\multirow{5}{*}{$\big[0,\tfrac12\big]^{\pm1}_{\tfrac{1+\Gamma^{(1,\pm 1)}}{2}}$}&\multirow{5}{*}{$\nicefrac{(1+\Gamma^{(1,\pm 1)})}{2}$} & $\nicefrac{(3+\Gamma^{(1,\pm 1)})}{2}$ & $2+\Gamma^{(1,\pm 1)}$ & $1$ & $\big(0,\nicefrac{1}{2}\big)^{\pm1}$ \\
	&& $\nicefrac{(2+\Gamma^{(1,\pm 1)})}{2}$ & $\dfrac{3}{2}+\Gamma^{(1,\pm 1)}$ & $\nicefrac{1}{2}$ & $\big(\nicefrac{1}{2},0\big)^{\pm1}+\big(\nicefrac{1}{2},1\big)^{\pm1}$  \\
	&& $\nicefrac{(1+\Gamma^{(1,\pm 1)})}{2}$ & $1+\Gamma^{(1,\pm 1)}$ & $0$ & $\big(0,\nicefrac{3}{2}\big)^{\pm1}+\big(0,\nicefrac{1}{2}\big)^{\pm1}+\big(1,\nicefrac{1}{2}\big)^{\pm1}$  \\
	&& $\nicefrac{\Gamma^{(1,\pm 1)}}{2}$ & $\dfrac{1}{2}+\Gamma^{(1,\pm 1)}$ & $-\nicefrac{1}{2}$ & $\big(\nicefrac{1}{2},0\big)^{\pm1}+\big(\nicefrac{1}{2},1\big)^{\pm1}$  \\
	&& $\nicefrac{(-1+\Gamma^{(1,\pm 1)})}{2}$ & $\Gamma^{(1,\pm 1)}$ & $-1$ & $\big(0,\nicefrac{1}{2}\big)^{\pm1}$\\\cmidrule{1-6}
	\multirow{5}{*}{$\big[0,\tfrac12\big]^{\pm3}_{\tfrac{1+\Gamma^{(1,\pm 3)}}{2}}$}&\multirow{5}{*}{$\nicefrac{(1+\Gamma^{(1,\pm 3)})}{2}$} & $\nicefrac{(3+\Gamma^{(1,\pm 3)})}{2}$ & $2+\Gamma^{(1,\pm 3)}$ & $1$ & $\big(0,\nicefrac{1}{2}\big)^{\pm3}$ \\
	&& $\nicefrac{(2+\Gamma^{(1,\pm 3)})}{2}$ & $\dfrac{3}{2}+\Gamma^{(1,\pm 3)}$ & $\nicefrac{1}{2}$ & $\big(\nicefrac{1}{2},0\big)^{\pm3}+\big(\nicefrac{1}{2},1\big)^{\pm3}$  \\
	&& $\nicefrac{(1+\Gamma^{(1,\pm 3)})}{2}$ & $1+\Gamma^{(1,\pm 3)}$ & $0$ & $\big(0,\nicefrac{3}{2}\big)^{\pm3}+\big(0,\nicefrac{1}{2}\big)^{\pm3}+\big(1,\nicefrac{1}{2}\big)^{\pm3}$  \\
	&& $\nicefrac{\Gamma^{(1,\pm 3)}}{2}$ & $\dfrac{1}{2}+\Gamma^{(1,\pm 3)}$ & $-\nicefrac{1}{2}$ & $\big(\nicefrac{1}{2},0\big)^{\pm3}+\big(\nicefrac{1}{2},1\big)^{\pm3}$  \\
	&& $\nicefrac{(-1+\Gamma^{(1,\pm 3)})}{2}$ & $\Gamma^{(1,\pm 3)}$ & $-1$ & $\big(0,\nicefrac{1}{2}\big)^{\pm3}$
  \end{tabular}}
  \caption{Full spectrum at level 1 for the family of supersymmetric vacua satisfying \eqref{eq:susylocus}, with $\Gamma^{(1,\pm 1)}=\sqrt{3+e^{2\omega}}$ and $\Gamma^{(1,\pm 3)}=\sqrt{-5+9\,e^{2\omega}}$. The ${\rm SL}(2,\mathbb{R})_{\rm R}$-dimension of the superconformal primaries, $h$, follow from \eqref{eq:lowestdelta0} and are omitted.}
  \label{tab:multsusyeta1}
\end{table}

\begin{table}[H]
\vspace*{-1.5cm}
 \renewcommand{\arraystretch}{1.2}
 \footnotesize
 \centering
 \centerline{\begin{tabular}{cccccc}
	&$\Delta_{\rm L}$ & $\Delta_{\rm R}$ & $\Delta$ & $s$ & $\Big(\SO(3)_{\rm diag}\times\SO(3)_{\rm R}\Big)^{\SO(2)_{\rm L}}$ \\\cmidrule{1-6}\morecmidrules\cmidrule{1-6}
	\multirow{5}{*}{$\big[0,1\big]^{0}_{3}$} & \multirow{5}{*}{$3$} & $3$ & $6$ & $0$ & $\big(0,1\big)^{0}$ \\
	&& $\nicefrac{5}{2}$ & $\nicefrac{11}{2}$ & $-\nicefrac{1}{2}$ & $\big(\nicefrac{1}{2},\nicefrac{1}{2}\big)^{0}+\big(\nicefrac{1}{2},\nicefrac{3}{2}\big)^{0}$  \\
	&& $2$ & $5$ & $-1$ & $\big(0,0\big)^{0}+\big(0,1\big)^{0}+\big(1,1\big)^{0}+\big(0,2\big)^{0}$  \\
	&& $\nicefrac{3}{2}$ & $\nicefrac{9}{2}$ & $-\nicefrac{3}{2}$ & $\big(\nicefrac{1}{2},\nicefrac{1}{2}\big)^{0}+\big(\nicefrac{1}{2},\nicefrac{3}{2}\big)^{0}$  \\
	&& $1$ & $4$ & $-2$ & $\big(0,1\big)^{0}$\\\cmidrule{1-6}
	\multirow{5}{*}{$\big[0,1\big]^{\pm2}_{(3+\Gamma^{(2,\pm 2)})/2}$} & \multirow{5}{*}{$\nicefrac{(3+\Gamma^{(2,\pm 2)})}{2}$} & $\nicefrac{(3+\Gamma^{(2,\pm 2)})}{2}$ & $3+\Gamma^{(2,\pm 2)}$ & $0$ & $\big(0,1\big)^{\pm2}$ \\
	&& $\nicefrac{(2+\Gamma^{(2,\pm 2)})}{2}$ & $\dfrac{5}{2}+\Gamma^{(2,\pm 2)}$ & $-\nicefrac{1}{2}$ & $\big(\nicefrac{1}{2},\nicefrac{1}{2}\big)^{\pm2}+\big(\nicefrac{1}{2},\nicefrac{3}{2}\big)^{\pm2}$  \\
	&& $\nicefrac{(1+\Gamma^{(2,\pm 2)})}{2}$ & $2+\Gamma^{(2,\pm 2)}$ & $-1$ & $\big(0,0\big)^{\pm2}+\big(0,1\big)^{\pm2}+\big(1,1\big)^{\pm2}+\big(0,2\big)^{\pm2}$  \\
	&& $\nicefrac{\Gamma^{(2,\pm 2)}}{2}$ & $\dfrac{3}{2}+\Gamma^{(2,\pm 2)}$ & $-\nicefrac{3}{2}$ & $\big(\nicefrac{1}{2},\nicefrac{1}{2}\big)^{\pm2}+\big(\nicefrac{1}{2},\nicefrac{3}{2}\big)^{\pm2}$  \\
	&& $\nicefrac{(-1+\Gamma^{(2,\pm 2)})}{2}$ & $1+\Gamma^{(2,\pm 2)}$ & $-2$ & $\big(0,1\big)^{\pm2}$\\\cmidrule{1-6}
	\multirow{5}{*}{$\big[0,1\big]^{0}_1$}&\multirow{5}{*}{$1$} & $3$ & $4$ & $2$ & $\big(0,1\big)^{0}$ \\
	&& $\nicefrac{5}{2}$ & $\nicefrac{7}{2}$ & $\nicefrac{3}{2}$ & $\big(\nicefrac{1}{2},\nicefrac{1}{2}\big)^{0}+\big(\nicefrac{1}{2},\nicefrac{3}{2}\big)^{0}$  \\
	&& $2$ & $3$ & $1$ & $\big(0,0\big)^{0}+\big(0,1\big)^{0}+\big(1,1\big)^{0}+\big(0,2\big)^{0}$  \\
	&& $\nicefrac{3}{2}$ & $\nicefrac{5}{2}$ & $\nicefrac{1}{2}$ & $\big(\nicefrac{1}{2},\nicefrac{1}{2}\big)^{0}+\big(\nicefrac{1}{2},\nicefrac{3}{2}\big)^{0}$  \\
	&& $1$ & $2$ & $0$ & $\big(0,1\big)^{0}$\\\cmidrule{1-6}
	\multirow{5}{*}{$\big[0,1\big]^{\pm2}_{\left(-1+\Gamma^{(2,\pm 2)}\right)/2}$}&\multirow{5}{*}{$\nicefrac{(-1+\Gamma^{(2,\pm 2)})}{2}$} & $\nicefrac{(3+\Gamma^{(2,\pm 2)})}{2}$ & $1+\Gamma^{(2,\pm 2)}$ & $2$ & $\big(0,1\big)^{\pm2}$ \\
	&& $\nicefrac{(2+\Gamma^{(2,\pm 2)})}{2}$ & $\dfrac{1}{2}+\Gamma^{(2,\pm 2)}$ & $\nicefrac{3}{2}$ & $\big(\nicefrac{1}{2},\nicefrac{1}{2}\big)^{\pm2}+\big(\nicefrac{1}{2},\nicefrac{3}{2}\big)^{\pm2}$  \\
	&& $\nicefrac{(1+\Gamma^{(2,\pm 2)})}{2}$ & $\Gamma^{(2,\pm 2)}$ & $1$ & $\big(0,0\big)^{\pm2}+\big(0,1\big)^{\pm2}+\big(1,1\big)^{\pm2}+\big(0,2\big)^{\pm2}$  \\
	&& $\nicefrac{\Gamma^{(2,\pm 2)}}{2}$ & $-\dfrac{1}{2}+\Gamma^{(2,\pm 2)}$ & $\nicefrac{1}{2}$ & $\big(\nicefrac{1}{2},\nicefrac{1}{2}\big)^{\pm2}+\big(\nicefrac{1}{2},\nicefrac{3}{2}\big)^{\pm2}$  \\
	&& $\nicefrac{(-1+\Gamma^{(2,\pm 2)})}{2}$ & $-1+\Gamma^{(2,\pm 2)}$ & $0$ & $\big(0,1\big)^{\pm2}$\\\cmidrule{1-6}
	\multirow{5}{*}{$\big[0,1\big]^{0}_2$}&\multirow{5}{*}{$2$} & $3$ & $5$ & $1$ & $\big(0,1\big)^{0}$ \\
	&& $\nicefrac{5}{2}$ & $\nicefrac{5}{2}$ & $\nicefrac{1}{2}$ & $\big(\nicefrac{1}{2},\nicefrac{1}{2}\big)^{0}+\big(\nicefrac{1}{2},\nicefrac{3}{2}\big)^{0}$  \\
	&& $2$ & $4$ & $0$ & $\big(0,0\big)^{0}+\big(0,1\big)^{0}+\big(1,1\big)^{0}+\big(0,2\big)^{0}$  \\
	&& $\nicefrac{3}{2}$ & $\nicefrac{5}{2}$ & $-\nicefrac{1}{2}$ & $\big(\nicefrac{1}{2},\nicefrac{1}{2}\big)^{0}+\big(\nicefrac{1}{2},\nicefrac{3}{2}\big)^{0}$  \\
	&& $1$ & $3$ & $-1$ & $\big(0,1\big)^{0}$\\\cmidrule{1-6}
	\multirow{5}{*}{$\big[0,1\big]^{\pm 2}_{\left(1+\Gamma^{(2,\pm 2)}\right)/2}$}&\multirow{5}{*}{$\nicefrac{(1+\Gamma^{(2,\pm 2)})}{2}$} & $\nicefrac{(3+\Gamma^{(2,\pm 2)})}{2}$ & $2+\Gamma^{(2,\pm 2)}$ & $1$ & $\big(0,1\big)^{\pm 2}$ \\
	&& $\nicefrac{(2+\Gamma^{(2,\pm 2)})}{2}$ & $\dfrac{3}{2}+\Gamma^{(2,\pm 2)}$ & $\nicefrac{1}{2}$ & $\big(\nicefrac{1}{2},\nicefrac{1}{2}\big)^{\pm 2}+\big(\nicefrac{1}{2},\nicefrac{3}{2}\big)^{\pm 2}$  \\
	&& $\nicefrac{(1+\Gamma^{(2,\pm 2)})}{2}$ & $1+\Gamma^{(2,\pm 2)}$ & $0$ & $\big(0,0\big)^{\pm 2}+\big(0,1\big)^{\pm 2}+\big(1,1\big)^{\pm 2}+\big(0,2\big)^{\pm 2}$  \\
	&& $\nicefrac{\Gamma^{(2,\pm 2)}}{2}$ & $\dfrac{1}{2}+\Gamma^{(2,\pm 2)}$ & $-\nicefrac{1}{2}$ & $\big(\nicefrac{1}{2},\nicefrac{1}{2}\big)^{\pm 2}+\big(\nicefrac{1}{2},\nicefrac{3}{2}\big)^{\pm 2}$  \\
	&& $\nicefrac{(-1+\Gamma^{(2,\pm 2)})}{2}$ & $\Gamma^{(2,\pm 2)}$ & $-1$ & $\big(0,1\big)^{\pm 2}$\\\cmidrule{1-6}
	\multirow{5}{*}{$\big[0,1\big]^{\pm 4}_{\left(1+\Gamma^{(2,\pm 4)}\right)/2}$}&\multirow{5}{*}{$\nicefrac{(1+\Gamma^{(2,\pm 4)})}{2}$} & $\nicefrac{(3+\Gamma^{(2,\pm 4)})}{2}$ & $2+\Gamma^{(2,\pm 4)}$ & $1$ & $\big(0,1\big)^{\pm 4}$ \\
	&& $\nicefrac{(2+\Gamma^{(2,\pm 4)})}{2}$ & $\dfrac{3}{2}+\Gamma^{(2,\pm 4)}$ & $\nicefrac{1}{2}$ & $\big(\nicefrac{1}{2},\nicefrac{1}{2}\big)^{\pm 4}+\big(\nicefrac{1}{2},\nicefrac{3}{2}\big)^{\pm 4}$  \\
	&& $\nicefrac{(1+\Gamma^{(2,\pm 4)})}{2}$ & $1+\Gamma^{(2,\pm 4)}$ & $0$ & $\big(0,0\big)^{\pm 4}+\big(0,1\big)^{\pm 4}+\big(1,1\big)^{\pm 4}+\big(0,2\big)^{\pm 4}$  \\
	&& $\nicefrac{\Gamma^{(2,\pm 4)}}{2}$ & $\dfrac{1}{2}+\Gamma^{(2,\pm 4)}$ & $-\nicefrac{1}{2}$ & $\big(\nicefrac{1}{2},\nicefrac{1}{2}\big)^{\pm 4}+\big(\nicefrac{1}{2},\nicefrac{3}{2}\big)^{\pm 4}$  \\
	&& $\nicefrac{(-1+\Gamma^{(2,\pm 4)})}{2}$ & $\Gamma^{(2,\pm 4)}$ & $-1$ & $\big(0,1\big)^{\pm 4}$\\\cmidrule{1-6}
	\multirow{5}{*}{$\big[0,1\big]^{0}_2$}&\multirow{5}{*}{$2$} & $3$ & $5$ & $1$ & $\big(0,1\big)^{0}$ \\
	&& $\nicefrac{5}{2}$ & $\nicefrac{9}{2}$ & $\nicefrac{1}{2}$ & $\big(\nicefrac{1}{2},\nicefrac{1}{2}\big)^{0}+\big(\nicefrac{1}{2},\nicefrac{3}{2}\big)^{0}$  \\
	&& $2$ & $4$ & $0$ & $\big(0,0\big)^{0}+\big(0,1\big)^{0}+\big(1,1\big)^{0}+\big(0,2\big)^{0}$  \\
	&& $\nicefrac{3}{2}$ & $\nicefrac{7}{2}$ & $-\nicefrac{1}{2}$ & $\big(\nicefrac{1}{2},\nicefrac{1}{2}\big)^{0}+\big(\nicefrac{1}{2},\nicefrac{3}{2}\big)^{0}$  \\
	&& $1$ & $3$ & $-1$ & $\big(0,1\big)^{0}$
  \end{tabular}}
  \caption{Full spectrum at level 2 for the family of supersymmetric vacua satisfying \eqref{eq:susylocus}. The ${\rm SL}(2,\mathbb{R})_{\rm R}$-dimension of the superconformal primaries, $h$, follow from \eqref{eq:lowestdelta0} and are omitted. The multiplets with zero SO(2)$_{\rm L}$-charge are understood as a shorthand for the combination in \eqref{eq: breakingrule}.}
  \label{tab:multsusyeta2}
\end{table}


\providecommand{\href}[2]{#2}\begingroup\raggedright\endgroup

\end{document}